\documentclass[letterpaper,11pt]{article}
\pdfoutput=1
\usepackage{jheppub}
\usepackage{mathtools,amssymb,amsmath,braket,mathrsfs,tikz,subfig,xspace,xcolor,float,color,siunitx,comment,afterpage,multirow,array,hhline,amsthm,graphbox,url}
\usepackage[utf8]{inputenc}

\DeclareRobustCommand{\Sec}[1]{Sec.~\ref{#1}}

\DeclareRobustCommand{\App}[1]{App.~\ref{#1}}
\DeclareRobustCommand{\Tab}[1]{Table~\ref{#1}}

\DeclareRobustCommand{\Fig}[1]{Fig.~\ref{#1}}
\DeclareRobustCommand{\Figs}[2]{Figs.~\ref{#1} and \ref{#2}}

\DeclareRobustCommand{\Eq}[1]{Eq.~(\ref{#1})}
\DeclareRobustCommand{\Eqs}[2]{Eqs.~(\ref{#1}) and (\ref{#2})}

\DeclareRobustCommand{\Ref}[1]{Ref.~\cite{#1}}
\DeclareRobustCommand{\Refs}[1]{Refs.~\cite{#1}}

\newtheorem*{obdecomp}{Observable Decomposition}
\newtheorem*{ircsafedecomp}{IRC-Safe Observable Decomposition}
\newtheorem*{dsthm}{Deep Sets Theorem~\cite{\deepsets}}

\newcommand{\pythia}{\textsc{Pythia}\xspace}
\newcommand{\deepsets}{DBLP:conf/nips/ZaheerKRPSS17}

\definecolor{jdtcolor}{rgb}{0.8392,0.1529,0.1569}
\definecolor{emmcolor}{rgb}{0.1725,0.6275,0.1725}
\definecolor{ptkcolor}{rgb}{0.1216,0.4666,0.7059}

\title{Energy Flow Networks: Deep Sets for Particle Jets}

\author{Patrick T. Komiske,}
\author{Eric M. Metodiev,}
\author{and Jesse Thaler}

\affiliation{Center for Theoretical Physics, Massachusetts Institute of Technology, \\ 77 Massachusetts Avenue, Cambridge, MA 02139, U.S.A.\vspace{1mm}}
\affiliation{Department of Physics, Harvard University, \\ 17 Oxford Street, Cambridge, MA 02138, U.S.A.}

\emailAdd{pkomiske@mit.edu}
\emailAdd{metodiev@mit.edu}
\emailAdd{jthaler@mit.edu}

\preprint{MIT--CTP 5064}

\abstract{
A key question for machine learning approaches in particle physics is how to best represent and learn from collider events.
As an event is intrinsically a variable-length unordered set of particles, we build upon recent machine learning efforts to learn directly from sets of features or ``point clouds''.
Adapting and specializing the ``Deep Sets'' framework to particle physics, we introduce Energy Flow Networks, which respect infrared and collinear safety by construction.
We also develop Particle Flow Networks, which allow for general energy dependence and the inclusion of additional particle-level information such as charge and flavor.
These networks feature a per-particle internal (latent) representation, and summing over all particles yields an overall event-level latent representation.
We show how this latent space decomposition unifies existing event representations based on detector images and radiation moments.
To demonstrate the power and simplicity of this set-based approach, we apply these networks to the collider task of discriminating quark jets from gluon jets, finding similar or improved performance compared to existing methods.
We also show how the learned event representation can be directly visualized, providing insight into the inner workings of the model.
These architectures lend themselves to efficiently processing and analyzing events for a wide variety of tasks at the Large Hadron Collider. 
Implementations and examples of our architectures are available online in our \href{https://energyflow.network}{\tt EnergyFlow} package.
}

\begin{document}

\flushbottom
\maketitle

\clearpage

\section{Introduction}

Collisions at accelerators like the Large Hadron Collider (LHC) produce multitudes of particles.
Particles are the fundamental objects of interest in collider physics and provide an interface between theoretical calculations and experimental measurements, often reconstructed experimentally via ``particle flow'' algorithms~\cite{Beaudette:2014cea,Sirunyan:2017ulk,Aaboud:2017aca}.
Analyses of collider data rely on observables to distill these complicated multiparticle events and capture essential aspects of the underlying physics.
Because each collision event consists of a variable-length list of particles with no intrinsic ordering, collider observables must be sensibly defined as functions of \emph{sets} of particles.
In this paper, we develop a novel architecture for processing and learning from collider events in their natural set-based representation.

Recently, modern machine learning techniques have been used to achieve excellent performance on a variety of collider tasks by learning specalized functions of the events, which can be viewed as observables in their own right.
For instance, hadronic jet classification has been thoroughly studied using low-level~\cite{Cogan:2014oua,deOliveira:2015xxd,Baldi:2016fql,Barnard:2016qma,Komiske:2016rsd,ATL-PHYS-PUB-2017-017,Kasieczka:2017nvn,Bhimji:2017qvb,Macaluso:2018tck,Guo:2018hbv,Dreyer:2018nbf,Guest:2016iqz,Louppe:2017ipp,Cheng:2017rdo,Egan:2017ojy,Fraser:2018ieu,Almeida:2015jua,Pearkes:2017hku,Butter:2017cot,Roxlo:2018adx} and high-level~\cite{Datta:2017rhs,Aguilar-Saavedra:2017rzt,Luo:2017ncs,Moore:2018lsr,Datta:2017lxt,Komiske:2017aww} input observables.
Additional tasks include the removal of pileup~\cite{Komiske:2017ubm}, model-independent new physics searches~\cite{Collins:2018epr,DAgnolo:2018cun,DeSimone:2018efk,Hajer:2018kqm,Farina:2018fyg,Heimel:2018mkt}, constraining effective field theories~\cite{Brehmer:2018kdj,Brehmer:2018eca,DHondt:2018cww}, probabilistic and generative modeling of physics processes~\cite{deOliveira:2017pjk,Paganini:2017hrr,deOliveira:2017rwa,Paganini:2017dwg,Andreassen:2018apy}, and enhancing existing physics analyses~\cite{Baldi:2014kfa,Baldi:2014pta,Searcy:2015apa,Santos:2016kno,Barberio:2017ngd,Duarte:2018ite,Abdughani:2018wrw,Lin:2018cin,Lai:2018ixk}.
See \Refs{Larkoski:2017jix,Guest:2018yhq,Albertsson:2018maf,Radovic:2018dip,Sadowski2018} for more detailed reviews of machine learning in high-energy physics.

Two key choices must be made when using machine learning for a collider task: how to represent the event and how to analyze that representation.
These choices are often made together, with examples from collider physics including calorimeter images paired with convolutional neural networks (CNNs)~\cite{deOliveira:2015xxd,Baldi:2016fql,Barnard:2016qma,Komiske:2016rsd,ATL-PHYS-PUB-2017-017,Kasieczka:2017nvn,Bhimji:2017qvb,Macaluso:2018tck,Guo:2018hbv,Dreyer:2018nbf}, particle lists paired with recurrent/recursive neural networks (RNNs)~\cite{Guest:2016iqz,Louppe:2017ipp,Cheng:2017rdo,Egan:2017ojy,Fraser:2018ieu}, collections of ordered inputs paired with dense neural networks (DNNs)~\cite{Almeida:2015jua,Pearkes:2017hku,Butter:2017cot,Roxlo:2018adx,Datta:2017rhs,Aguilar-Saavedra:2017rzt,Luo:2017ncs,Moore:2018lsr}, and Energy Flow Polynomials (EFPs) paired with linear methods~\cite{Komiske:2017aww}.
One lesson that emerges from this body of work is that any two sufficiently general models, given access to complete enough information, achieve similar performance.
In light of this, criteria such as understandability of the model and closeness to theoretical and experimental constructs are of central importance.

Given that events are fundamentally sets of particles, particle-level inputs such as those used in \Refs{Guest:2016iqz,Louppe:2017ipp,Cheng:2017rdo,Egan:2017ojy,Fraser:2018ieu,Almeida:2015jua,Pearkes:2017hku,Butter:2017cot} are a desirable way of representing an event for use in a model.
That said, RNNs and DNNs, the two architectures typically used with particle-level inputs, each fail to be fully satisfactory methods for processing events:  DNNs because they require a fixed-size input and RNNs because they are explicitly dependent on the ordering of the inputs.
Though ad hoc workarounds for these problems exist, such as zero padding for DNNs or ordering particles by their transverse momentum ($p_T$) or clustering history for RNNs, an ideal architecture would manifestly respect the permutation symmetry of the problem.
Such an architecture would be able to handle variable-length inputs while being inherently symmetric with respect to the ordering of the input particles.

The machine learning community has recently developed (and continues to explore) technology which is ideally suited for representing sets of objects for a model~\cite{DBLP:conf/acl/IyyerMBD15,DBLP:conf/cvpr/QiSMG17,DBLP:conf/iccv/RezatofighiGMAD17,DBLP:conf/nips/QiYSG17,DBLP:conf/nips/ZaheerKRPSS17,DBLP:journals/corr/abs-1709-03019,DBLP:journals/corr/abs-1712-07262,DBLP:conf/aaai/RezatofighiMSD018,DBLP:journals/corr/abs-1805-00613,DBLP:journals/corr/abs-1806-00050}.
One context where this appears is learning from \emph{point clouds}, sets of data points in space.
For instance, the output of spatial sensors such as lidar, relevant for self-driving car technologies, is often in the form of a point cloud.
As point clouds share the variable-length and permutation-symmetric properties with collider events, it is worthwhile to understand and expand upon point cloud techniques for particle physics applications.

The Deep Sets framework for point clouds, recently developed in \Ref{DBLP:conf/nips/ZaheerKRPSS17}, demonstrates how permutation-invariant functions of variable-length inputs can be parametrized in a fully general way.
In \Ref{DBLP:conf/nips/ZaheerKRPSS17}, the method was applied to a wide variety of problems including red-shift estimation of galaxy clusters, finding terms associated with a set of words, and detecting anomalous faces in a set of images.
The key observation is that \emph{summation}, which is clearly symmetric with respect to the order of the arguments, is general enough to encapsulate all symmetric functions if one is allowed a large enough internal (\emph{latent}) space.

\begin{figure}[t]
\centering
\includegraphics[width=0.9\columnwidth]{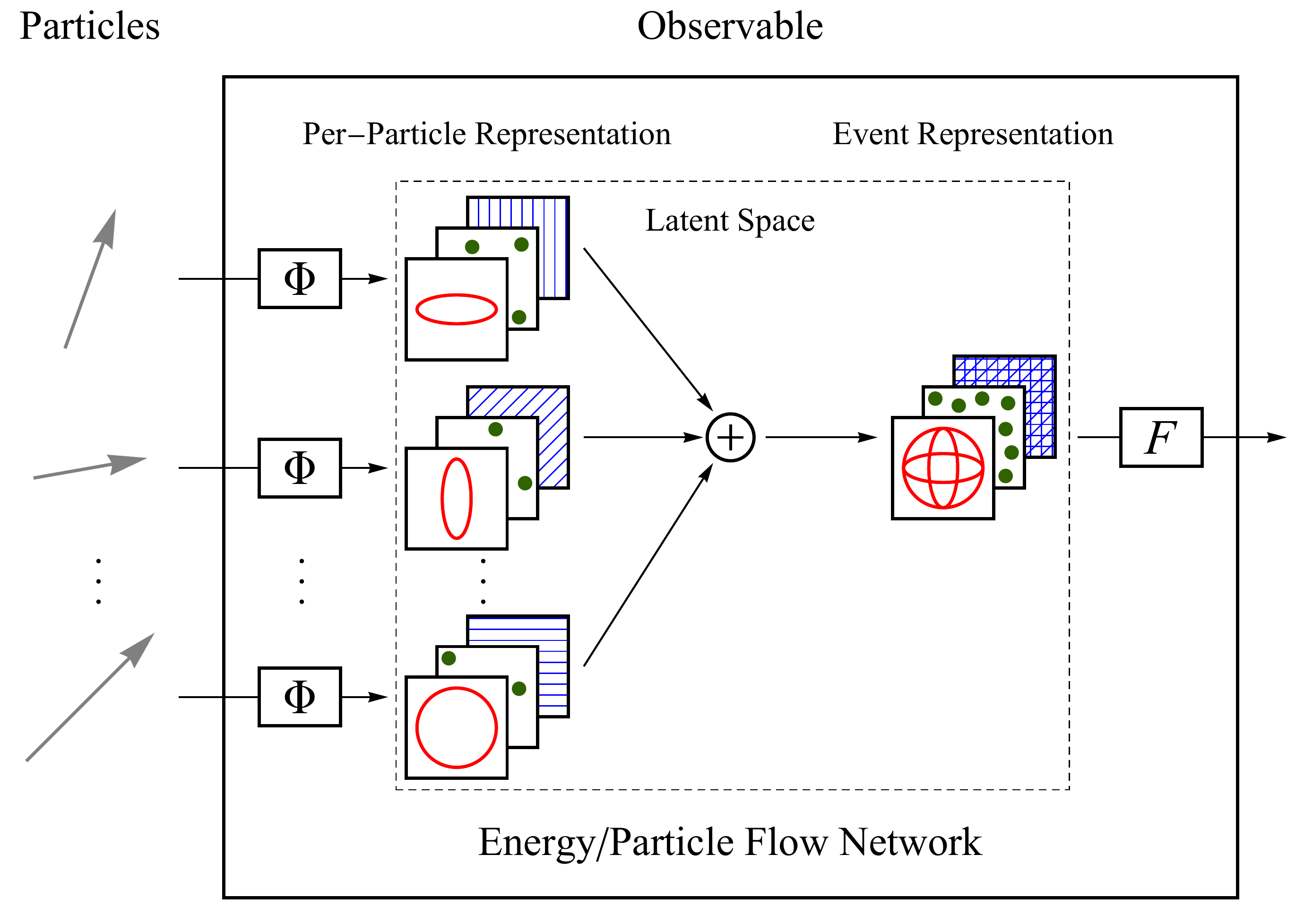}
\caption{
A visualization of the decomposition of an observable via \Eq{eq:obdecomp}.
Each particle in the event is mapped by $\Phi$ to an internal (latent) particle representation, shown here as three abstract illustrations for a latent space of dimension three.
The latent representation is then summed over all particles to arrive at a latent event representation, which is mapped by $F$ to the value of the observable.
For the IRC-safe case of \Eq{eq:ircsafeobdecomp}, $\Phi$ takes in the angular information of the particle and the sum is weighted by the particle energies or transverse momenta.}
\label{fig:diagram}
\end{figure}

In the context of a physics observable $\mathcal O$ that is a symmetric function of an arbitrary number of particles each with $d$ features, the result from \Ref{DBLP:conf/nips/ZaheerKRPSS17} can be stated as:

\begin{obdecomp}
An observable $\mathcal O$ can be approximated arbitrarily well as:
\begin{equation}
\label{eq:obdecomp}
\mathcal O(\{p_1,\ldots,p_M\}) = F\left(\sum_{i=1}^M \Phi (p_i)\right),
\end{equation}
where $\Phi: \mathbb R^d \to \mathbb R^\ell$ is a per-particle mapping and $F:\mathbb R^\ell\to \mathbb R$ is a continuous function.
\end{obdecomp}
\noindent A schematic representation of \Eq{eq:obdecomp} is shown in \Fig{fig:diagram}.
Inherent in the decomposition of \Eq{eq:obdecomp} is a latent space of dimension $\ell$ that serves to embed the particles such that an overall latent event representation is obtained when the sum is carried out.
One should think of the $d$ features for each particle as possibly being kinematic information, such as the particle's $p_T$, rapidity $y$, and azimuthal angle $\phi$, or other quantum numbers such as the particle's charge or flavor.
\Sec{sec:math} contains additional mathematical details regarding this decomposition.

With a suitable modification of \Eq{eq:obdecomp}, we can restrict the decomposition to infrared- and collinear-safe (IRC-safe) observables:
\begin{ircsafedecomp}
An IRC-safe observable $\mathcal O$ can be approximated arbitrarily well as:
\begin{equation}
\label{eq:ircsafeobdecomp}
\mathcal O(\{p_1,\ldots,p_M\}) = F\left(\sum_{i=1}^M z_i\Phi (\hat p_i)\right),
\end{equation}
where $z_i$ is the energy (or $p_T$) and $\hat p_i$ the angular information of particle $i$.
\end{ircsafedecomp}
\noindent
The energy-weighting factors $z_i$ as well as the energy-independent $\hat p_i$ in \Eq{eq:ircsafeobdecomp} ensure that the event representation in the latent space is IRC-safe.

In this paper, we show that many common observables are naturally encompassed by simple choices of $\Phi$ and $F$ from \Eqs{eq:obdecomp}{eq:ircsafeobdecomp}.
Furthermore, we can parametrize $\Phi$ and $F$ by neural network layers, capable of learning essentially any function, in order to explore more complicated observables.
In keeping with the naming convention of \Ref{Komiske:2017aww} for methods involving IRC-safe observables, we term a network architecture implementing \Eq{eq:ircsafeobdecomp} an \emph{Energy Flow Network} (EFN).
By contrast, we refer to the more general case of an architecture that implements \Eq{eq:obdecomp} as a \emph{Particle Flow Network} (PFN).
These two network architectures can be mathematically summarized as:
\begin{align}
&\text{EFN:}\quad F\left(\sum_{i=1}^M z_i \Phi(\hat p_i)\right), &\text{PFN:}\quad F\left(\sum_{i=1}^M \Phi(p_i)\right).
\end{align}
Our framework manifestly respects the variable length and permutation invariance of particle sets, achieves performance competitive with existing techniques on key collider tasks, and provides a platform for visualizing the information learned by the model.
Beyond this, we demonstrate how our framework unifies the existing event representations of calorimeter images and radiation moments, and we showcase the extraction of novel analytic observables from the trained model.

One ever-present collider phenomenon that involves complicated multiparticle final states is the formation and observation of \emph{jets}, sprays of color-neutral hadrons resulting from the fragmentation of high-energy quarks and gluons in quantum chromodynamics (QCD).
Numerous individual observables have been proposed to study jets including the jet mass, constituent multiplicity, image activity~\cite{Pumplin:1991kc}, $N$-subjettiness~\cite{Thaler:2010tr,Thaler:2011gf}, track-based observables~\cite{Krohn:2012fg,Chang:2013rca}, generalized angularities~\cite{Larkoski:2014pca}, (generalized) energy correlation functions~\cite{Larkoski:2013eya,Moult:2016cvt}, soft drop multiplicity~\cite{Larkoski:2014wba,Frye:2017yrw}, and many more (see \Refs{Abdesselam:2010pt,Altheimer:2012mn,Altheimer:2013yza,Adams:2015hiv,Larkoski:2017jix,Asquith:2018igt} for reviews).
Machine learning methods have found tremendous applicability to jet classification tasks, greatly outperforming individual standard observables.
Jet classification provides an ideal case study for the Deep Sets method in a collider setting since jets, like events, are fundamentally variably sized and invariant under reorderings of their constituents.

Many existing collider observables ranging from $e^+e^-$ event shapes to jet substructure observables naturally fit into the decomposition of \Eq{eq:obdecomp}.
Observables that are defined directly in terms of the particles themselves (i.e.\ not algorithmically) can often be exactly encompassed.
Several examples of such observables are summarized in \Tab{tab:sumobs}, with the associated functions $\Phi$ and $F$ listed for each observable.
The fact that the decomposition holds exactly in these familiar cases indicates that the Observable Decomposition indeed captures an essential aspect of particle-level collider observables.

\begin{table}[t]
\centering
\begin{tabular}{|ll|l|l|}
\hline
\bf Observable $\mathcal O$\hspace{1mm} & & \bf Map $\Phi$ & \bf Function $F$ \\
\hline \hline
Mass & $m$  & $p^\mu$ & $F(x^\mu) = \sqrt{x^\mu x_\mu}$ \\ 
Multiplicity & $M$ & $1$ & $F(x) = x$ \\
Track Mass & $m_\text{track}$ & $p^\mu \mathbb I_\text{track}$ & $F(x^\mu) = \sqrt{x^\mu x_\mu}$ \\
Track Multiplicity & $M_\text{track}$ & $\mathbb I_\text{track}$ & $F(x) = x$ \\
Jet Charge~\cite{Krohn:2012fg} & $\mathcal Q_\kappa$ & $(p_{T}, Q\, p_T^\kappa)$ & $F(x,y) = y/x^\kappa$ \\
Eventropy~\cite{Larkoski:2014pca} & $z \ln z$ & $(p_{T}, p_{T} \ln p_{T})$ & $F(x,y) = y/x - \ln x$\\
Momentum Dispersion~\cite{CMS:2013kfa} & $p_T^D$ & $(p_{T}, p_{T}^2)$ & $F(x,y) = \sqrt{y/x^2} $\\
$C$ parameter~\cite{Parisi:1978eg} & $C$ & $(|\vec p\,|, \vec p\otimes\vec p/|\vec p\,|)$ & $F(x, Y) = \frac{3}{2x^2}[(\text{Tr}\, Y)^2 - \text{Tr}\, Y^2]$ \\
\hline
\end{tabular}
\caption{
A variety of common collider observables decomposed into per-particle maps $\Phi$ and functions $F$ according to \Eq{eq:obdecomp}. 
Here $\mathbb I_\text{track}$ is an indicator function over charged tracks.
In the last column, the arguments of $F$ are placeholders for the summed output of $\Phi$.
}
\label{tab:sumobs}
\end{table}

To showcase the efficacy of EFNs and PFNs, we apply them to the task of distinguishing light-quark jets from gluon jets~\cite{Gallicchio:2011xq,Gallicchio:2012ez,Aad:2014gea,Gras:2017jty}, finding that they achieve excellent classification performance.
In general, the PFN model outperforms the EFN model, indicating that IRC-unsafe information is helpful for discriminating quark and gluon jets.
%
%
Additionally, including particle identification information improves the classification performance of the PFN.
It would be interesting to apply all of these methods in a fully-data driven way~\cite{Metodiev:2017vrx,Komiske:2018oaa,Komiske:2018vkc} to test these conclusions beyond the limited scope of parton shower generators.

One fascinating aspect of EFNs is that they enable a natural visualization of the learned latent space, providing insights as to what exactly the machine is learning.
In particular, since the function $\Phi$ of an EFN typically takes the two-dimensional angular information of a particle as input, this two-dimensional space is easily visualized.
In the context of quark/gluon discrimination, we observe that the EFN learns a latent representation that ``pixelates'' the rapidity-azimuth plane, dynamically sizing the pixels to be smaller near the core of the jet and larger farther out.
We also find qualitative and quantitative evidence that the EFN has in a sense ``understood'' the collinear singularity structure of QCD.

The rest of this paper is organized as follows.
\Sec{sec:math} provides a detailed mathematical discussion of the observable decompositions and explores \Eqs{eq:obdecomp}{eq:ircsafeobdecomp} in the context of specific observables and event representations. 
\Sec{sec:implement} discusses the implementation details of our EFN and PFN architectures, with other models discussed in \App{app:models}.
\Sec{sec:qg} contains the case study discriminating quark- and gluon-initiated jets and demonstrates our new techniques for visualizing and analyzing the learned information.
Conclusions are presented in \Sec{sec:conc}.
A supplementary top jet tagging study is presented in \App{app:toptag}, and additional visualizations of the models are provided in \App{app:addvis}. 
The EFN and PFN architectures are available online as part of our \href{https://energyflow.network}{\tt EnergyFlow} package~\cite{energyflow} along with example code.

\section{A general framework for observables}
\label{sec:math}

Events consist of variable numbers of particles with no intrinsic ordering, so observables are described mathematically as functions of sets of particles.
Such a mathematical formulation allows for a direct exploration of the space of observables.
For instance, \Ref{Komiske:2017aww} exploited IRC safety to construct a linear approximating basis of all IRC-safe observables.
Here, we treat the entire space of observables (both with and without IRC safety), using their mathematical structure to arrive at a general decomposition relevant for theoretically categorizing observables as well as developing machine learning techniques.

\subsection{Observables as functions of sets of particles}
\label{sec:funcofsets}

The key mathematical fact that we exploit, due to \Ref{DBLP:conf/nips/ZaheerKRPSS17}, is that a generic function of a set of objects can be decomposed to arbitrarily good approximation in a practical and intuitive way.
We state this result explicitly below:
\begin{dsthm}
\label{thm:ds}
Let $\mathfrak X\subset \mathbb{R}^d$ be compact, $X\subset 2^{\mathfrak X}$ be the space of sets with bounded cardinality of elements in $\mathfrak X$, and $Y\subset\mathbb{R}$ be a bounded interval.
Consider a continuous function $f:X\to Y$ that is invariant under permutations of its inputs, i.e.\ $f(x_1,\ldots,x_M)=f(x_{\pi(1)},\ldots,x_{\pi(M)})$ for all $x_i\in\mathfrak X$ and $\pi\in S_M$.
Then there exists a sufficiently large integer $\ell$ and continuous functions $\Phi:\mathfrak X\to \mathbb{R}^\ell$, $F:\mathbb{R}^\ell\to Y$ such that the following holds to an arbitrarily good approximation:\footnote{It is formally necessary to restrict the domains and ranges of the functions to be compact because the proof of the Deep Sets Theorem, given fully in \Ref{DBLP:conf/nips/ZaheerKRPSS17}, makes use of the Stone-Weierstrass polynomial approximation theorem~\cite{Stone:1948gen}, which applies for compact spaces. After the expansion in polynomials of the features, the result follows by careful application of the fundamental theorem of symmetric polynomials.}
\begin{equation}\label{eq:dsthm}
f(\{x_1,\ldots,x_M\}) = F\left(\sum_{i=1}^M \Phi(x_i)\right).
\end{equation}
\end{dsthm}
\noindent We only rely on the Deep Sets Theorem to justify the generality of \Eq{eq:dsthm}, which can otherwise be regarded as an interesting, manifestly permutation-invariant parameterization.

The Deep Sets Theorem can be immediately applied to the collider physics context where observables are viewed as functions of sets of particles.
We denote an event with $M$ particles as $\{p_i\}_{i=1}^M$, where $p_i\in \mathbb{R}^d$ contains the relevant attributes of particle $i$ (momentum, charge, flavor, etc.).
Phrased in the collider physics language, it states that an observable $\mathcal O$ can be approximated arbitrarily well as:
\begin{equation}
\label{eq:obdecomp2}
\mathcal O(\{p_1,\ldots,p_M\}) = F\left(\sum_{i=1}^M \Phi (p_i)\right),
\end{equation}
where $\Phi: \mathbb R^d\to \mathbb R^\ell$ is a per-particle mapping and $F:\mathbb R^\ell\to Y$ is a continuous function.
This provides a mathematical justification for the Observable Decomposition stated in \Eq{eq:obdecomp}.

The content of the Observable Decomposition is that any observable can be viewed as linearly summing over the particles in some internal space and then mapping the result to an output space.
We refer to $\mathbb R^\ell$ as the \emph{latent space} and each component of the per-particle mapping $\Phi(p_i)$ as a \emph{filter}.
The latent space could be, for example, the pixel values of a detector image or a moment decomposition of the radiation pattern.
Summing $\Phi(p_i)$ over the particles induces a latent description of the entire event, which is mapped by the function $F$ to the value of the observable.

\subsection{Enforcing infrared and collinear safety}

We can formulate the Observable Decomposition specifically for a class of observables of particular theoretical interest, namely IRC-safe observables~\cite{Kinoshita:1962ur,Lee:1964is,sterman1995handbook,Weinberg:1995mt}.
IRC safety corresponds to robustness of the observable under collinear splittings of a particle or additions of soft particles, which makes the observable tractable in perturbative quantum field theory as well as robust to experimental resolution effects.

Remarkably, building IRC safety into the latent representation simply corresponds to energy-weighting the contributions of each particle and restricting $\Phi$ to only depend on the particle geometry $\hat p_i$.
The energy-weighting $z_i$ and geometry $\hat p_i$ for particle $i$ depends on the collider context.
At an $e^+e^-$ collider, it is natural to take $z_i = E_i$ and $\hat p_i = p^\mu_i/E_i$, where $E_i$ is the energy and $p^\mu_i$ the four-momentum.
At a hadron collider, it is natural to take $z_i = p_{T,i}$ and $\hat p_i=(y_i,\phi_i)$, where $p_{T,i}$ is the transverse momentum, $y_i$ is the rapidity, and $\phi_i$ the azimuthal angle.\footnote{As discussed in \Ref{Komiske:2017aww}, another sensible choice for the angular measure is $\hat p_i = p^\mu_i/p_{T,i}$. Particle mass information, if present, can be passed to a PFN via flavor information.}
In practice, we typically focus on dimensionless observables and use the appropriate normalized weights: $z_i=E_i/\sum_jE_j$  or $z_i=p_{T,i}/\sum_jp_{T,j}$.

Any IRC-safe observable $\mathcal O$ can be approximated arbitrarily well by the decomposition:
\begin{equation}
\label{eq:ircsafeobdecomp2}
\mathcal O\left(\{p_i\}_{i=1}^M\right) = F\left(\sum_{i=1}^M z_i\, \Phi (\hat p_i)\right),
\end{equation}
where $\Phi: \mathbb R^d \to \mathbb R^\ell$ is a per-particle angular mapping and $F:\mathbb R^\ell\to \mathbb R$ is continuous.
All observables of the form in \Eq{eq:ircsafeobdecomp2} are manifestly IRC safe due to the energy-weighted linear sum structure, the dependence of $\Phi$ on purely geometric inputs $\hat p_i$, and the fact that continuous functions of IRC-safe observables are IRC safe.\footnote{Ratios of IRC-safe observables are not necessarily IRC safe~\cite{Larkoski:2013paa,Larkoski:2015lea} since division is discontinuous at zero.}

The fact that the energy-weighted decomposition in \Eq{eq:ircsafeobdecomp2} suffices to approximate all IRC-safe observables is intuitive from the fact that a continuous function of a sufficiently high-resolution calorimeter image can be used to approximate an IRC-safe observable arbitrarily well~\cite{Tkachov:1995kk,Sveshnikov:1995vi,Cherzor:1997ak}.
As discussed in \Sec{sec:encompass}, an image of the calorimeter deposits is exactly encompassed by the energy-weighted observable decomposition.

Here, we provide a direct argument to arrive at \Eq{eq:ircsafeobdecomp2}, building off the Deep Sets Theorem and following similar logic as \Ref{Komiske:2017aww}.
Given the decomposition of an IRC-safe observable $\mathcal O$ into $F$ and $\Phi$ via \Eq{eq:obdecomp2}, the IRC safety of the observable $\mathcal O$ corresponds to the following statements:
\begin{align}
\text{IR safety}:&\quad F\left(\sum_{i=1}^M\Phi(z_i,\hat p_i)\right)=F\left(\Phi(0,\hat p_0)+\sum_{i=1}^M\Phi(z_i,\hat p_i)\right),\label{eq:irsafe}\\
\text{C safety}:&\quad F\left(\sum_{i=1}^M\Phi(z_i,\hat p_i)\right)=F\left(\Phi(\lambda z_1,\hat p_1)+\Phi((1-\lambda)z_1,\hat p_1)+\sum_{i=2}^M\Phi(z_i,\hat p_i)\right),\label{eq:csafe}
\end{align}
where \Eq{eq:irsafe} holds for all directions $\hat p_0$ that a soft particle could be emitted and \Eq{eq:csafe} holds for all energy fractions $\lambda\in[0,1]$ of the collinear splitting.
In \Eq{eq:csafe}, we have selected particle 1 to undergo the collinear splitting but the statement holds for any of the particles by permutation symmetry.
The equations here only hold to a specified accuracy of approximation in the Observable Decomposition, which we leave implicit since it does not alter the structure of our argument.

We now make the following suggestive redefinition of $\Phi$ to ensure that the latent representation of a particle vanishes if the particle has zero energy:
\begin{equation}
\label{eq:phiredef1}
\Phi(z,\hat p)\to\Phi(z,\hat p)-\Phi(0,\hat p).
\end{equation}
Infrared safety via \Eq{eq:irsafe} ensures that the value of the observable is unchanged under this redefinition, so without loss of generality we may take $\Phi$ to vanish on arbitrarily soft particles.

Making another convenient redefinition of $\Phi$, we choose a $\lambda\in[0,1]$ and let:
\begin{equation}
\label{eq:phiredef2}
\Phi(z,\hat p)\to\Phi(\lambda z,\hat p)+\Phi((1-\lambda)z,\hat p).
\end{equation}
Collinear safety via \Eq{eq:csafe} ensures that the value of the observable is unchanged under such a redefinition, which holds for any $\lambda\in[0,1]$.

We now show that the freedom to redefine the mapping $\Phi$ using \Eqs{eq:phiredef1}{eq:phiredef2} for an IRC-safe observable leads to the IRC-safe Observable Decomposition in \Eq{eq:ircsafeobdecomp2}.
To see this, consider approximating $\Phi$ in the energy argument $z$ via the Stone-Weierstrass theorem.
Calling the angular coefficients of each term $C_n(\hat p)$ yields:
\begin{equation}
\label{eq:phiexpand}
\Phi(z,\hat p)=\sum_{n=0}^{\mathcal N}z^n\,C_n(\hat p) = C_0(\hat p) + z\, C_1(\hat p)  + \sum_{n=2}^{\mathcal N} z^n\, C_n(\hat p),
\end{equation}
for some large but finite $\mathcal N$.
How large $\mathcal N$  must be depends on the specified precision that we have been leaving implicit.

Invoking the soft redefinition in \Eq{eq:phiredef1}, $\Phi$ may be taken to vanish on arbitrarily soft particles, which allows $C_0(\hat p)$ to be set to zero without changing the value of the observable.
Implementing the collinear redefinition in \Eq{eq:phiredef2} after the expansion in energy, we obtain:
\begin{equation}
\label{eq:expandwlambda}
\Phi(z,\hat p)=z\,C_1(\hat p)+\sum_{n=2}^{\mathcal N}(\lambda^n+(1-\lambda)^n)z^nC_n(\hat p).
\end{equation}
From this equation, we seek to argue that $C_n(\hat p)$ for $n\ge2$ may be taken to vanish.
For $\lambda\in (0,1)$, this redefinition decreases the higher-order coefficients $C_n(\hat p)$ by a factor of $\lambda^n+ (1-\lambda)^n<1$ without changing the corresponding observable.
Iterated application of this fact allows the higher-order coefficients to be removed while keeping the term linear in the energy.
Thus, to an arbitrarily good approximation, we can take $\Phi(z,\hat p)=z\,C_1(\hat p)$ for some angular function $C_1(\hat p)$, which we subsequently rename to $\Phi(\hat p)$.

To summarize, the Deep Sets Theorem, combined with IRC safety, shows that the map $\Phi$ can be taken to be linear in energy without loss of generality.
Collinear safety was critical in arguing that $\Phi$ could be taken to be affine linear in the energy and infrared safety was critical in arguing that the constant piece could be set to zero without loss of generality.
This is exactly the result needed to justify the IRC-safe Observable Decomposition in \Eq{eq:ircsafeobdecomp2}, thereby completing the argument.
Beyond potential applications for building IRC safety directly into models, such an observable decomposition for IRC-safe observables may be useful for shedding light on the structure of these important observables more broadly.

\subsection{Encompassing image and moment representations}
\label{sec:encompass}

Beyond the single observables tabulated in \Tab{tab:sumobs}, entire event representations can be encompassed in the Observable Decomposition framework as well.

One common event representation is to view events as images by treating the energy deposits in a calorimeter as pixel intensities~\cite{Cogan:2014oua,deOliveira:2015xxd,Baldi:2016fql,Barnard:2016qma,Komiske:2016rsd,ATL-PHYS-PUB-2017-017}.
Since typical pixelizations for jet classification are $33\times 33 \simeq 1000$, the images are quite sparse, with an order of magnitude more pixels than particles.
Treating the detector as a camera and events as images allows for modern image recognition technology to be applied to collider physics problems.
These images are typically fed into a convolutional neural network, which is trained to learn a function of the images optimized for a specific task.

\begin{figure}[t]
\centering
\includegraphics[width=0.7\columnwidth]{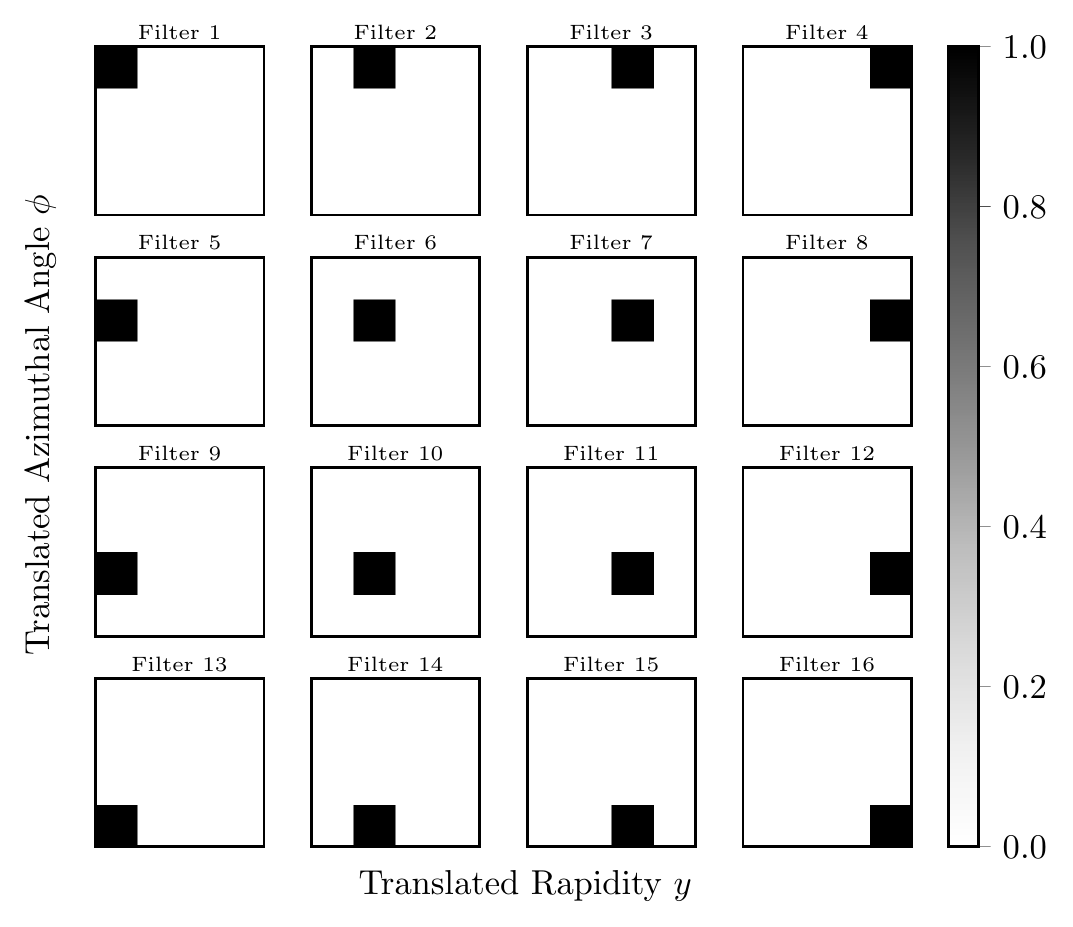}
\caption{The calorimeter image representation decomposed into a collection of $\Phi(y,\phi)$ filters according to the IRC-safe Observable Decomposition, shown here for the illustrative case of a $4\times4$ image.
The energy deposits in each pixel can be decomposed via \Eq{eq:ircsafeobdecomp} into an indicator function $\Phi(y,\phi)$ determining whether a particle in position $(y,\phi)$ hits the pixel.}
\label{fig:jetimagefilters}
\end{figure}

The image-based event representation of a jet as a collection of pixels fits naturally into the Observable Decomposition.
The energy (or transverse momentum) deposited in each pixel is simply a sum over the energies $z_i$ of the particles hitting that pixel.
Letting $\mathbb I_{j,k}(y,\phi)$ be an indicator function of pixel $(j,k)$ in the rapidity-azimuth plane, we have that the intensity $P_{j,k}$ of pixel $(j,k)$ is:
\begin{equation}\label{eq:img}
P_{j,k} = \sum_i z_i \, \mathbb I_{j,k}(y_i,\phi_i).
\end{equation}
Thus, having $\Phi$ be an indicator function for the location of the pixel directly allows the latent representation of the IRC-safe Observable Decomposition to be a detector image.
We illustrate this in \Fig{fig:jetimagefilters} for the rapidity-azimuth plane relevant for a hadron collider.
Here, the filters are a collection of localized square bumps evenly spaced throughout the rapidity-azimuth plane.

\begin{figure}[t]
\centering
\includegraphics[width=0.7\columnwidth]{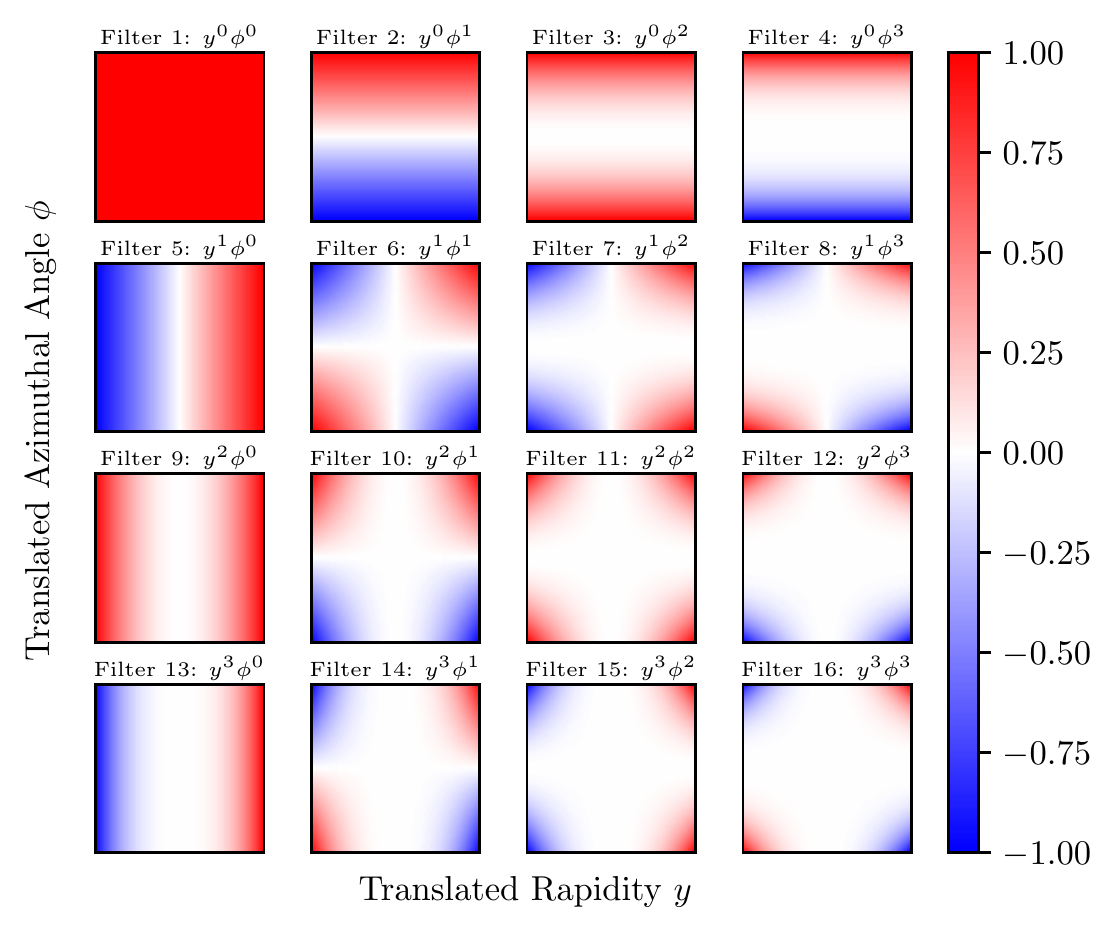}
\caption{\label{fig:momentfilters}
The radiation moment representation decomposed into a collection of $\Phi(y,\phi)$ filters according to the IRC-safe Observable Decomposition.
The $(m,n)$ moment of the energy distribution in the rapidity-azimuth plane can be decomposed via \Eq{eq:ircsafeobdecomp} into $\Phi(y,\phi)=y^m\phi^n$, shown here with increasing $m$ downward and increasing $n$ to the right.
}
\end{figure}

Another way to represent an event or jet is as a collection of moments of its radiation pattern.
Moments (or tensors) have been considered for analyzing hadronic energy flow patterns both for $e^+e^-$ and hadron colliders~\cite{Fox:1978vu,Donoghue:1979vi,GurAri:2011vx}.
A moment-based representation has yet to be directly exploited for machine learning applications in collider physics, though is closely related to the EFPs~\cite{Komiske:2017aww}.\footnote{There is a rich connection between the moments of the event radiation pattern and multiparticle energy correlators, a detailed discussion of which we leave to future work.  See footnote 8 of \Ref{Komiske:2017aww}.}
Here we restrict to the collimated case of jets, but a similar discussion holds at the event level.
The moments $I_{m,n}$ of the radiation pattern in the rapidity-azimuth plane are:
\begin{equation}\label{eq:moments}
I_{m,n} = \sum_i z_{i} \, y^m_i \phi^n_i.
\end{equation}
This can be manifestly decomposed according to the IRC-safe Observable Decomposition by simply making each filter $\Phi(y,\phi)=y^m\phi^n$, as illustrated in \Fig{fig:momentfilters}.
Here, the filters are a collection of non-localized functions which weight different parts of the event differently.

More generally, we can visualize $\Phi(y,\phi)$ for learned IRC-safe latent spaces, where the model itself learns its own event representation.
In interpreting these visualizations, it is worth keeping in mind that localized filters like \Fig{fig:jetimagefilters} correspond to an image-like representation, while global filters like \Fig{fig:momentfilters} correspond to a moment-like representation.
The flexibility of the IRC-safe Observable Decomposition allows for more complicated filters as well.
As we will see in \Sec{sec:visualizeqcd}, visualizing the latent space is extremely useful in understanding the behavior of EFNs.
Moreover, similar (albeit higher-dimensional) visualizations can be performed in the general PFN case of $\Phi(p)$ and have been explored in the point cloud context~\cite{DBLP:conf/cvpr/QiSMG17}.

\section{Network implementation}
\label{sec:implement}

In this section, we describe our implementation and adaptation of the Deep Sets decomposition for use in a particle physics context.
In light of the quark versus gluon jet case study presented in \Sec{sec:qg}, we focus here on inputting individual jets to the model, though we emphasize that the method is broadly applicable at the event level.

\subsection{Preprocessing}
\label{sec:preproc}

The goal of preprocessing inputs is to assist the model in its effort to solve an optimization problem.
Typically, preprocessing steps are optional, but are applied in order to improve the numerical convergence of the model, given the practical limitations of finite dataset and model size, as well as the particular choice of parameter initialization.
The preprocessing described in this section was found to be helpful, and sometimes necessary, for achieving a well-trained EFN or PFN model for the applications considered in \Sec{sec:qg}.
It is likely that for further applications of EFNs or PFNs, such as event-level studies, the appropriate preprocessing steps may change.

For the models we construct, kinematic information---transverse momentum $p_T$, rapidity $y$, and azimuthal angle $\phi$---are always given for each particle.
We preprocess these features as follows:  the transverse momenta are divided by the total scalar sum $p_T$ and the rapidities and azimuthal angles are centered based on the rapidity and azimuthal angle of the jet, using the $E$-scheme jet axis.
In terms of the four-momentum of each particle, this preprocessing step can be cast into the following suggestive form:
\begin{equation}\label{eq:preproc}
p_{T,i} \to \frac{p_{T,i}}{\sum_j p_{T,j}},\quad y_i \to y_i -\left( \sum_j p_{T,j}\, \hat p_j\right)_y, \quad\phi_i\to\phi_i-\left( \sum_j p_{T,j}\, \hat p_j\right)_\phi,
\end{equation}
with $\hat p_i = p^\mu_i/p_{T,i}$, where the subscripts indicate the rapidity and azimuth of the jet four-vector.
This notation makes clear that the per-particle preprocessing of \Eq{eq:preproc} solely relies on the scalar sum $p_T$, rapidity, and azimuth of the jet, which itself can be written in terms of an IRC-safe Observable Decomposition with $\Phi(\hat p) = (1, \hat p)$.
Alternative jet centerings, such as those based on the $p_T$-weighted centroid, also fit nicely into this framework.%
\footnote{These observations motivate an iterative local-global architecture which learns an event representation, applies it per-particle, and repeats.
Such an architecture could explicitly or learnably fold in this preprocessing as a first step.
We leave further developments in this direction to future work.}

Optionally, the inputs may also include particle identification (ID) information.
Though typically encoded using the Particle Data Group (PDG) particle numbering scheme~\cite{Tanabashi:2018oca}, the large and irregular integer values employed therein are not ideal inputs to a model expecting inputs roughly in the numerical range $[-1,1]$.
Therefore, a mapping from PDG IDs to small floats is performed for each particle (the details of which are provided below).
While this approach, which only uses a single feature to encode the particle ID information, should be sufficient to input this information to the model, alternative approaches using multiple categorical features may be easier for the model to interpret, since particle ID is inherently discrete rather than continuous.
For instance, using two additional features per particle, one feature could indicate the charge of the particle $\{-1,0,+1\}$ and the other one could indicate $\{h,\gamma,e,\mu\}$ (where $h$ corresponds to a hadron, one of $\pi,\,K,\,n,\,p$), covering an experimentally realistic particle ID scheme.
One-hot encoding of the particle ID is another option.

In order to explore how particle identification is helpful to a model,\footnote{We perform this comparison at particle level without detector simulation. Detector effects may change or degrade the information available in the different particle types. Doing such an exploration with detector simulation (or in data) is an interesting avenue for additional exploration.} we use it in four different ways, each with a PFN architecture.
We describe each of the different models and levels of information used throughout \Sec{sec:qg} below:
\begin{itemize}
\item{\bf PFN-ID}: PFN, adding in the full particle ID information. 
For the case study in \Sec{sec:qg}, particles are indicated as being one of $\gamma$, $\pi^+$, $\pi^-$, $K^+$, $K^-$, $K_L$, $n$, $\bar n$, $p$, $\bar p$, $e^-$, $e^+$, $\mu^-$, $\mu^+$, which are represented to the model as a single float value starting at 0 and increasing by 0.1 for each distinct type, respectively.\footnote{Note that $\pi^0$ is absent since we include its decay, usually into two photons.}
\item{\bf PFN-Ex}: PFN, adding in experimentally realistic particle ID information. 
For the case study in \Sec{sec:qg}, particles are indicated as being one of $\gamma$, $h^+$, $h^-$, $h^0$, $e^-$, $e^+$, $\mu^-$, $\mu^+$, which are represented to the model analogously to the PFN-ID case.\footnote{These categories are based on particle flow reconstruction algorithms at ATLAS and CMS~\cite{Beaudette:2014cea,Sirunyan:2017ulk,Aaboud:2017aca}, where $h^\pm = \pi^\pm/K^\pm/p/\bar{p}$ and $h^0 = K_L/n/\bar{n}$. Additional experimental information, such as $\pi/K/p$ separation, feasible at ALICE and LHCb (or at ATLAS and CMS at low $p_T$), can carry added information, as could exclusive hadron reconstruction. Particle ID information is typically captured in likelihood ratios for different particle hypotheses, which fits naturally into a categorical encoding scheme where there is a feature for each particle-type likelihood ratio.
}
\item{\bf PFN-Ch}: PFN, adding in the electric charge of the particles as an additional feature.
\item{\bf PFN}: The particle flow network using only three-momentum information via \Eq{eq:obdecomp}.
\item{\bf EFN}: The energy flow network using only IRC-safe latent space information via \Eq{eq:ircsafeobdecomp}.
\end{itemize}

\subsection{Network architecture}
\label{sec:arch}

So far, there has not yet been any machine learning in our effort to apply the decompositions in \Eqs{eq:obdecomp}{eq:ircsafeobdecomp} to collider data.
The machine learning enters by choosing to approximate the functions $\Phi$ and $F$ with neural networks.\footnote{\Ref{\deepsets} describes two types of architectures in the Deep Sets framework, termed \emph{invariant} and \emph{equivariant}. Equivariance corresponds to producing per-particle outputs that respect permutation symmetry. For this paper, our interest is in the invariant case, but we leave for future work an exploration of the potential particle physics applications of an equivariant architecture.}
Neural networks are a natural choice to use because sufficiently large neural networks can approximate any well-behaved function.

\begin{figure}[t]
\centering
\subfloat[]{\includegraphics[scale=0.58]{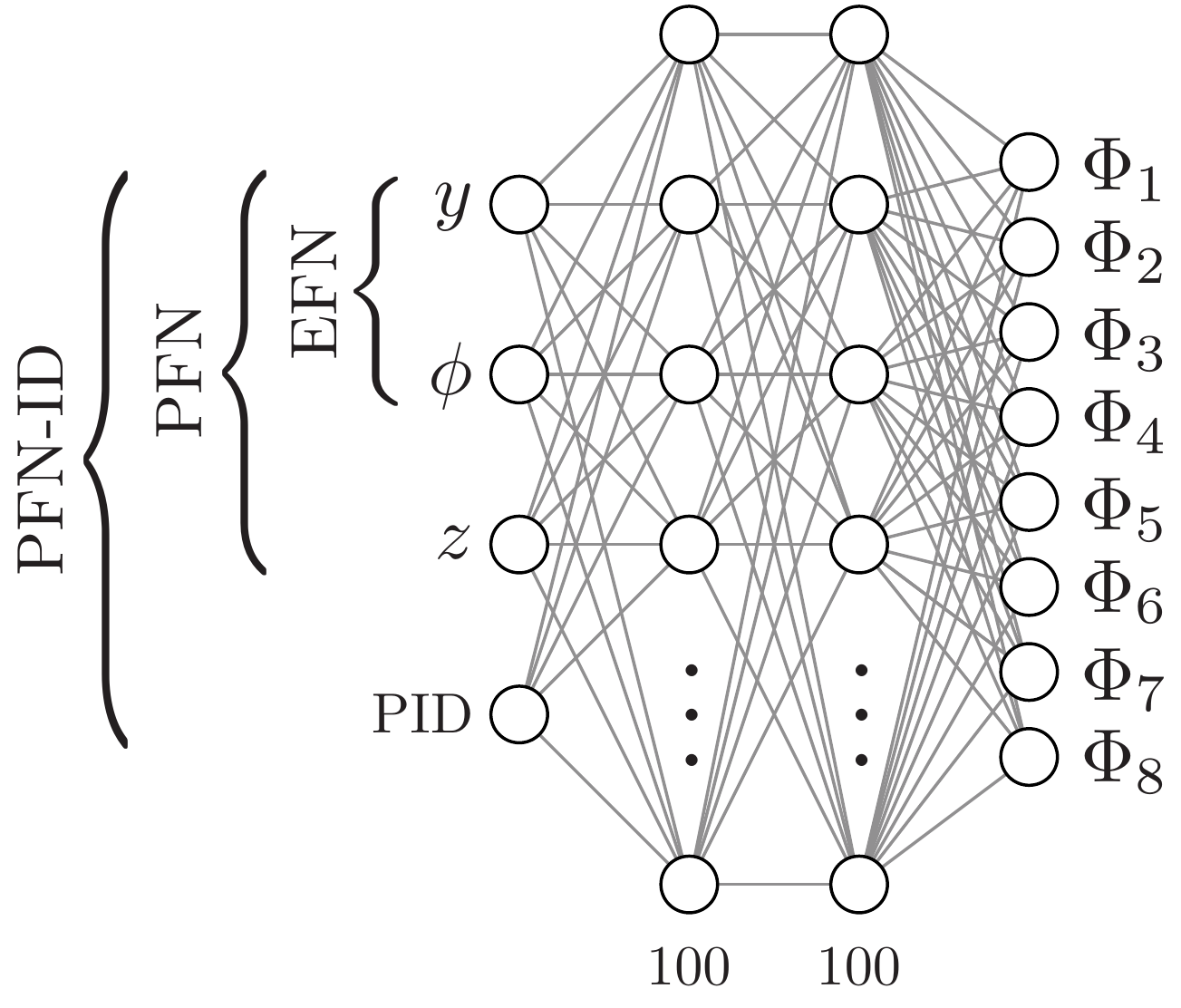}}\hspace{10mm}
\subfloat[]{\includegraphics[scale=0.58]{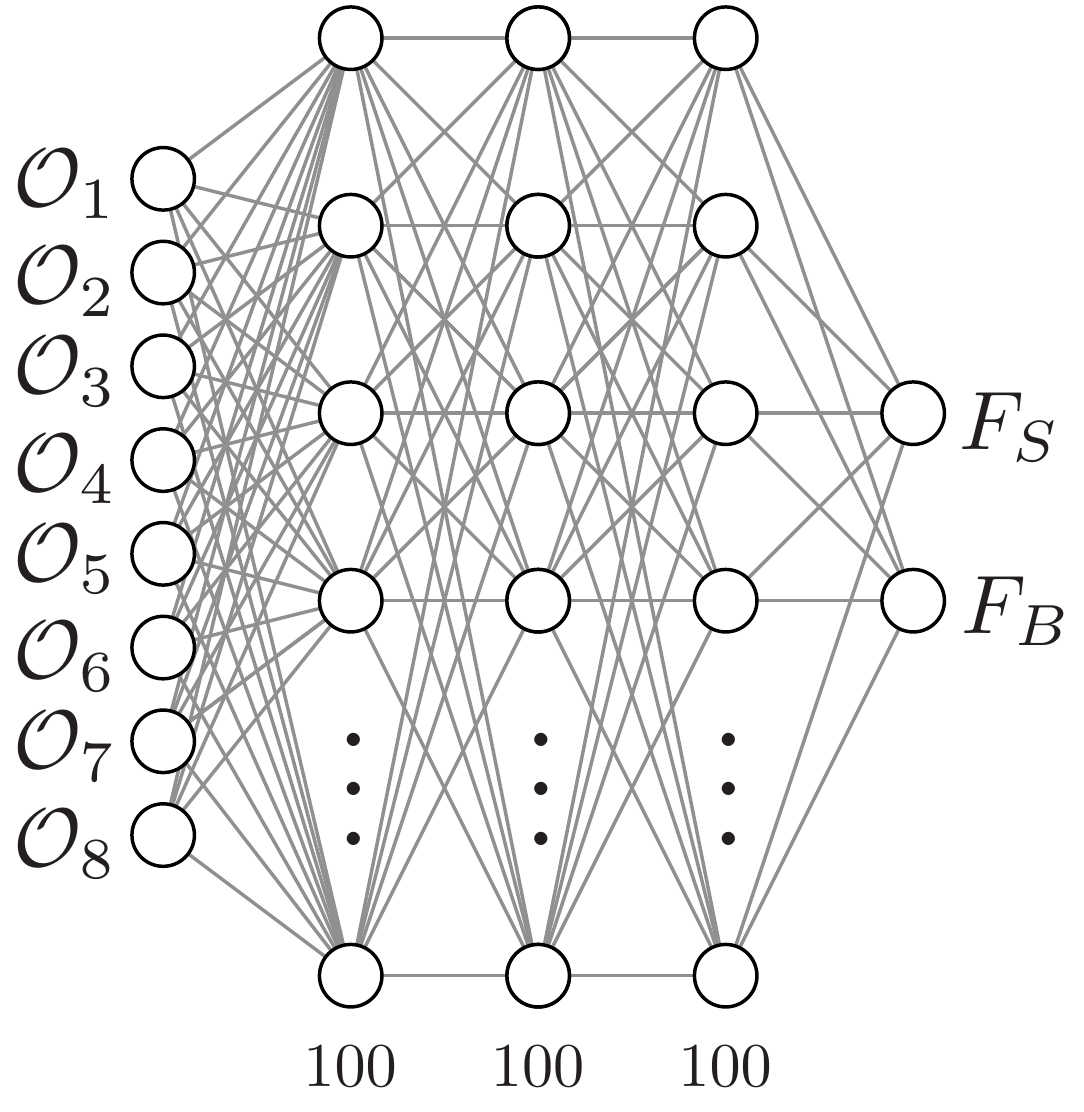}}
\caption{\label{fig:explicitNNs}
The particular dense networks used here to parametrize (a) the per-particle mapping $\Phi$ and (b) the function $F$, shown for the case of a latent space of dimension $\ell = 8$.
For the EFN, the latent observable is $\mathcal{O}_a = \sum_i z_i \, \Phi_a(y_i, \phi_i)$.
For the PFN family, the latent observable is $\mathcal{O}_a = \sum_i \, \Phi_a(y_i, \phi_i, z_i, \textsc{pid}_i)$, with different levels of particle-ID (PID) information.
The output of $F$ is a softmaxed signal ($S$) versus background ($B$) discriminant.
}
\end{figure}

To parametrize the functions $\Phi$ and $F$ in a sufficiently general way, we use several dense neural network layers as universal approximators, as shown in \Fig{fig:explicitNNs}.
For $\Phi$, we employ three dense layers with 100, 100, and $\ell$ nodes, respectively, where $\ell$ is the latent dimension that will be varied in powers of 2 up to 256.
For $F$, we use three dense layers, each with 100 nodes.
We confirmed that several network architectures with more or fewer layers and nodes achieved similar performance.
Each dense layer uses the ReLU activation function~\cite{relu} and He-uniform parameter initialization~\cite{heuniform}.
A two-unit layer with a softmax activation function is used as the output layer of the classifier.
See \App{app:models} for additional details regarding the implementations of the EFN, PFN, and other networks.
The \href{https://energyflow.network}{\tt EnergyFlow} Python package~\cite{energyflow} contains implementations and examples of EFN and PFN architectures.

\section{Discriminating quark and gluon jets}
\label{sec:qg}

To demonstrate the EFN architecture in a realistic setting, we implement and train an EFN and several PFN variants to discriminate light-quark from gluon initiated jets~\cite{Gallicchio:2011xq,Gallicchio:2012ez,Aad:2014gea,Gras:2017jty}, a problem relevant for new physics searches as well as precision measurements.
See \App{app:toptag} for a similar study on classifying top jets from QCD jets using samples based on \Ref{Butter:2017cot}.

\subsection{Event generation}
\label{sec:eventgen}

The samples used for this study were $Z(\to\nu\bar\nu)+g$ and $Z(\to\nu\bar\nu)+(u,d,s)$ events generated with \pythia 8.226~\cite{Sjostrand:2006za,Sjostrand:2014zea} at $\sqrt{s}=14$ TeV using the {\tt WeakBosonAndParton:qqbar2gmZg} and {\tt WeakBosonAndParton:qg2gmZq} processes, ignoring the photon contribution and requiring the $Z$ to decay invisibly to neutrinos.
Hadronization and multiple parton interactions (i.e.\ underlying event) were turned on and the default tunings and shower parameters were used.
Final state non-neutrino particles were clustered into $R=0.4$ anti-$k_T$ jets~\cite{Cacciari:2008gp} using \textsc{FastJet} 3.3.0~\cite{Cacciari:2011ma}.
Jets with $p_T\in[500,550]$ GeV and $|y|<2.0$ were kept.
No detector simulation was performed.\footnote{In the context of experimental applications, it is worth noting that the different resolutions of different particle types can be naturally accomodated in our framework.}
While labeling these jets using quark/gluon parton labels is manifestly unphysical, applications of these techniques at colliders could rely on an operational jet flavor definition~\cite{Komiske:2018vkc} and weak supervision techniques for training directly on data~\cite{Metodiev:2017vrx,Komiske:2018oaa} (see also \Refs{Dery:2017fap,Cohen:2017exh,blanchard2016classification,blanchard2018corrigendum,Metodiev:2018ftz}).

\subsection{Classification performance}
\label{sec:performance}

\begin{figure}[t]
\centering
\includegraphics[width=0.6\columnwidth]{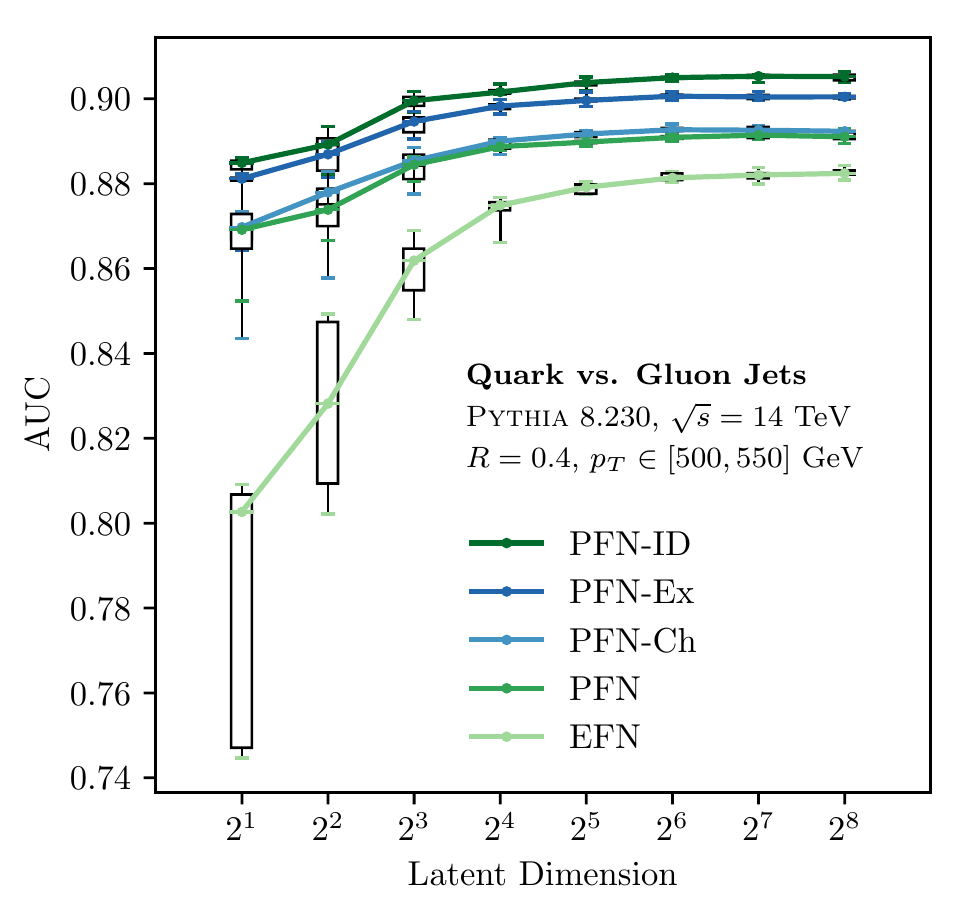}
\caption{The AUC performance of the EFN and PFN models as a function of the latent dimension of the model, which is varied from 2 to 256 in powers of 2.
The spread in values is due to training the model ten times with different initializations.
The performance generally increases with larger latent dimensions, with saturation observed by latent dimension 256.
The best model is PFN-ID, which uses full particle-type information, followed closely by PFN-Ex, which uses experimentally realistic particle-type information.
The PFN without any extra information performs roughly the same as the PFN-Ch, which uses charge information.
The fact that the EFN is lowest on this plot indicates that there is discrimination power to be found in IRC-unsafe information.}
\label{fig:qgdimsweep}
\end{figure}

A standard tool to analyze a classifier is the receiver operating characteristic (ROC) curve, obtained from the true positive $\varepsilon_s$ and false positive $\varepsilon_b$ rates as the decision threshold is varied.
This may also be plotted as a Significance Improvement (SI) curve~\cite{Gallicchio:2012ez}, namely $\varepsilon_s/\sqrt{\varepsilon_b}$ as a function of $\varepsilon_s$.
To condense the performance of a classifier into a single quantity, the area under the ROC curve (AUC) is commonly used, which is also the probability that the classifier correctly sorts randomly drawn signal (quark jet) and background (gluon jet) samples.
An AUC of 0.5 corresponds to a random classifier and an AUC of 1.0 corresponds to a perfect classifier.
We also report the background rejection at 50\% signal efficiency ($1/\varepsilon_b$ at $\varepsilon_s=50\%$) as an alternative performance metric.

For each of the models, we sweep the latent dimension $\ell$ of the internal representation from 2 to 256 in powers of 2.
As discussed in \Sec{sec:preproc}, four PFN models were trained each with different particle-type information.
Models are trained ten times each to give a sense of the variation and stability of the training.
The resulting model performances as quantified by the AUC are shown in \Fig{fig:qgdimsweep}.
As anticipated, the performance of each model increases as the latent dimension increases, with good performance achieved by $\ell=16$.
The higher variance at low latent dimensions arises because some of the filters fail to train to non-zero values in those cases.
The performance of the models appears to saturate by the larger latent dimensions, which justifies our use of $\ell=256$ as our benchmark latent dimension size for additional explorations.

\begin{figure}[t]
\centering
\subfloat[]{\includegraphics[width=0.5\columnwidth]{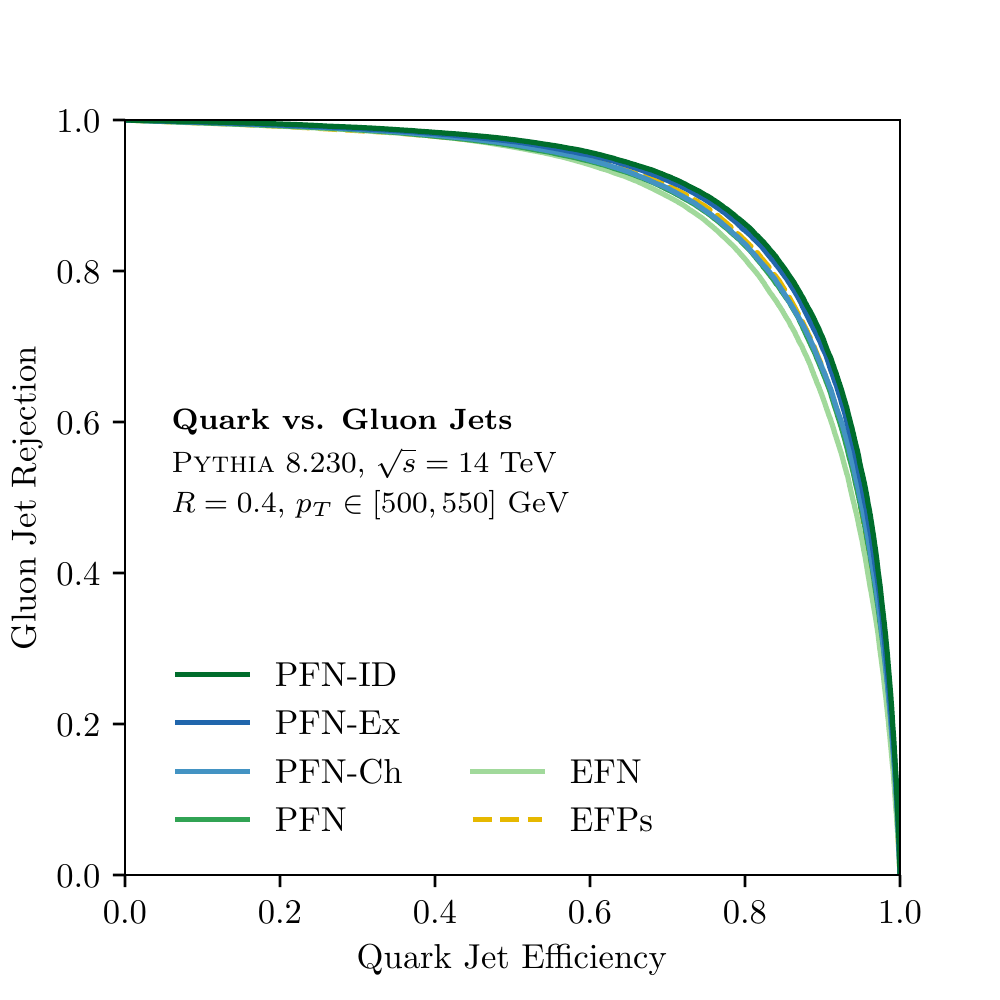}}
\subfloat[]{\includegraphics[width=0.5\columnwidth]{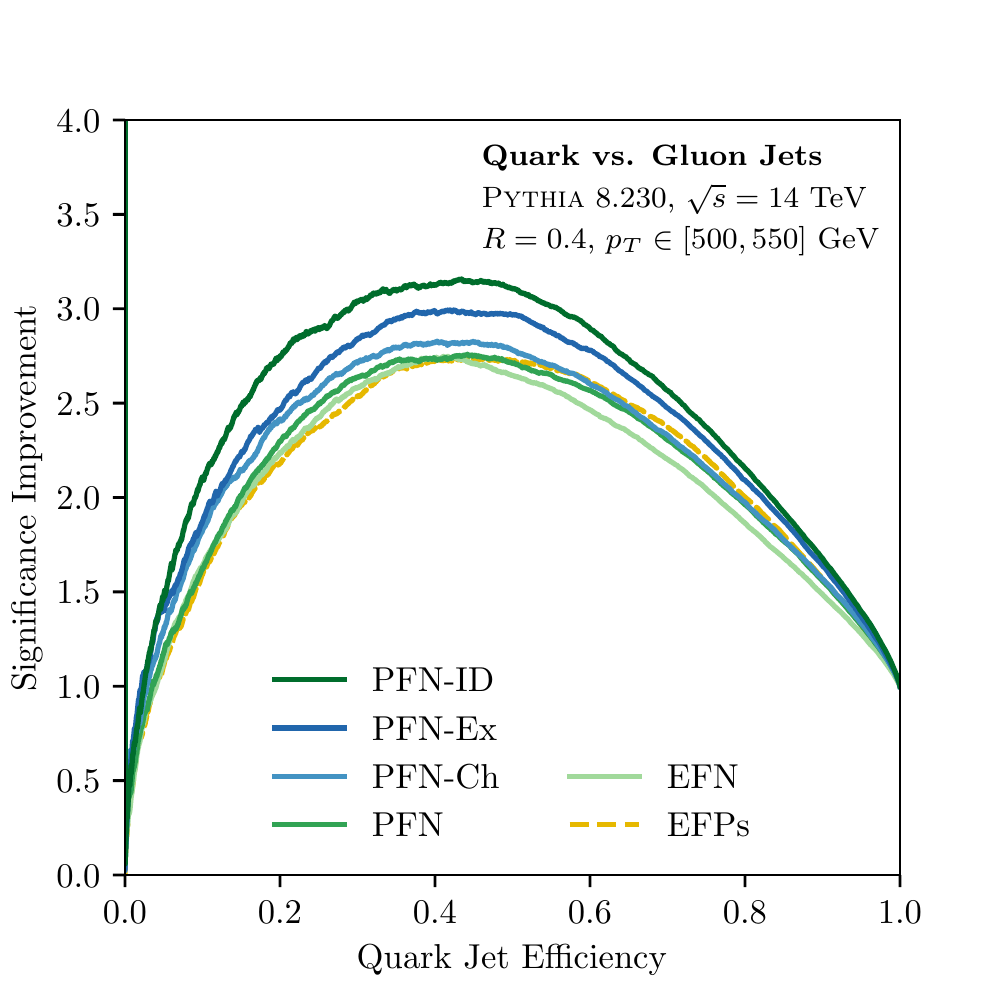}}
\caption{The (a) ROC and (b) SI curves of the median (selected by AUC) EFN and PFN models with latent dimensions of 256.
The linear EFP model is shown for comparison.
The PFN-ID with full particle ID yields the best performance of all models, followed by the PFN-Ex using experimentally realistic particle ID.
The EFN and EFP models perform comparably in terms of maximum SI, indicating that the available IRC-safe information is being captured consistently by these very different architectures.}
\label{fig:qgroccomp1}
\end{figure}

In \Fig{fig:qgroccomp1}, we show the full ROC and SI curves of these models with latent dimension 256.
The best model performance of all tested techniques and models was the PFN-ID with full particle ID, followed closely by the PFN-Ex with experimentally realistic particle ID.
\Figs{fig:qgdimsweep}{fig:qgroccomp1} show a well-defined hierarchy of model performances at all latent dimension sizes based on the information available to the model.
The fact that the PFNs outperform the EFN indicates that IRC-unsafe information is helpful for discrimination, which is not surprising in light of the fact that the constituent multiplicity is IRC unsafe and is known to be a good quark/gluon discriminant~\cite{Gallicchio:2012ez}.
Though IRC-unsafe information is helpful, it is instructive to test both EFNs and PFNs to probe how different kinds of information are used by the classifier.
Furthermore, sometimes an IRC-safe model is desired as it may be more robust to detector effects or mismodeling of infrared physics such as hadronization in simulated training data.

\subsection{Comparison to other architectures}
\label{subsec:compare}

\begin{table}[p]
\centering
\begin{tabular}{|c|l|l|}
\hline
 Symbol & Name &Short Description  \\ \hline \hline
PFN-ID &  Particle Flow Network w.\ ID & PFN with full particle ID  \\
PFN-Ex &  Particle Flow Network w.\ PF ID& PFN with realistic particle ID \\
PFN-Ch &  Particle Flow Network w.\ charge & PFN with charge information \\
PFN & Particle Flow Network & Using three-momentum information\\
EFN & Energy Flow Network & Using IRC-safe information\\
\hline
RNN-ID & Recurrent Neural Network w.\ ID & RNN with full particle ID \\
RNN & Recurrent Neural Network & Using three-momentum information \\
EFP & Energy Flow Polynomials & A linear basis for IRC-safe information \\
DNN & Dense Neural Network & Trained on an $N$-subjettiness basis \\
CNN & Convolutional Neural Network & Trained on $33\times 33$ grayscale jet images \\
\hline
$M$ & Constituent Multiplicity & Number of particles in the jet \\
$n_\text{SD}$ & Soft Drop Multiplicity & Probes number of perturbative emissions \\
$m$ & Jet Mass & Mass of the jet \\ \hline
\end{tabular}
\caption{
The (top) PFN/EFN architectures, (middle) other machine learning models, and (bottom) jet substructure observables used in comparisons of quark/gluon discrimination performance, along with their corresponding symbols and short descriptions.
A detailed discussion of model implementation and observable computation is given in \App{app:models}.}
\label{tab:obsandmodels}
\end{table}

\begin{figure}[p]
\centering
\subfloat[]{\includegraphics[width=0.5\columnwidth]{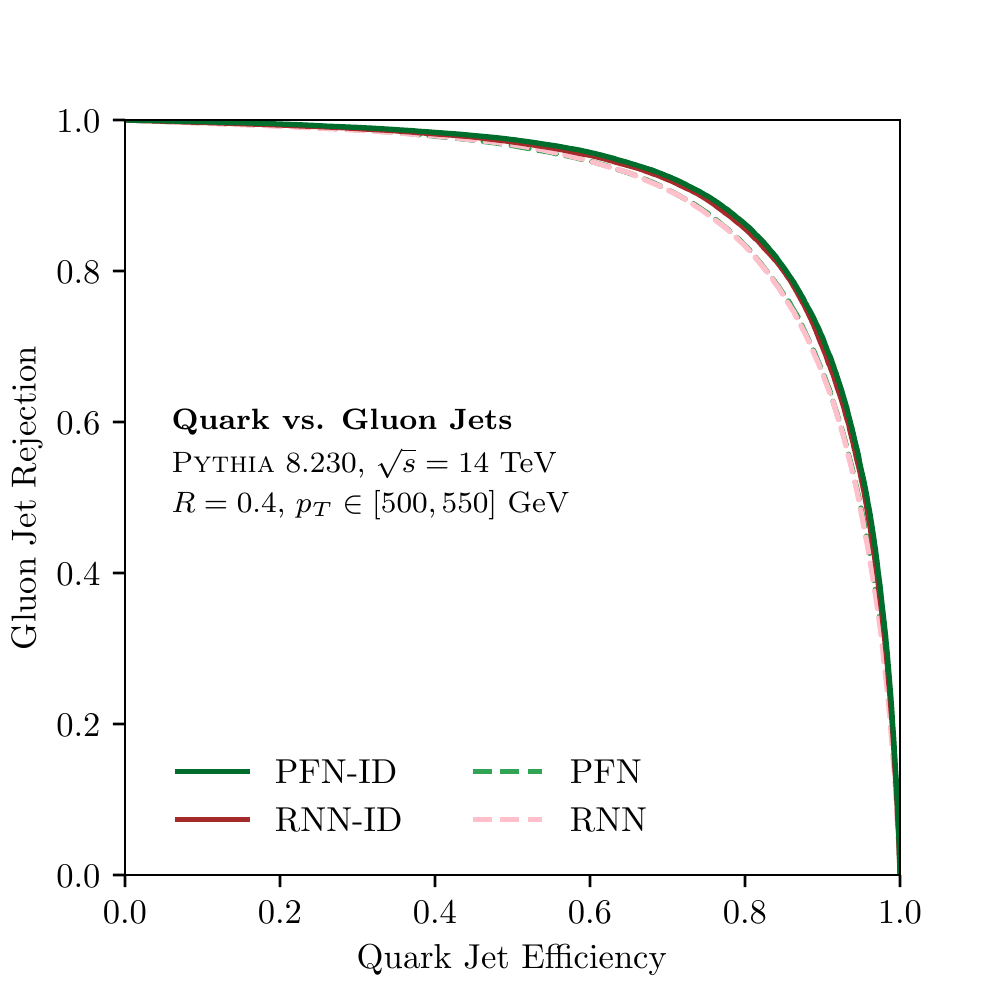}}
\subfloat[]{\includegraphics[width=0.5\columnwidth]{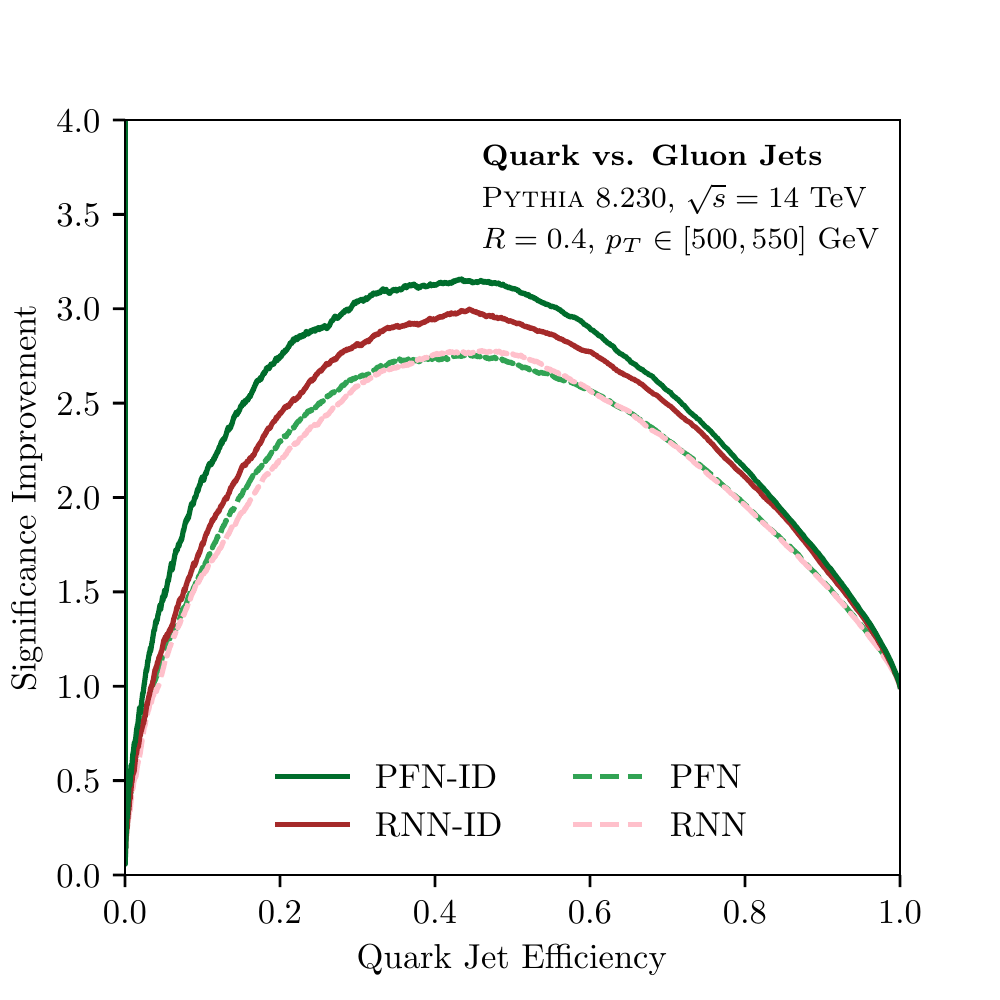}}
\caption{
The (a) ROC and (b) SI curve classification performances of PFN and RNN models both with and without full particle ID information.
From the SI curve, it appears that the PFN-ID model is doing better than the RNN-ID model, whereas the PFN and RNN models perform roughly equally.}
\label{fig:qgroccomp2}
\end{figure}

\begin{table}[p]
\centering
\begin{tabular}{|c|c|c|}
\hline
\multicolumn{1}{|c|}{\textbf{Model}}  & {\bf AUC} & {\bf $\boldsymbol{1/\varepsilon_g}$ at  $\boldsymbol{\varepsilon_q=50\%}$}  \\
\hline \hline
PFN-ID & ${\bf 0.9052}\pm0.0007$ & ${\bf 37.4}\pm0.7$ \\
PFN-Ex & $0.9005\pm0.0003$ & $34.7\pm 0.4$ \\
PFN-Ch & $0.8924\pm0.0001$ & $31.2\pm0.3$ \\
PFN & $0.8911\pm0.0008$ & $30.8\pm0.4$ \\
EFN & $0.8824\pm0.0005$ & $28.6\pm0.3$  \\ 
\hline
RNN-ID & 0.9010 & 34.4 \\
RNN & 0.8899 & 30.5 \\
EFP & 0.8919 & 29.7 \\
DNN & 0.8849 & 26.4 \\
CNN & 0.8781 & 25.5 \\
\hline
$M$ & 0.8401 & 19.0 \\
$n_\text{SD}$ & 0.8297 &  14.2\\
$m$ & 0.7401 & 7.2\\
\hline
\end{tabular}
\caption{The classification performances, quantified by the AUC and background rejection at 50\% signal efficiency, for each of the models and observables in \Tab{tab:obsandmodels}.
Reported uncertainties on the EFN and PFN family of models are half of the interquartile range over ten trainings.
Performance uniformly improves with the inclusion of more particle-type information.
}
\label{tab:qgtabcomp}
\end{table}

\begin{figure}[p]
\subfloat[]{\includegraphics[width=0.5\columnwidth]{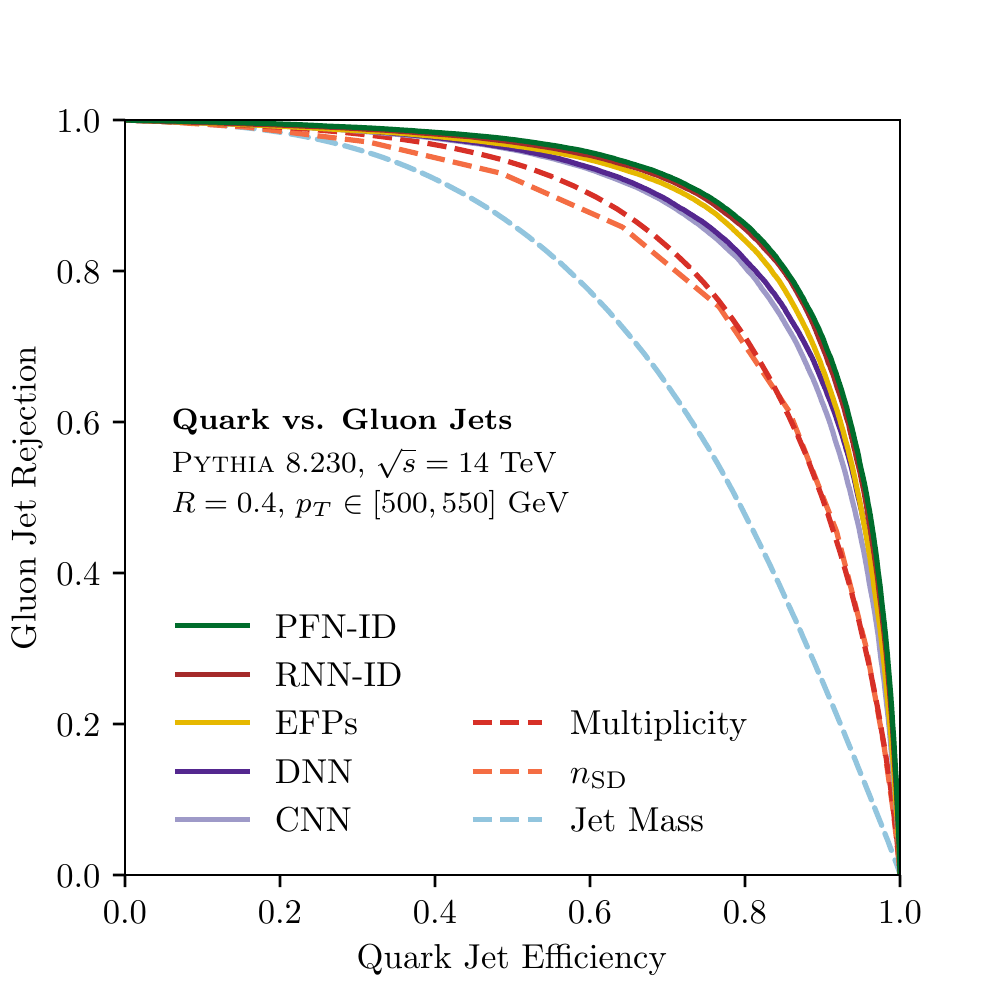}}
\subfloat[]{\includegraphics[width=0.5\columnwidth]{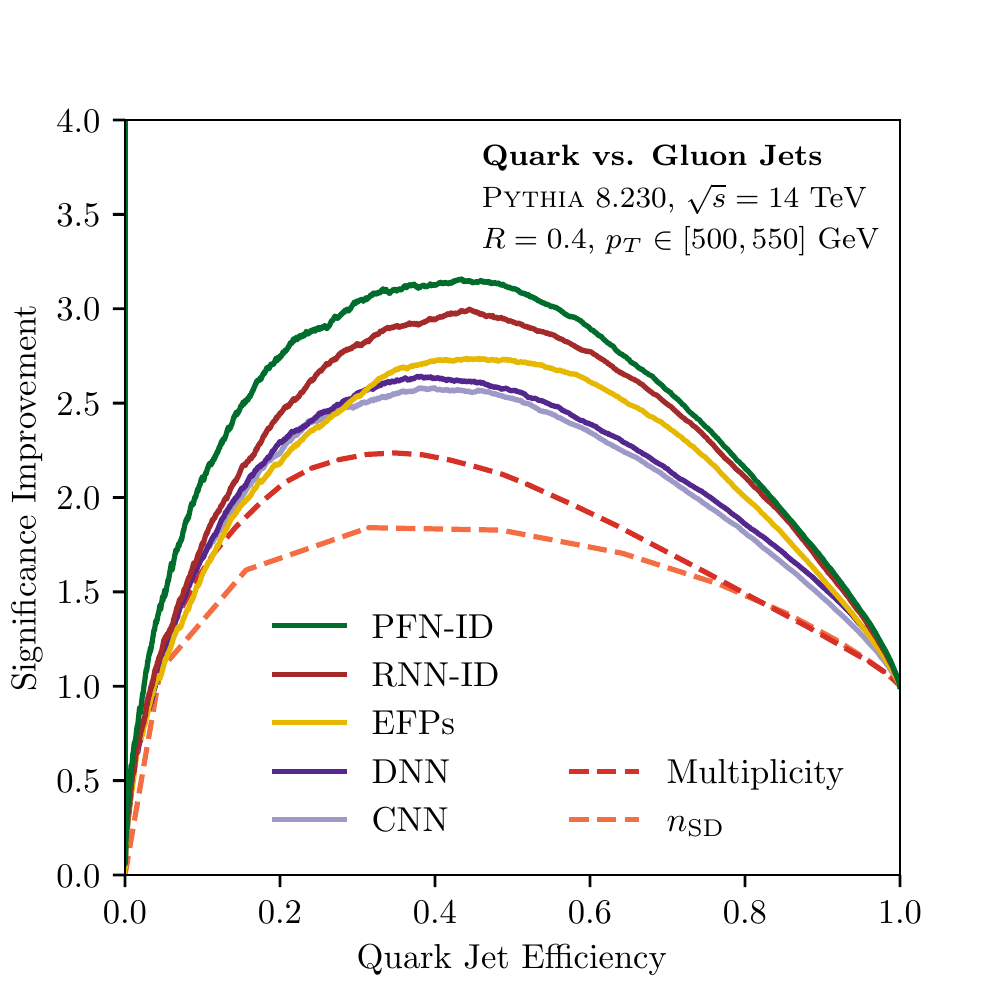}}
\caption{The (a) ROC and (b) SI curve classification performances of several different models and observables.
The PFN-ID and RNN-ID curves are shown in order to facilitate comparison with \Figs{fig:qgroccomp1}{fig:qgroccomp2}.
The PFN-ID architecture compares well to existing techniques, often notably outperforming them.}
\label{fig:qgroccomp3}
\end{figure}

Besides comparing the EFN and PFN architectures to each other, we also compare the $\ell=256$ models to a variety of other classifiers, summarized in \Tab{tab:obsandmodels} and described in more detail in \App{app:models}.

Of particular interest are the RNN-ID and RNN models, which also take particles as input (with and without full particle ID, respectively), but process them in a way which is dependent on the order the particles were fed into the network (decreasing $p_T$ ordering was used).
In \Fig{fig:qgroccomp2}, ROC and SI curves are shown for the RNN-ID and RNN architectures, as well as their natural counterparts, PFN-ID and PFN.
We see that PFN-ID slightly outperforms RNN-ID whereas the PFN and RNN are comparable, though we emphasize that making broad conclusions based on this one result is difficult given the variety of different RNN architectures we could have chosen.
Since PFNs are less expressive than RNNs, which can learn order-dependent functions, it is satisfying that both the PFN and RNN architectures achieve comparable classification performance with similar information.

The other machine learning architectures we compare to are a DNN trained on the $N$-subjettiness basis~\cite{Thaler:2010tr,Thaler:2011gf,Datta:2017rhs}, a CNN trained on jet images~\cite{Cogan:2014oua,deOliveira:2015xxd,Komiske:2016rsd}, and a linear classifier trained on the energy flow basis~\cite{Komiske:2017aww}.
Their performance, as given by their AUC and background rejection at 50\% signal efficiency, is summarized in \Tab{tab:qgtabcomp}.
Classification improves with the addition of IRC-unsafe information, as seen in the gain that the various PFN and RNN models have over the EFN and EFP models.
There is also a boost in performance from providing the model with ever-more specific particle-type information.

\Fig{fig:qgroccomp3} shows ROC and SI curves for all of these models, as well as some common jet substructure observables.
The best model is PFN-ID, followed by RNN-ID, and then (as shown in this figure) linear EFPs, which, somewhat remarkably, is the best architecture by AUC and SI curve height that does not take particles as direct inputs.
We note that for the CNN, one can in principle include particle ID information via additional channels, though training a 14-channel CNN is computationally challenging as each channel comes with $O(1000)$ additional numbers, most of which are zero.
Similar to RNNs and CNNs, the EFN and PFN architectures endeavor to be efficient by reducing the number of trainable parameters using weight sharing by applying the same $\Phi$ network to each particle.
Adding particle-type information to the EFPs or the $N$-subjettiness DNNs might be possible through a suitable generalization of jet charge \cite{Krohn:2012fg}, though we know of no concrete implementation of this in the literature.
The fact that PFNs naturally incorporate particle ID information is a important aspect of this architecture.

\afterpage{\clearpage}

\subsection{Visualizing the singularity structure of QCD}
\label{sec:visualizeqcd}

Beyond their excellent classification performance, the EFN and PFN architectures have the additional benefit that the learned function $\Phi$ can be directly explored.
As discussed in \Sec{sec:encompass}, this is particularly true of the EFNs, where $\Phi(\hat p)$ is a two-dimensional function of the angular information and thus can be directly visualized in the rapidity-azimuth plane.

\begin{figure}[t]
\centering
\includegraphics[width=0.7\columnwidth]{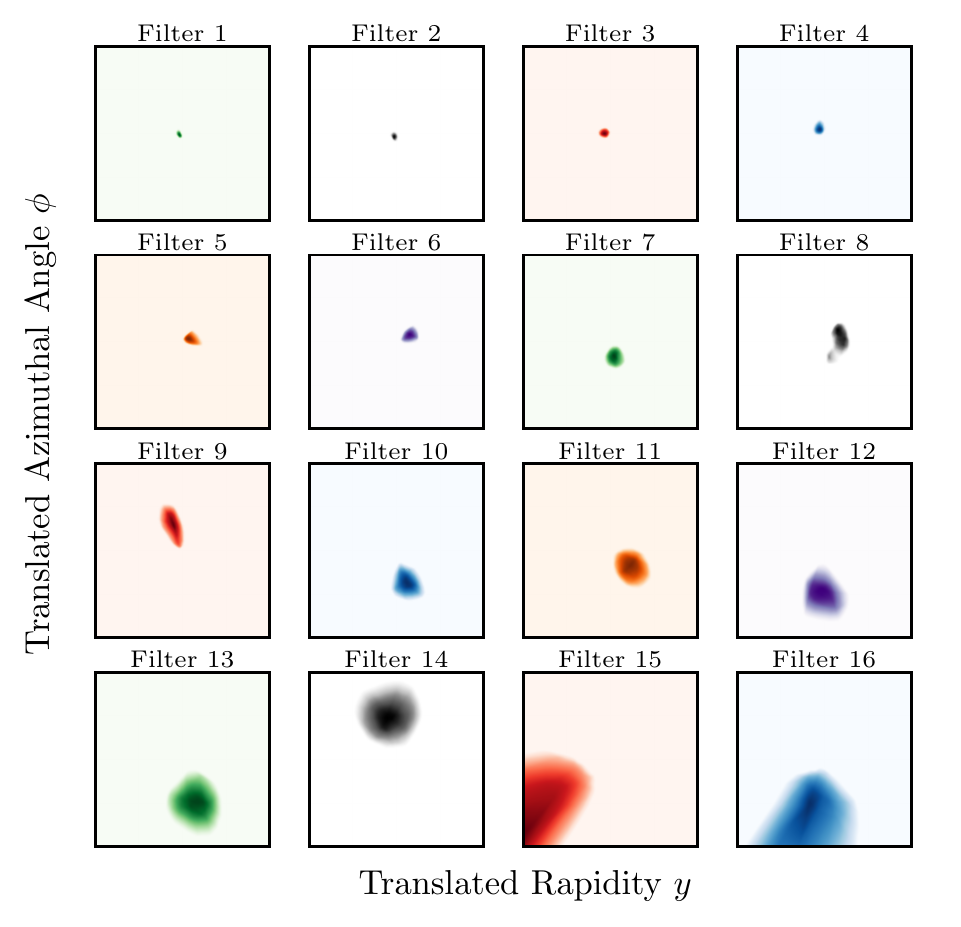}
\caption{Visualizations of 16 of the 256 filters learned by the $\ell=256$ EFN, with the filters sorted by their activated area.
The domain is the rapidity-azimuth plane from $-R$ to $R$ in both $y$ and $\phi$, since the jets have been preprocessed by centering them at (0,0).
The localized nature of the filters leads to our interpretation that the model has learned an image-like ``pixelization'' of the rapidity-azimuth plane, albeit one that is not square as in \Fig{fig:jetimagefilters}.
}
\label{fig:qgfilters}
\end{figure}

We take the learned $\Phi: \mathbb R^2 \to \mathbb R^\ell$ network from the best EFN model, as determined by the AUC, and evaluate it at many rapidity-azimuth points $(y,\phi)$ in the range $y,\phi\in[-R,R]$ to form a set of $\ell$ filters representing the learned latent space.
We show several of these filters from the $\ell=256$ EFN models in \Fig{fig:qgfilters}.
These can be directly compared with the corresponding filters for the detector image representation in \Fig{fig:jetimagefilters} and for the radiation moment representation in \Fig{fig:momentfilters}.
Like the image representation, we see that the learned filters are localized bumps in the rapidity-azimuth plane, and thus we say that the model appears to have learned a ``pixelization'' of the rapidity-azimuth plane.\footnote{Note that the ReLU activation function that we used in the model forces the filter values to be positive and allows the model to easily turn off regions of the inputs. Different activation functions may result in different learned latent representations.}
Unlike the image representation, the ``pixels'' learned by the model are smaller near the core of the jet and larger near the edge of the jet.

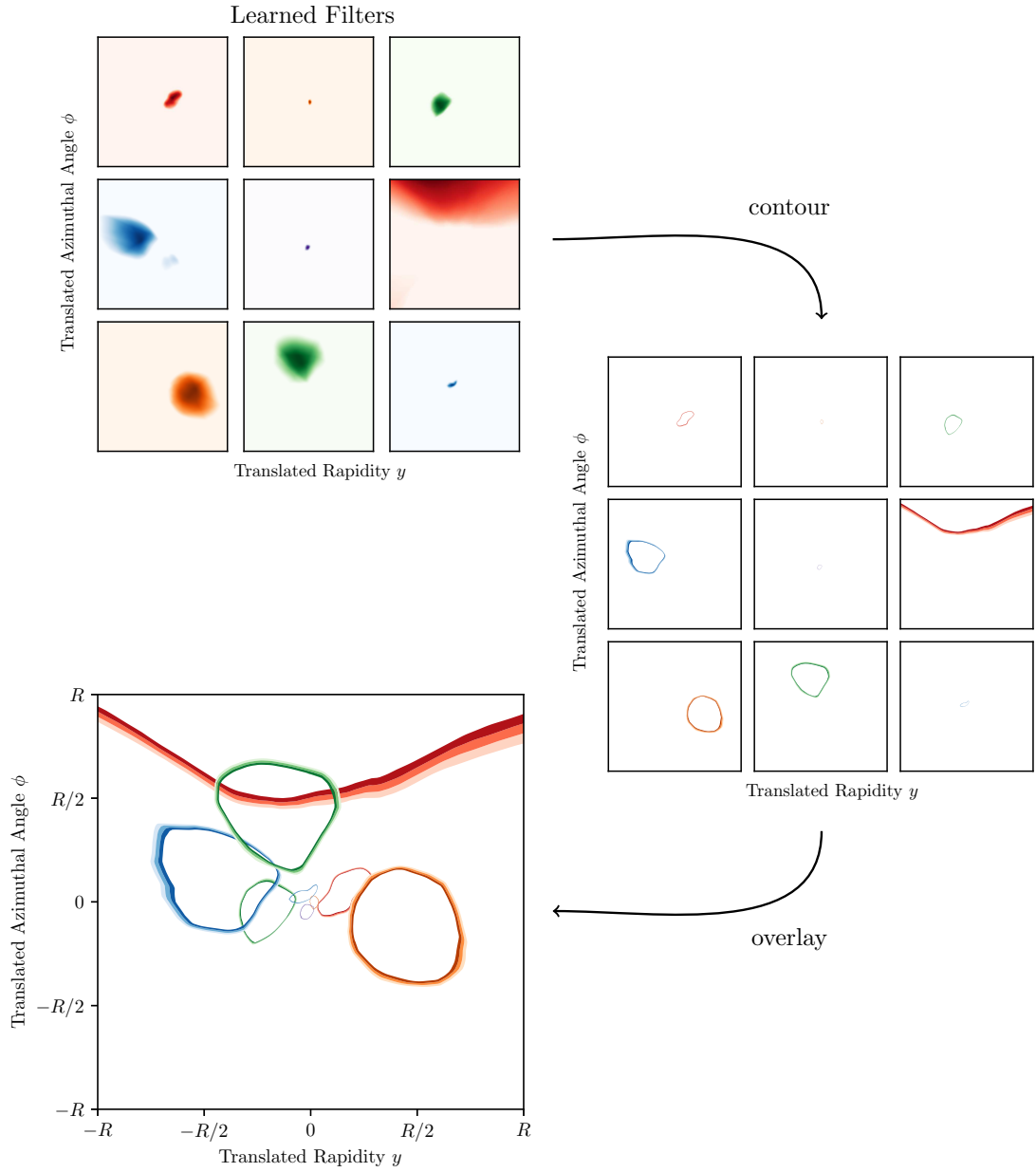
\begin{figure}[p]
\resizebox{\textwidth}{!}{
{{
\begin{tikzpicture}
\node[] at (-1,0)  {\includegraphics[width=0.5\columnwidth]{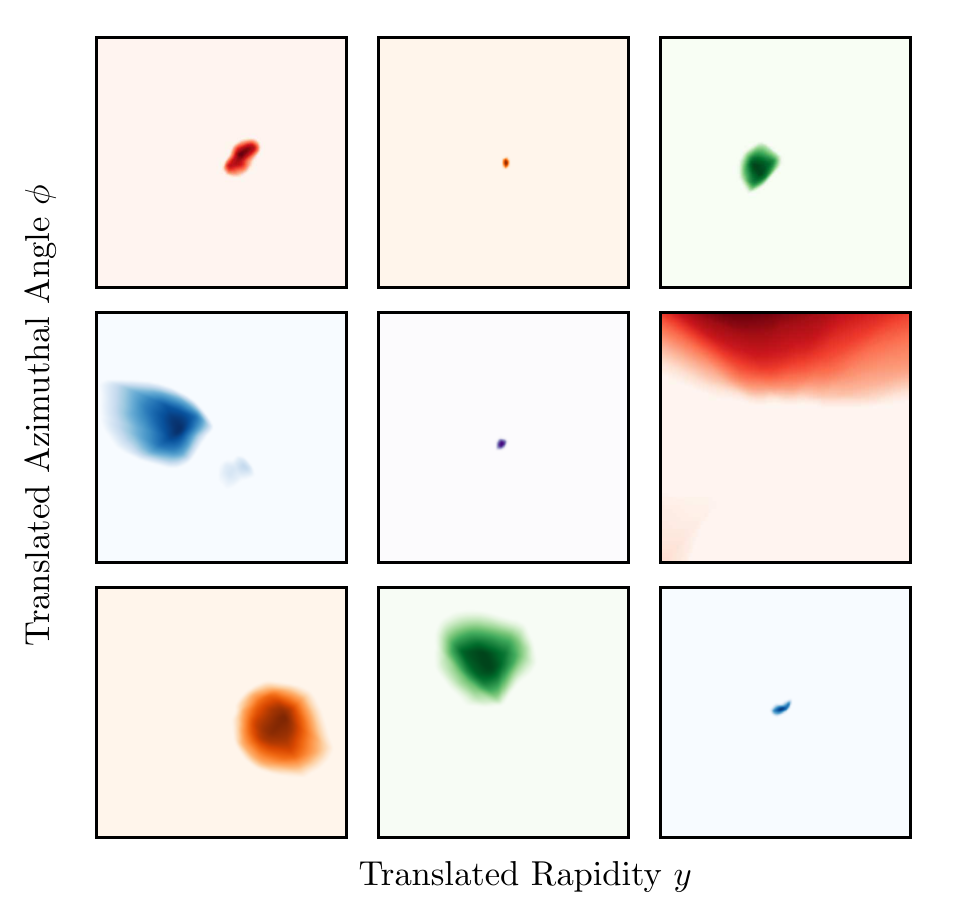}};
\node[] at (7,-5)  {\includegraphics[width=0.5\columnwidth]{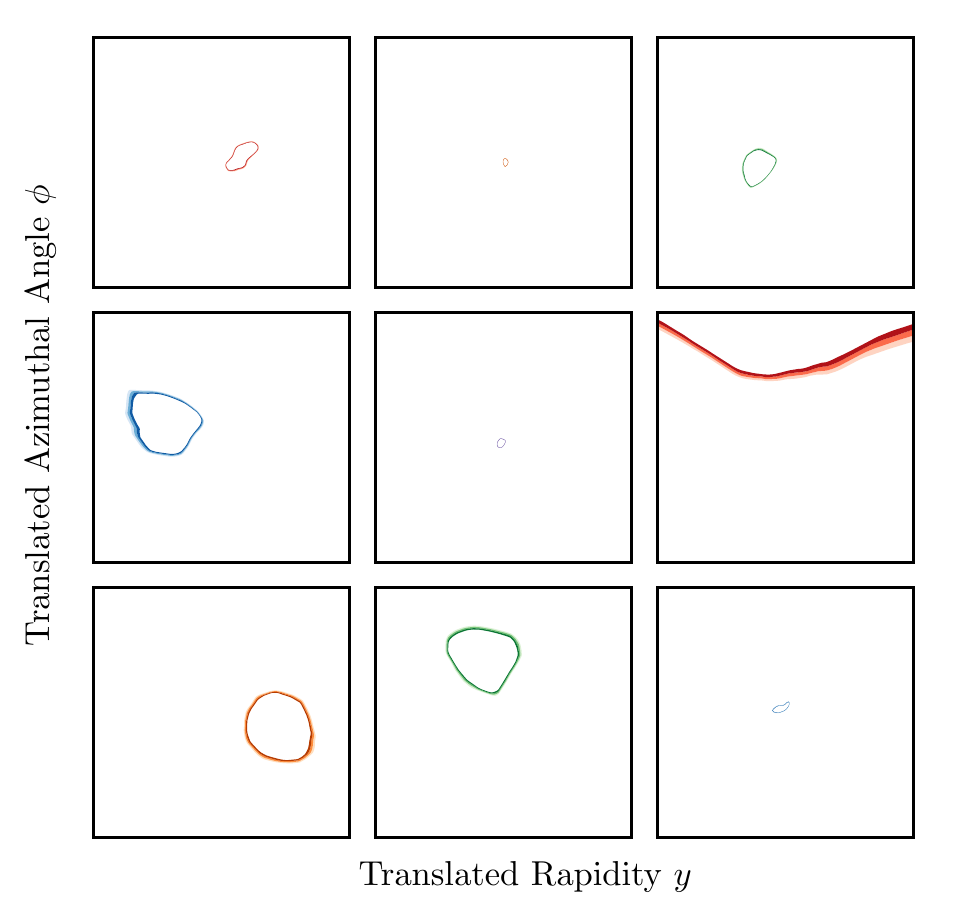}};
\node[] at (-1.4,-10.5)  {\includegraphics[width=0.55\columnwidth]{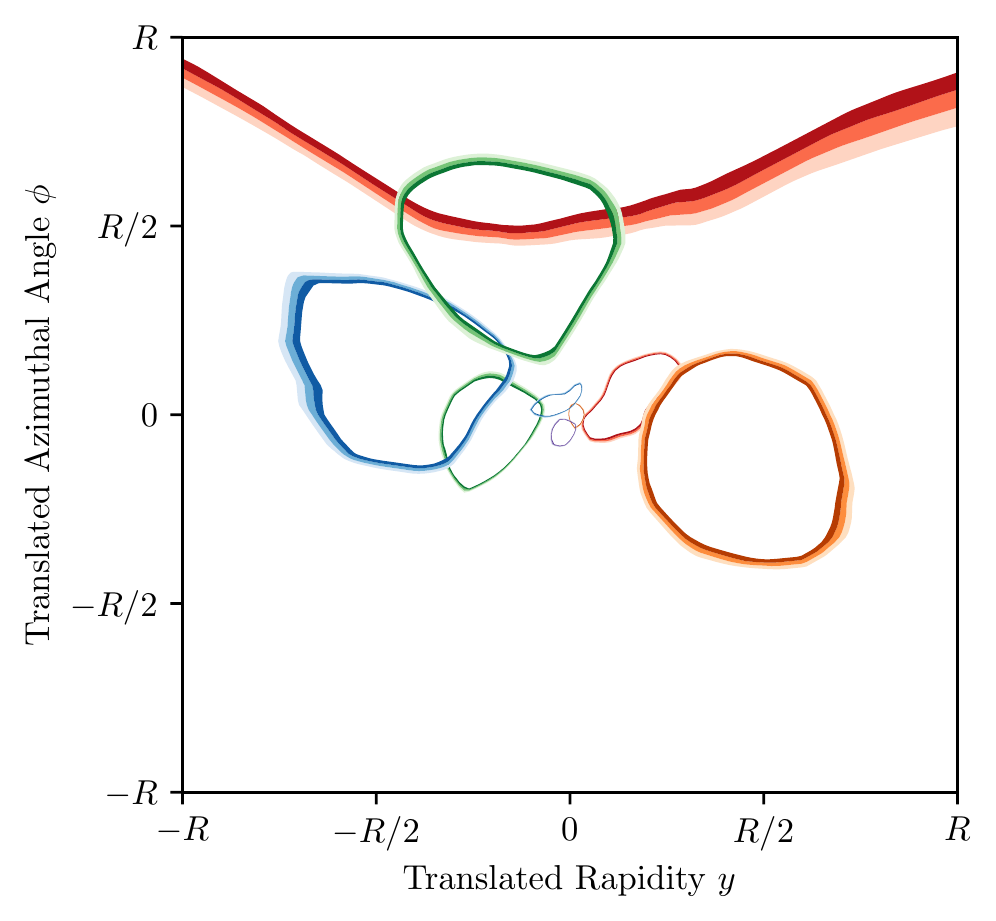}};
\draw [->,line width=1] (3.0,0.25) to [out = 0, in = 90] (7.2,-1);
\draw [->,line width=1] (7.2, -9.0) to [out = -90, in = 0] (3.0, -10.25);
\node [above] at (6.7, 0.5) {contour};
\node [above] at (6.7,-11) {overlay};
\node [above] at (-0.75,3.5) {Learned Filters};
\end{tikzpicture}
}}
}
\caption{
An illustration of our simultaneous visualization procedure for example EFN filters.
Contours of each filter are shown from 45\% to 55\% of its maximum value.
These contours are then overlaid on the same figure with different colors.
The resulting contour plot shows the dynamical pixelization of the plane determined by the model.
}
\label{fig:visualdemo}
\end{figure}

Beyond showing individual filters, it is informative to attempt to visualize an entire EFN latent space at once.
We achieve this by finding the boundary of each of the learned pixels (corresponding to one component of $\Phi$) and showing these boundaries together.
Plotting the boundary contours simultaneously allows for a direct visualization of the latent space representation learned by the model on a single figure.
In this way, we arrive at a proxy for the ``image'' that the model projects each jet into, which empirically emerges as a dynamically-sized calorimeter image.
Larger latent space dimensions correspond roughly to higher resolution images.
This strategy is illustrated in \Fig{fig:visualdemo}, where each filter is contoured around its 50\% value and the contours are overlaid.

\begin{figure}[p]
\centering
\includegraphics[width=0.9\columnwidth]{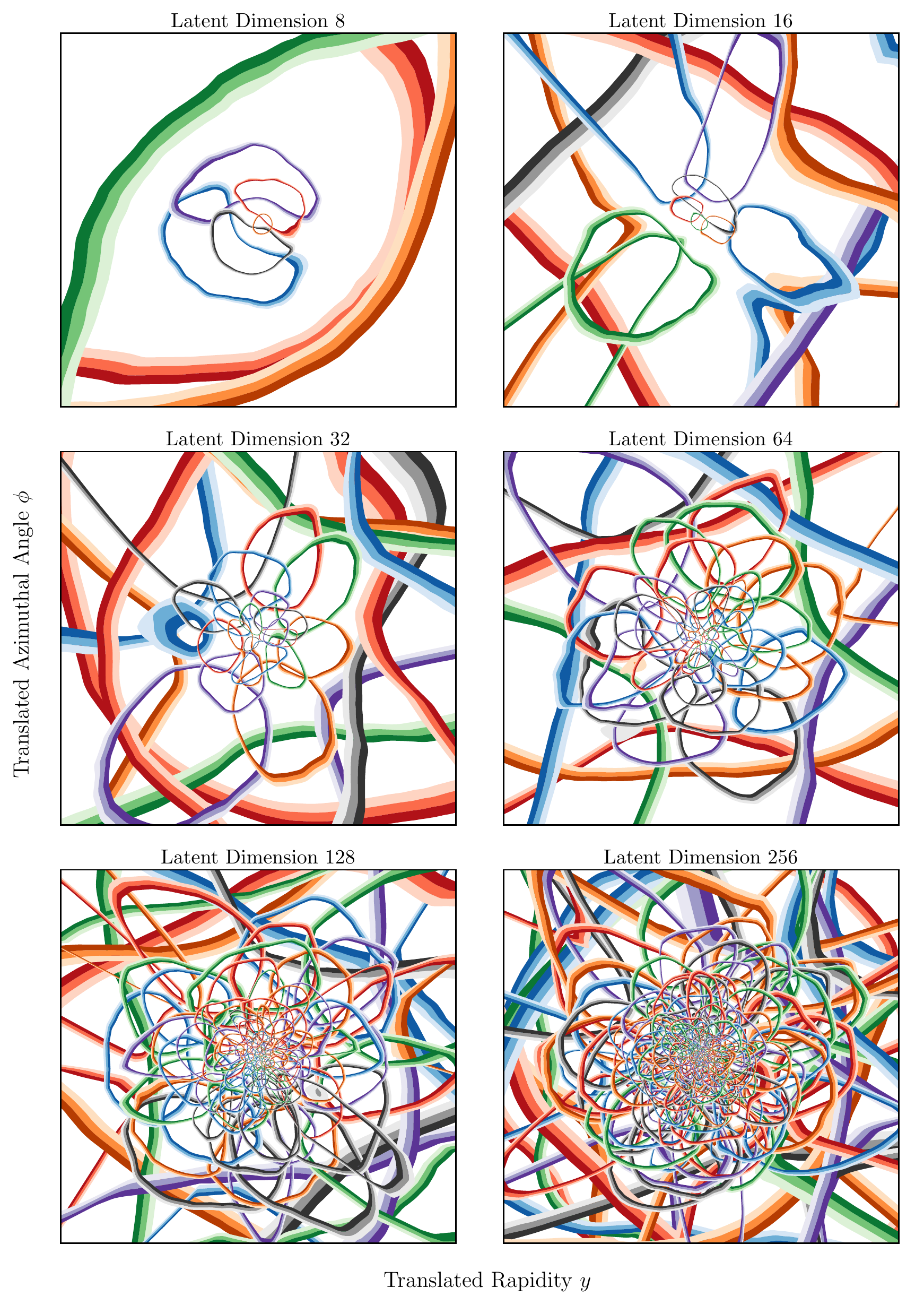}
\caption{The learned EFN pixelization of the rapidity-azimuth plane around the jet center with latent dimensions between 8 and 256 in powers of 2.
The learned filters are dynamically sized, with smaller filters probing the core of the jet and larger filters in the periphery.
A large version of the last panel is shown in \Fig{fig:qg256visual}.}
\label{fig:qgvisual}
\end{figure}

\begin{figure}[p]
\centering
\includegraphics[width=0.9\columnwidth]{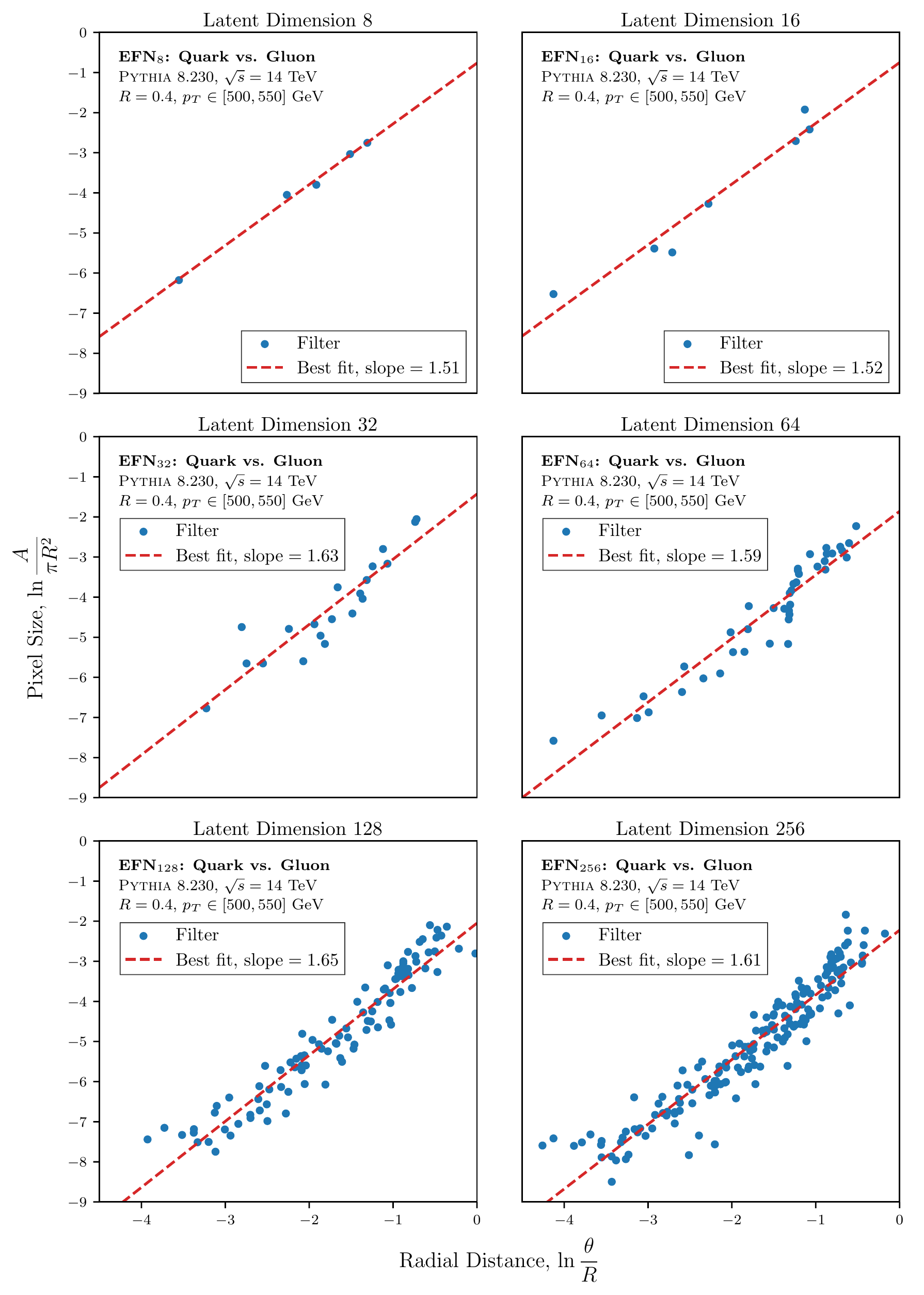}
\caption{The size of the EFN filters as a function of their distance from the origin.
The tendency of small filters to be located near the core of the jet and larger ones to be farther out is clearly visible.
The best fit slope is around 2, which is the scale-invariant expectation from \Eq{eq:slope_two}.}
\label{fig:qgquant}
\end{figure}

In \Fig{fig:qgvisual}, we show this visualization for EFN models with latent dimension varying from 8 to 256 in powers of 2.
Some of the filters are zero in the region of interest, perhaps as a result of dying ReLUs, so these are not shown.
It is evident from the simultaneous overlay of the filters that their sizes are correlated with their distance from the origin, which is especially clear for the larger latent dimensions.
As quark and gluon jets are (approximately) fractal objects with radiation singularly enhanced near the core of the jet as a result of the collinear singularity of QCD, the dynamically-sized pixelization learned by the EFN suggests that the model, in a sense, has understood this fact and adjusted itself accordingly.

To quantify the tendency of the filters to change size as they approach the center of the jet, we plot the area of each filter as a function of its distance from the origin.
To define the area of a filter, $A$, we integrate the region of the rapidity-azimuth for which the filter is greater than or equal to half of its maximum.
To capture a notion of distance from the origin, $\theta$, we take the distance from the origin to the maximum value of the filter.
We exclude filters that have centers outside of the jet radius.
The resulting plots of the filters in this space are shown in \Fig{fig:qgquant} for the models with latent space dimension from 8 to 256 in powers of 2.
There is a clear linear relationship between the (log) pixel size and the (log) distance to the jet core.
In particular, the slope between $\ln A$ and $\ln \theta$ is around 1.6 in the cases studied.

We can attempt to understand why the slopes in \Fig{fig:qgquant} are around 2 by considering a uniform pixelization in $(\ln \frac{R}{\theta}, \varphi)$, where $\theta$ is the distance from the jet axis and $\varphi$ is the azimuthal angle around the jet axis (not to be confused with $\phi$).
As discussed in \Ref{Dreyer:2018nbf}, this is the natural emission space of the jet.
Translating an area element from this natural emission space to the rapidity-azimuth $(y,\phi)$ plane yields:
\begin{equation}
\left|\text{d}\ln\frac{R}{\theta}\, \text{d}\varphi \right| = \frac{\text{d}\theta}{\theta}\,\text{d}\varphi = \frac{\theta\, \text{d}\theta\,\text{d}\varphi}{\theta^2} =\theta^{-2}\,\text{d}y\,\text{d}\phi. 
\end{equation}
Thus, a uniform pixelization in $(\ln \frac{R}{\theta}, \varphi)$ yields the following relationship between the area element (or pixel) size in the rapidity-azimuth plane and its distance from the origin:
\begin{equation}
\label{eq:slope_two}
\ln \frac{A}{\pi R^2} = 2\ln \frac{\theta}{R} + \text{const},
\end{equation}
explaining the slopes around 2 observed empirically in \Fig{fig:qgquant}.
This emergent behavior suggests an interesting connection with recent work on machine learning directly in the emission space of the jet~\cite{Dreyer:2018nbf}.
Deviations from the scale-invariant expectation of 2 are largest near the core of the jet, where non-perturbative physics or axis-recoil effects~\cite{Larkoski:2014uqa} become important.
The emission plane is visualized directly in \App{app:addvis}, where the pixelization is indeed seen to be highly uniform and regular in that space.

\afterpage{\clearpage}

\subsection{Extracting new observables from the model}
\label{sec:newobs}

Given that we are able to examine $\Phi$ for a trained EFN by visualizing its components, we can attempt to go further and obtain a quantitative description of both $\Phi$ and $F$ as closed-form observables.
Obtaining novel jet substructure observables from machine learning methods has been approached previously by parameterizing an observable and learning the optimal parameters for a particular task~\cite{Datta:2017lxt}.
Here, we go in a different direction and look directly at the latent observables learned by an EFN.
This represents a first, concrete step towards gaining a full analytic understanding what is being learned by the model.

To make this tractable, we focus on the simple case of a two-dimensional latent space.
A trained $\ell=2$ EFN has two learned filters, $\Phi_1(y,\phi)$ and $\Phi_2(y,\phi)$, and a learned function $F(\mathcal O_1, \mathcal O_2)$.
The filters can be visualized in the rapidity-azimuth plane and the function $F$ can be viewed in the $(\mathcal O_1,\mathcal O_2)$ phase space.
By studying these visualizations and noting their emergent properties, we can construct observables that reproduce the behavior and predictive power of the trained EFN.

\begin{figure}[t]
\centering
\subfloat[]{\includegraphics[width=0.5\columnwidth]{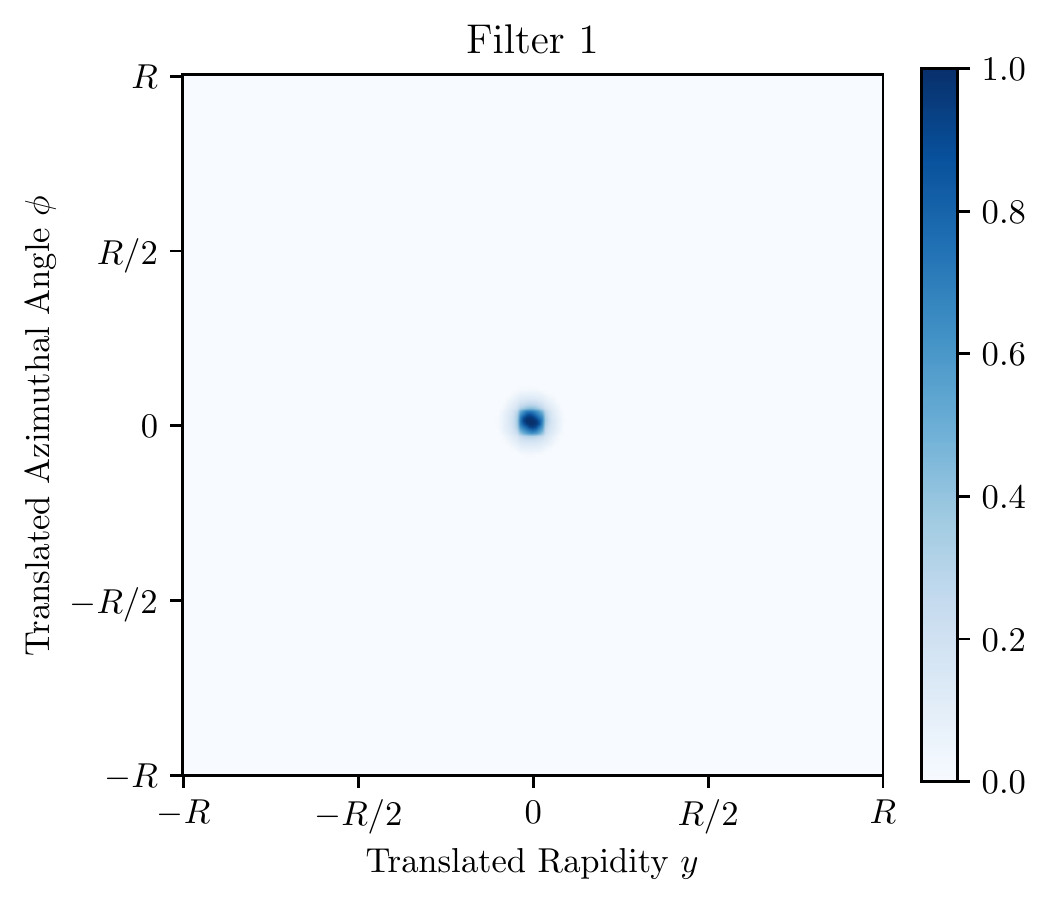}\label{fig:EFN2filt1}}
\subfloat[]{\includegraphics[width=0.5\columnwidth]{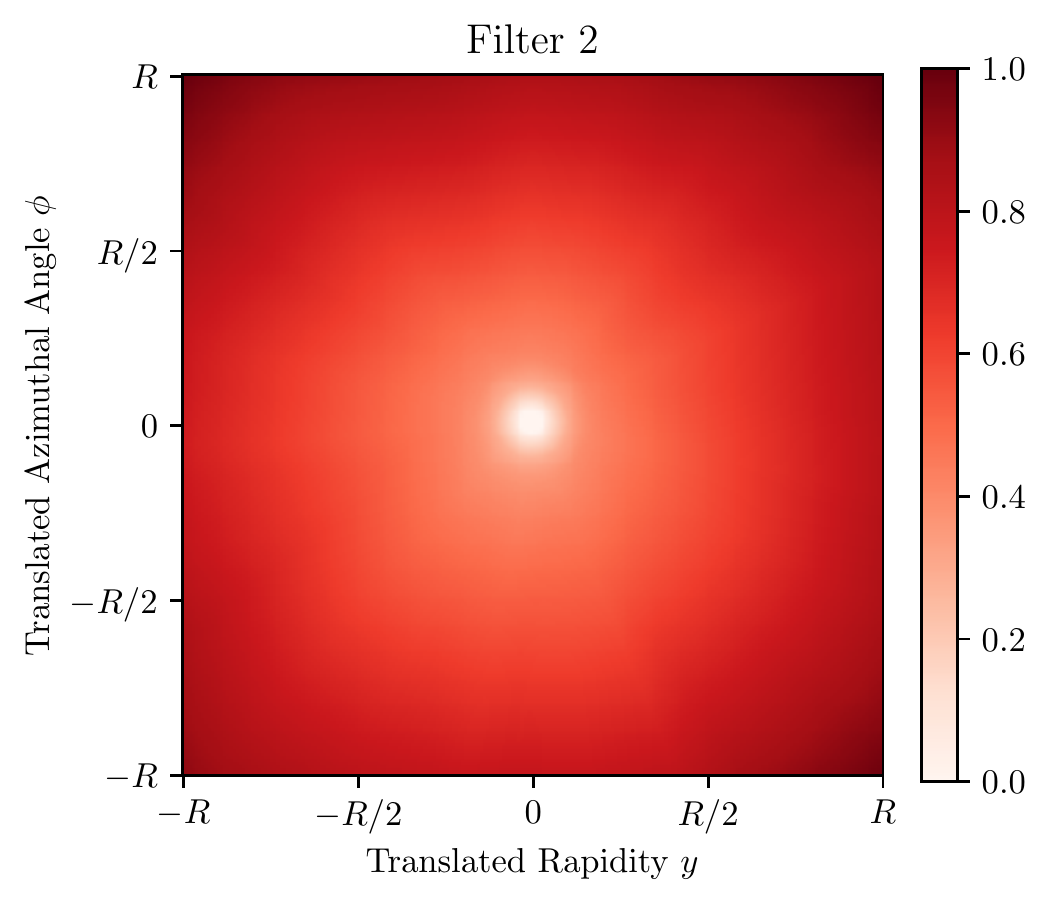}\label{fig:EFN2filt2}}

\subfloat[]{\includegraphics[width=0.6\columnwidth]{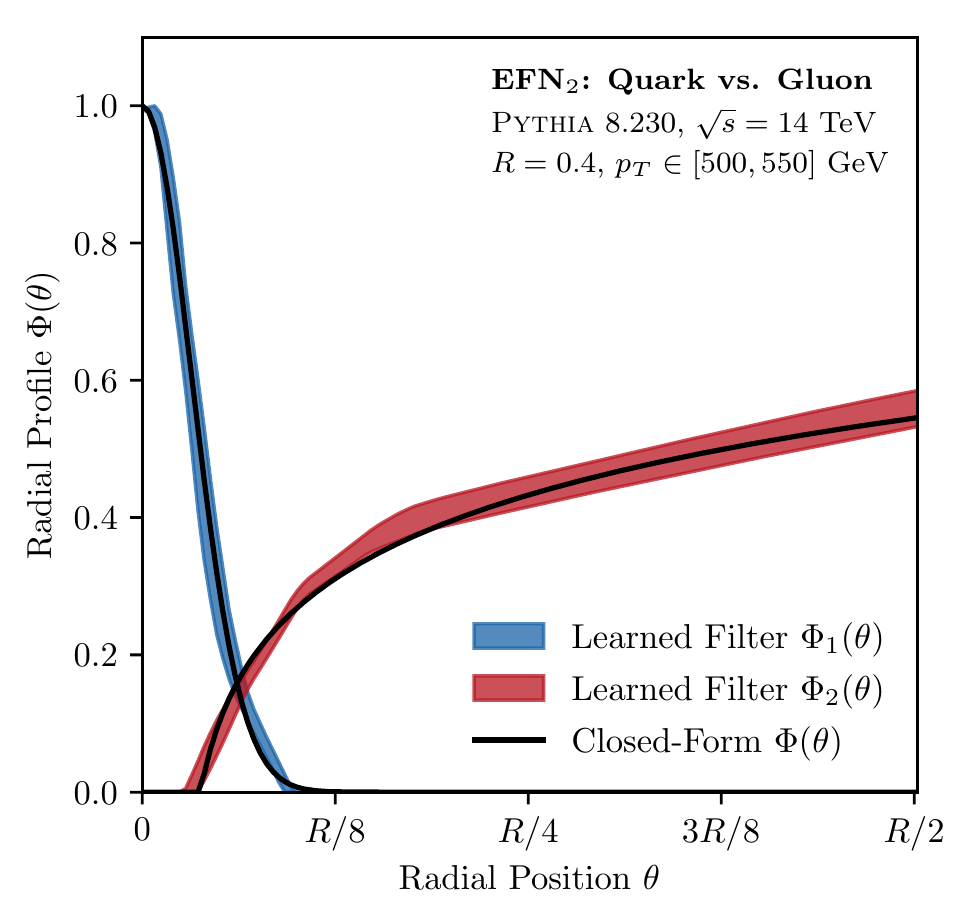}\label{fig:EFN2filtcf}}
\caption{(a, b) The two filters learned by an $\ell=2$ EFN, normalized to have a maximum value of 1.
The rotational symmetry of the filters about the jet axis is evident, with one filter probing radiation near the core of the jet and the other probing wide-angle radiation.
(c) Radial behavior of the two filters, from the center of the jet along the vertical and horizontal directions.
The analytic forms of \Eq{eq:newangs} are shown as black lines, with $B_{r_1,\beta}$ scaled by 0.15.
}
\label{fig:learnedEFN2}
\end{figure}


\begin{figure}[t]
\centering
\subfloat[]{\includegraphics[width=0.5\columnwidth]{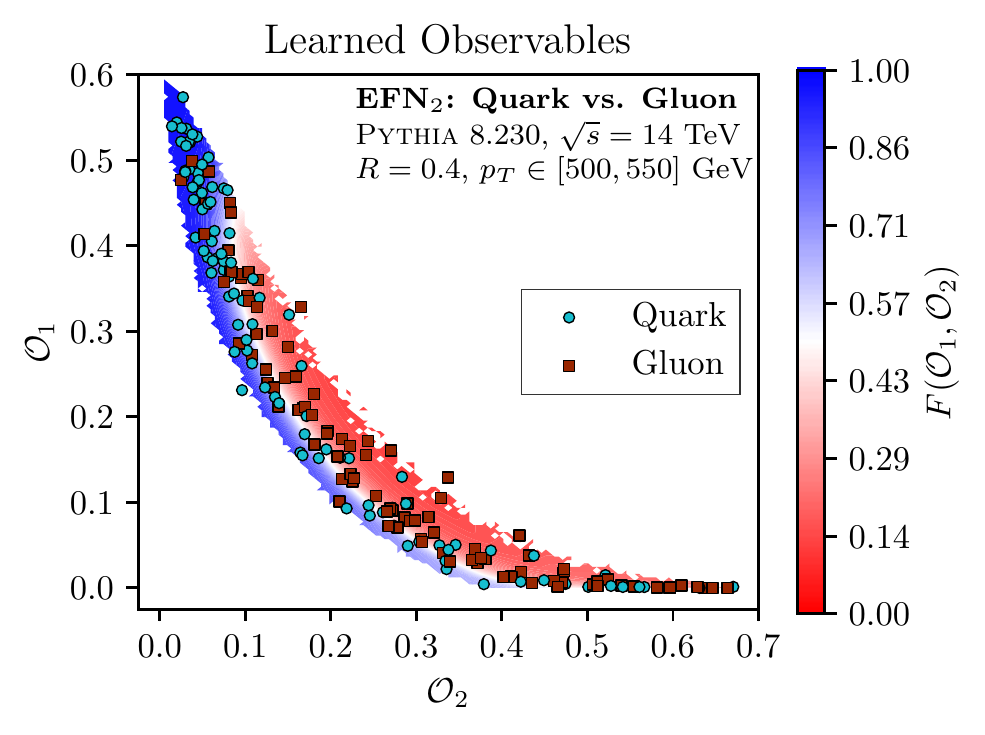}\label{fig:modelscatter}}
\subfloat[]{\includegraphics[width=0.5\columnwidth]{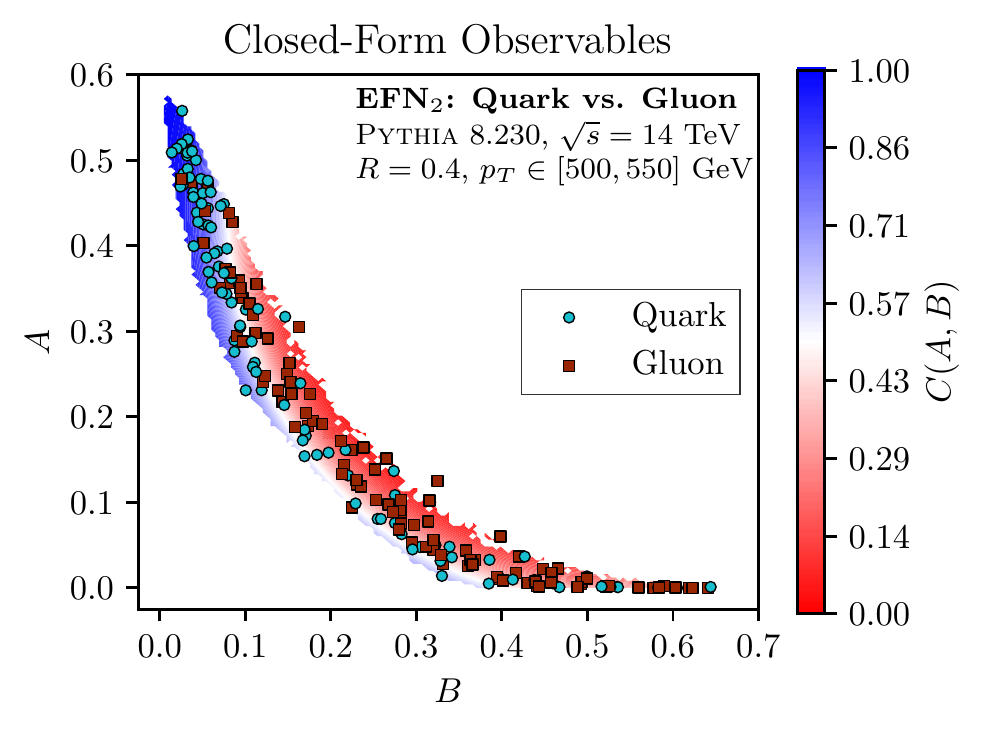}\label{fig:analyticscatter}}
\caption{
(a) The EFN model output $F(\mathcal O_1,\mathcal O_2)$ in the plane spanned by the learned latent space observables $\mathcal O_1$ and $\mathcal O_2$.
(b) The closed-form function $C(A,B)$ in the plane of the analytic observables $A_{r_0}$ and $B_{r_1,\beta}$.
One hundred quark jets (light blue circles) and gluon jets (dark red squares) are indicated to highlight the separation power.
The distribution of the closed-form observables and output value faithfully reproduce those of the trained EFN.}
\label{fig:learnedEFN2radial}
\end{figure}

\begin{figure}[t]
\centering
\subfloat[]{\includegraphics[width=0.5\columnwidth]{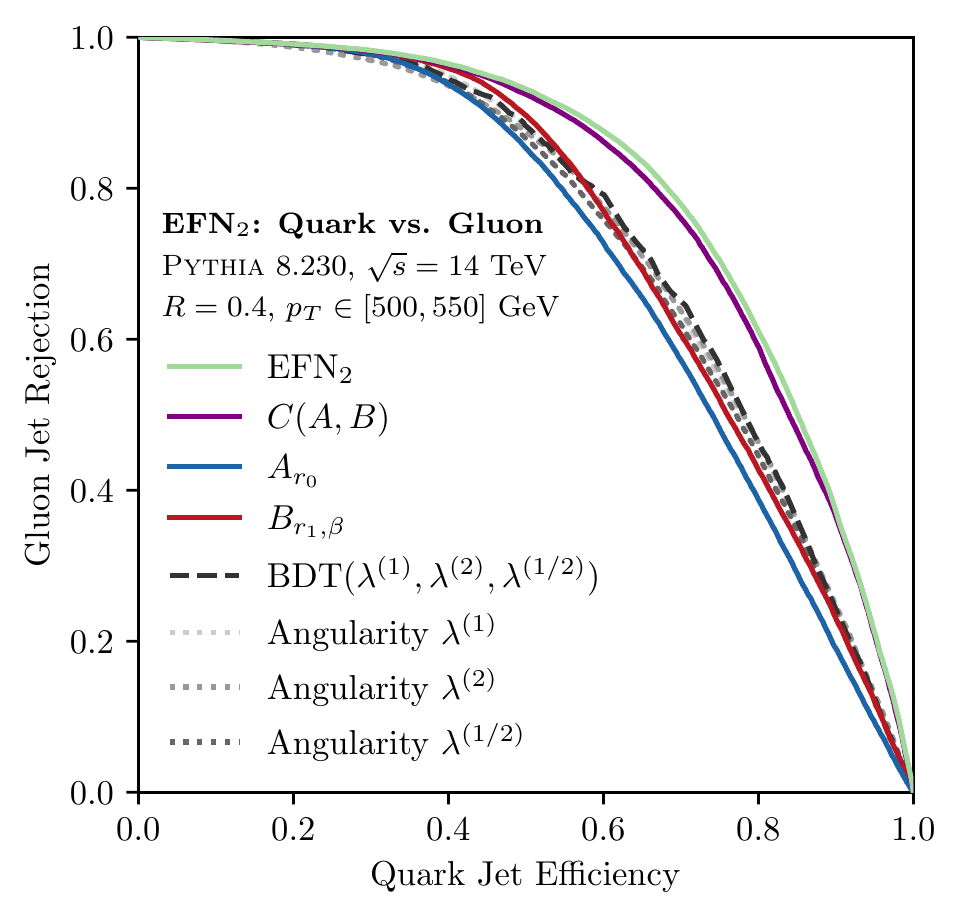}}
\subfloat[]{\includegraphics[width=0.5\columnwidth]{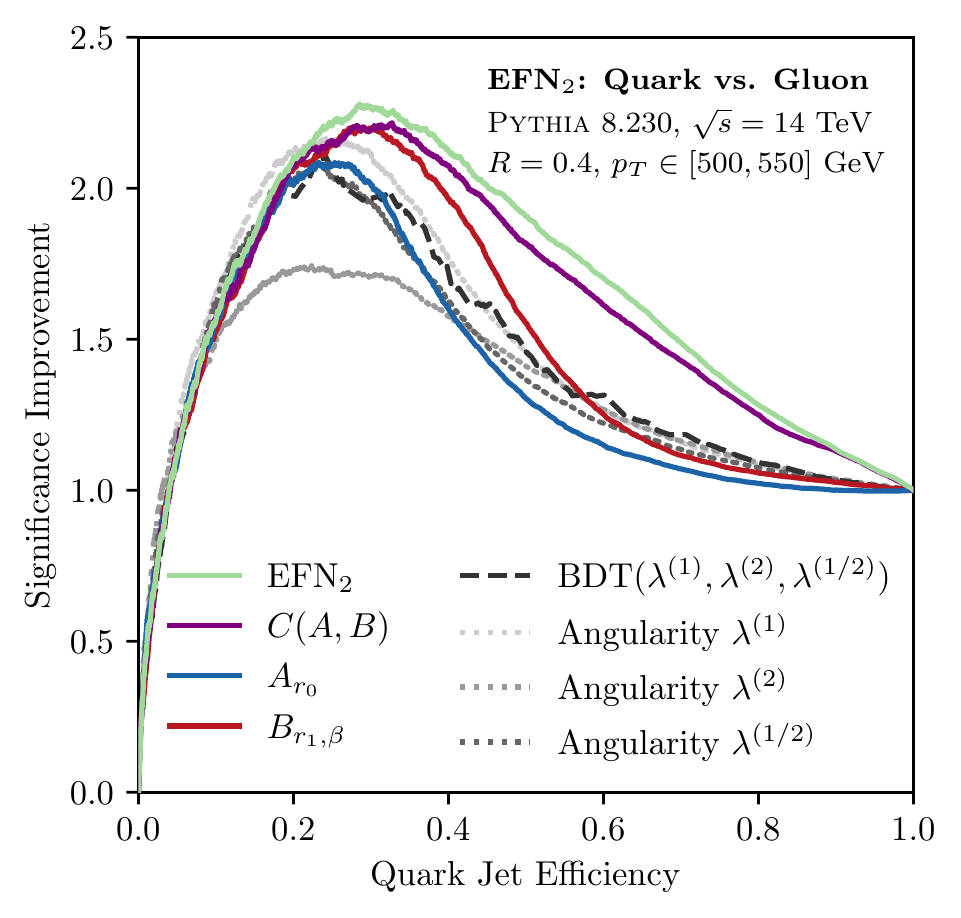}}
\caption{
The (a) ROC and (b) SI curves for the two closed-form observables $A_{r_0}$ and $B_{r_1,\beta}$ as well as their combination $C(A,B)$, compared to the trained $\ell = 2$ EFN model.
Three angularities are also shown for comparison, along with their corresponding performance when combined with a BDT.
While the two learned observables perform similarly to the angularities on an individual basis, they are evidently more informative than the angularities when combined.
The output of the trained EFN model and the closed-form estimate achieve similar performance.}
\label{fig:EFN2ROCComp}
\end{figure}

In \Figs{fig:EFN2filt1}{fig:EFN2filt2}, we show the learned filters $\Phi_1$ and $\Phi_2$ of a trained $\ell=2$ EFN.
It is evident that the filters exhibit approximate radial symmetry, with one of the filters concentrated at the center of the jet and the other activated at larger angular distances.
Thus, we can restrict our attention to functional forms which depend only on the rapidity-azimuth distance $\theta$ from the origin.
In particular, due to its built-in IRC-safety, the EFN model has learned filters that correspond to observables of the following approximate form:
\begin{align}\label{eq:learnedobsang}
&\mathcal O_1 = \sum_{i=1}^M z_i \, \Phi_1(\theta_i), &\mathcal O_2 = \sum_{i=1}^M z_i \, \Phi_2(\theta_i).
\end{align}
These are of the general form of IRC-safe angularities~\cite{Larkoski:2014pca} with a generic radially-symmetric angular weighting function~\cite{Gallicchio:2012ez}.\footnote{Following \Ref{Moult:2016cvt}, this section could alternatively be titled ``New angles on IRC-safe angularities.''}
To quantify the filters further, in \Fig{fig:EFN2filtcf} we plot the value of the learned filters as a function of the radial distance, taking an envelope over several radial slices.
The complementary central and wide-angle nature of the two filters are clearly evident.

By observing the properties of the curves in \Fig{fig:EFN2filtcf}, we fit two IRC-safe observables to the learned profiles of the following forms:
\begin{equation}
\label{eq:newangs}
A_{r_0} = \sum_i z_i \, e^{-\theta_i^2/r_0^2},\qquad B_{r_1,\beta} = \sum_i z_i \ln(1 + \beta (\theta_i - r_1) )\, \Theta(\theta_i - r_1),
\end{equation}
with values of $r_0=0.018$, $\beta=200$, and $r_1=0.015$ found to be suitable.
The observables in \Eq{eq:newangs} are then multiplied by overall factors of 0.60 and 0.18, respectively, to match the arbitrary normalization of the learned filters.
While the precise values and shapes of the observable profiles changed from training to training, these general forms emerged for several of the best-performing models.

The observables $A_{r_0}$ and $B_{r_1,\beta}$ in \Eq{eq:newangs} are IRC-safe angularities with a linear energy dependence and interesting angular weighting functions.
$A_{r_0}$ probes the collinear radiation near the core of the jet at angles $\theta \lesssim r_0$, and $B_{r_1,\beta}$ probes wide-angle radiation away from the core of the jet at angles $\theta > r_1$.
The separate treatment of collinear and wide-angle particles is unlike the behavior of the traditional angularities, which have explicit contributions from both collinear and wide-angle regions of phase space.
Though, as will be shown, each is individually a comparable quark/gluon jet classifier to the traditional angularities, the model is able to combine them in such a way as to achieve a significantly better performance.
It would be interesting to perform a first-principles QCD study to understand in what sense the separation of collinear and wide-angle behavior is beneficial for discrimination.

We now proceed to obtain a closed-form estimate of the learned function $F(\mathcal O_1,\mathcal O_2)$.
In \Fig{fig:modelscatter}, we populate the $(\mathcal O_1,\mathcal O_2)$ phase space by quark and gluon jets and color most of the populated space (with a mild threshold against outliers) according to the value of $F$ at that phase space point.
We also indicate the truth labels of one hundred quark and gluon jets to guide the eye.
Based on the transition from red to blue, we can see that the model selects a curved slice through the populated phase space region to obtain its predictions.\footnote{A similar strategy of fitting analytic functions to learned decision boundaries was carried out in \Ref{Likhomanenko:2015aba}.}
Correspondingly, we choose a closed-form estimate based on the simple (squared) Euclidean distance in phase space from a reference point $(a_0, b_0)$, namely:
\begin{equation}
\label{eq:estimatedF}
C(A,B) = (A-a_0)^2 + (B-b_0)^2,
\end{equation}
with $a_0=0.445$ and $b_0 = 0.545$, where, for ease of visualization, a sigmoid is also applied to $50\,(C(A,B) - 0.460^2)$ to monotonically rescale the predictions between 0 and 1 and approximate the value of the learned function.
We visualize $C(A,B)$ in \Fig{fig:analyticscatter}, finding a satisfying correspondence between the closed-form observables and the learned case.

Finally, we study the classification performance of these closed-form observables, compared to the trained $\ell=2$ EFN model and several IRC-safe angularities.
We consider three IRC-safe angularities $\lambda^{(\beta)} = \sum_i z_i \theta_i^\beta$ with $\beta=2$, $\beta=1$, and $\beta=1/2$, which are the jet mass, jet width, and Les Houches angularity, respectively \cite{Gras:2017jty}.
ROC curves for the various observables are shown in \Fig{fig:EFN2ROCComp}.
As single observables, the closed-form observables $A_{r_0}$ and $B_{r_1,\beta}$ of \Eq{eq:newangs} perform similarly to the individual angularities.
When combined via $C(A,B)$ of \Eq{eq:estimatedF}, they outperform all of the single angularities and approach the trained $\ell=2$ EFN performance.
By contrast, when the three angularities are combined with a Boosted Decision Tree (BDT), there is not a significant improvement in the classification performance.%
\footnote{The BDT classifier was implemented using scikit-learn~\cite{scikit-learn} using 100 estimators, trained and tested on 100,000 quark and gluon jets with a 50-50 train/test split.  We also tested combining $A_{r_0}$ and $B_{r_1,\beta}$ with a BDT, which resulted in similar performance to the closed-form and learned cases, verifying that the information is being adequately captured by $C(A,B)$.}
The fact that $A_{r_0}$ and $B_{r_1,\beta}$ can be combined to significantly increase the discrimination power indicates that, unlike the considered angularities, the new closed-form observables probe complementary information.

\section{Conclusions}
\label{sec:conc}

In this paper, we presented a new technique to learn from collider events in their natural representation as sets of particles.
Our approach relies on the Deep Sets theorem~\cite{DBLP:conf/nips/ZaheerKRPSS17}, which guarantees that a generic symmetric function can be represented by an additive latent space.
In the context of particle-level collider observables, each particle is mapped to a latent representation and then summed over, and observables are then functions on that latent space.
This decomposition encompassed a wide variety of existing event- and jet-level collider observables and representations, including image-based and moment-based methods.

While these observable decompositions are interesting in their own right, parameterizing them with neural networks yields a machine learning framework ideally suited for learning from variable-length unordered lists of particles.
We proposed two fundamental network variants.
The IRC-safe EFNs treat each latent space observable as an energy-weighted function of geometry, ensuring IRC safety in the latent space by construction.
The fully-general PFNs are able to incorporate additional particle-level information such as charge and flavor, maximizing the information available to achieve collider tasks.
We showcased the efficacy of these method in a quark/gluon discrimination case study, achieving favorable performance compared with many existing machine learning techniques for particle physics.

A fascinating aspect of the EFNs is that one can directly visualize the per-particle mapping into the latent space.
This allowed us to peer inside the network and discover that, in the context of quark/gluon discrimination, it had learned a dynamically-sized pixelization of the rapidity-azimuth plane.
This pixelization was reminiscent of a jet image but did not adhere to the strict rectilinear grid imposed by traditional jet images.
We also found some compelling evidence that the model had ``understood'' a key property of QCD, specifically its famous collinear singularity structure, since filters closer to the center of the jet were more finely resolved.
The relationship between the size and position of the filters could be roughly understood as a uniform pixelization in the natural angular emission plane.

We also presented a simple example of directly learning the physics used by the trained model and obtaining new closed-form observables.
The IRC-safe Observable Decomposition, combined with the visualizability of the EFN filters, provides a general way to access what the model has learned.
In the case of an $\ell = 2$ EFN, the model learned to separately probe collinear and wide-angle radiation and then use a phase space distance to classify quarks and gluons.
While the quark/gluon classification performance of $A_{r_0}$ and $B_{r_1,\beta}$ is not yet competitive with other simple observables such as the constituent multiplicity, together they notably outperform other IRC-safe angularities.
Even though our numerical choices of parameters are specific to these quark/gluon jet samples, it may be interesting to explore the $A_{r_0}$ and $B_{r_1,\beta}$ classes of observables further in the context of theoretical efforts to jointly explore and understand the correlations between two angularities~\cite{Larkoski:2014pca,Larkoski:2014tva,Procura:2018zpn}.

We conclude by discussing possible extensions and additional applications of these methods.
Pileup or underlying event mitigation could be an interesting avenue to explore with this architecture, similar to \Ref{Komiske:2017ubm} but also to PUPPI~\cite{Bertolini:2014bba} in that a per-particle correction factor could be applied.
Such an effort would need to make use of the equivariant designs of \Ref{\deepsets}, which have a separate output for each input.
One may also consider adding high-level information -- such as the jet four-momentum, individual substructure observables, or clustering histories -- directly to $\Phi$ or $F$ in order to improve the network performance.
Further, while we used jet-level examples for our case study, the problem of learning functions of sets applies at the event level more broadly.
At the event level, new challenges arise that the simple preprocessing performed here for the jet case studies may not completely solve.
An iterative deep sets technique, where multiple latent spaces are constructed, may help to capture the local/global intuition that collider physicists regularly employ to study the hierarchical relationships between events, jets, and particles.

\acknowledgments

We are grateful to Kyle Cranmer, Fr\'{e}d\'{e}ric Dreyer, Felice Frankel, Philip Harris, Michael Kagan, Gregor Kasieczka, Markus Klute, Andrew Larkoski, Sung Hak Lim, Benjamin Nachman, Siddharth Narayanan, Daniel Roberts, Daniel Whiteson, Michael Williams, and Manzil Zaheer for helpful conversations.
We are also grateful to Gregor Kasieczka for providing us with the top samples used in \App{app:toptag}.
We thank the Harvard Center for the Fundamental Laws of Nature for hospitality while this work was completed.
This work was supported by the Office of Nuclear Physics of the U.S. Department of Energy (DOE) under grant DE-SC-0011090 and by the DOE Office of High Energy Physics under grant DE-SC-0012567.
JT is supported by the Simons Foundation through a Simons Fellowship in Theoretical Physics.
Cloud computing resources were provided through a Microsoft Azure for Research award.

\appendix

\section{Details of models and observables}
\label{app:models}

In this appendix, we describe the details of the machine learning models and observables.
All of the models employed for comparison with the EFN/PFN models were applied to our quark/gluon samples in \Sec{sec:qg}, but only the linear EFP model will be applied to the top samples in \App{app:toptag}.
For the substructure observables, the jet mass $m$ and constituent multiplicity $n_\text{const}$ are easily obtained from the jet in \textsc{FastJet}. 
Using \textsc{FastJet contrib} 1.033~\cite{fjcontrib}, the \textsc{RecursiveTools} 2.0.0-beta1 module was used to calculate the soft drop multiplicity $n_\text{SD}$~\cite{Frye:2017yrw} with parameters $\beta=-1$, $z_\text{cut}=0.005$, and $\theta_\text{cut}=0$.
The \textsc{Nsubjettiness} 2.2.4 module was used to calculate all $N$-subjettiness values $\tau_N^{(\beta)}$~\cite{Thaler:2010tr,Thaler:2011gf}.
All neural networks were implemented in Keras~\cite{keras} with the TensorFlow~\cite{tensorflow} backend and trained on NVIDIA Tesla K80 GPUs on Microsoft Azure.

Here, we provide details for the machine learning models listed in \Tab{tab:obsandmodels}:

\begin{itemize}

\item \textbf{EFN}, \textbf{PFN}: Both the EFN and PFN architectures are contained as part of our \href{https://energyflow.network}{\tt EnergyFlow} Python package~\cite{energyflow}, which contains examples demonstrating their training, evaluation, and use.
Keras requires contiguous NumPy~\cite{numpy} arrays as input, so the events are padded with all-zero particles to ensure they have equal length.
This zero-padding is a technical trick, not a conceptual limitation.  It is distinct from the zero-padding necessary to use variable-length inputs in a DNN.
This operation can be done on a per-batch level, but to avoid repeating this procedure at each epoch, we zero-pad all the events once at the beginning.
The Keras {\tt Masking} layer is used to ignore all-zero particles during the training and evaluation of the network.
The Keras {\tt TimeDistributed} layer is used to apply the function $\Phi$ to each particle, which in the relevant language is a ``time step''.
To carry out the summation in the latent space, we use the Keras {\tt Dot} layer for the EFN architecture, to product and sum the transverse momentum fractions with the latent observables, and the {\tt Lambda} layer for the PFN architecture, to sum over the particles.
It may be interesting to consider alternative ``pooling'' functions to summation, such as max-pooling or average-pooling as in \Refs{DBLP:journals/corr/abs-1709-03019,DBLP:conf/cvpr/QiSMG17}.

\item \textbf{RNN-ID}, \textbf{RNN}: Recurrent Neural Networks provide an alternative way to process variable-length inputs, albeit one that is explicitly not agnostic to the order of the inputs.
We choose to order the particles by their transverse momenta and train two variants: one with only kinematic information (RNN) and one with kinematic as well as full particle ID information (RNN-ID).
The former should be compared with the PFN model and the latter to the PFN-ID model.
We did not consider alternative particle orderings for the RNN, though other investigations have found that performance is robust to choices in particle ordering~\cite{Louppe:2017ipp,Andreassen:2018apy}.
Our RNN architecture consists of an {\tt LSTM} layer with 64 nodes (the performance was insensitive to changing this value to 128 or 256) followed by three fully-connected layers each with 100 nodes and a fully-connected output layer with 2 nodes. 
Due to the {\tt Masking} layer employed, the significantly faster {\tt CuDNNLSTM} layer was not used and the batch size was taken to be 2000 to help speed up training.
An architecture using two or three {\tt SimpleRNN} layers was also tried but the performance was not as good as with an {\tt LSTM} layer.
The long training time of the RNN models was prohibitive in exploring additional hyperparameters.

\item \textbf{EFPs}: The energy flow basis~\cite{Komiske:2017aww} is a linear basis for the space of IRC-safe jet substructure observables, allowing linear methods to be applied to a set of Energy Flow Polynomials with good performance.
The \href{https://energyflow.network}{\texttt{EnergyFlow}} 0.8.2 Python package~\cite{energyflow} was used to compute all EFPs with degree $d\le7$ and complexity $\chi\le3$, using the normalized default hadronic measure with $\beta=0.5$.
The same EFPs with $\beta=1$ were also tested and found to perform slightly worse.
These 996 EFPs, including the trivial constant EFP, were used to train a Fisher's Linear Discriminant model from the scikit-learn package~\cite{scikit-learn}.

\item \textbf{DNN}: The $N$-subjettiness basis~\cite{Datta:2017rhs} is a $K$-body phase space basis consisting of the following $3K-4$ observables:
\begin{equation}
\label{eq:nsubbasis}
\left\{\tau_1^{(1/2)},\,\tau_1^{(1)},\,\tau_1^{(2)},\tau_2^{(1/2)},\,\tau_2^{(1)},\,\tau_2^{(2)},\,\ldots,\,\tau_{K-2}^{(1/2)},\,\tau_{K-2}^{(1)},\,\tau_{K-2}^{(2)},\tau_{K-1}^{(1/2)},\,\tau_{K-1}^{(1)}\right\}.
\end{equation}
We use $K=25$ and, following \Ref{Datta:2017rhs}, compute the $N$-subjettiness values with respect to $k_T$ axes.
A Dense Neural Network consisting of three fully-connected layers each with 100 nodes and one fully-connected output layer with 2 nodes was trained on this set of $N$-subjettiness values.

\item \textbf{CNN}: Following \Ref{Komiske:2016rsd}, we compute $33\times33$ one-channel (grayscale) jet images in a $2R\times2R$ patch of the rapidity-azimuth plane.
Images were preprocessed as in \Refs{Komiske:2016rsd,Komiske:2018vkc} by subtracting the mean image of the training set and dividing by the per-pixel standard deviation.
A Convolutional Neural Network architecture was trained on the jet images consisting of three convolutional layers with 48, 32, and 32 filters of size $8\times8$, $4\times4$, and $4\times4$, respectively. 
These layers were followed by two fully-connected layers each with 50 nodes and one fully-connected output layer with 2 nodes.
Maxpooling of size $2\times2$ and dropout with a rate of 0.1 was implemented after each convolutional layer.

\end{itemize}

\begin{table}[t]
\centering
\begin{tabular}{|c|c|c|}
\hline
\textbf{Model} & \textbf{Time per Epoch (sec)} & \textbf{Number of Epochs} \\
\hline\hline
PFN-ID & 73 & 47 \\
PFN-Ex & 67 & 37 \\
PFN-Ch & 64 & 51 \\
PFN & 71 & 34 \\
EFN & 74 & 54 \\
\hline
RNN-ID & 192 & 61 \\ 
RNN & 188 & 96 \\
DNN & 9 & 39 \\
CNN & 63 & 22 \\
\hline
\end{tabular}
\caption{
Median times per epoch and numbers of epochs during training of the considered machine learning models.
The linear EFP model is not shown because it is trained by a different technique.
Note the substantially longer training times for our RNN implementations compared to the EFN and PFN models.
}
\label{tab:timing}
\end{table}

For all quark/gluon models, one million jets were used for training, 200k for validation (except for the EFPs, which did not use validation), and 200k for testing.
Common properties to all neural networks include the use of ReLU~\cite{relu} activation functions for all non-output layers, a softmax activation function on the 2 node output layer, He-uniform weight initialization~\cite{heuniform}, the categorical crossentropy loss function, the Adam optimization algorithm~\cite{adam}, and a learning rate of 0.001.
All non-RNN networks had a batch size of 500; all non-CNN networks had a patience parameter of 8 with the CNN having a patience parameter of 5.
Each non-EFN/PFN model was trained twice and the model with the higher AUC on the test set is reported (except when $\varepsilon_g$ at $\varepsilon_q=50\%$ values are reported, in which case the model with the value of this statistic is reported).
In \Tab{tab:timing}, we report the typical training times for each of the models.

\section{Tagging top jets}
\label{app:toptag}

In this appendix, we train EFN and PFN models to discriminate top jets from the QCD jet background to provide an additional demonstration of the excellent performance of these models.
Since top jets do not have any singularities associated with radiation about their center, training EFN models on tops provides an important cross check about our conclusions in \Sec{sec:visualizeqcd} for how the model is learning to pixelize the rapidity-azimuth plane.

The top and QCD jets used in this study are based on the dataset used in \Ref{Butter:2017cot}, which were provided to us by those authors.
The jets are \pythia-generated, anti-$k_T$, $R=0.8$ jets at $\sqrt{s}=14$ TeV with no underlying event or pileup using a \textsc{Delphes}~\cite{deFavereau:2013fsa} detector fast-simulation of the ATLAS detector.
Jets are kept if they have transverse momentum $p_T\in [550,650]$ GeV and pseudorapidity $|\eta|<2$ and if they can be matched to a top parton to within a rapidity-azimuth distance of $\Delta R = 0.8$.
The jets are required to be fully merged, with the three quarks from the top decay contained within the jet radius.
See \Ref{Butter:2017cot} for additional information about the samples and their generation details, as well as information about other top-tagging algorithms.
There are 1.2 million training events, with 400k validation events and 400k test events.
The jet samples do not contain particle-level flavor or charge information, and thus we can only train EFN and PFN models that make use of kinematic information alone.
A linear EFP model is also trained for comparison.

\begin{figure}[t]
\centering
\includegraphics[width=0.6\columnwidth]{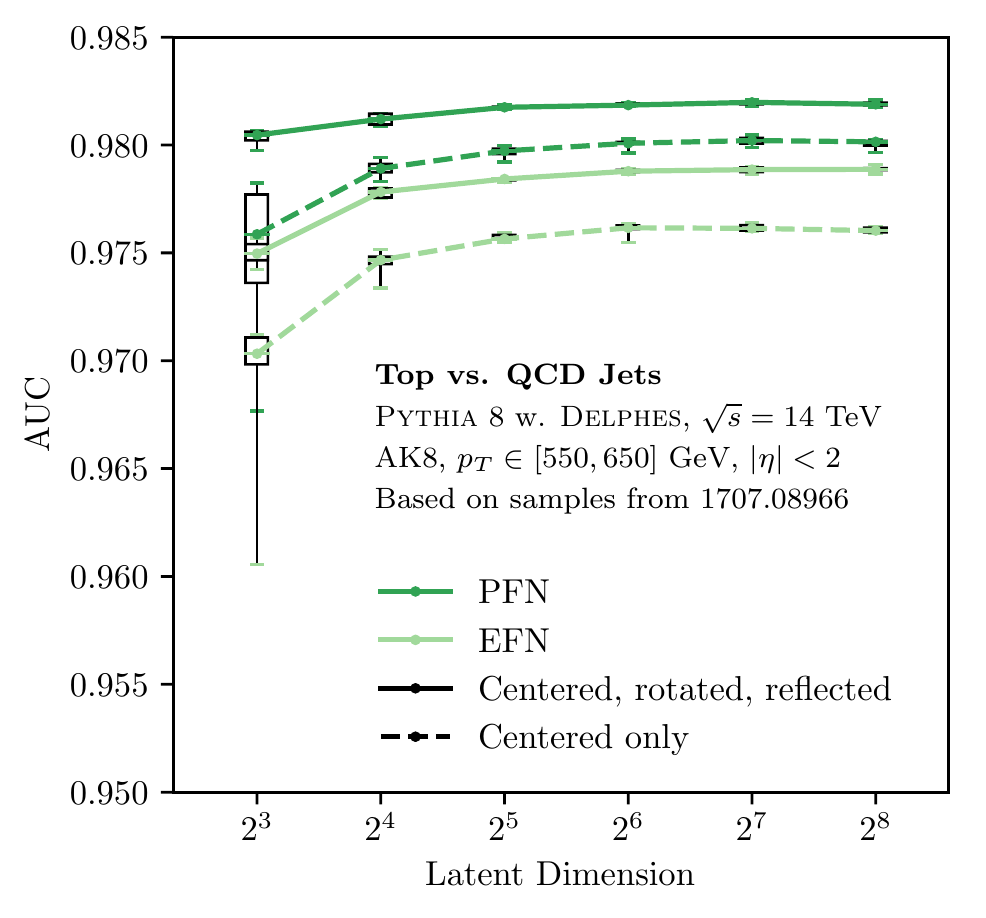}
\caption{The AUC performance of the EFN/PFN top tagging models as a function of the latent dimension, which is varied from 8 to 256 in powers of 2.
Including rotation and reflection in the rapidity-azimuth plane as preprocessing steps (solid curves) improves the model performance significantly compared to only centering the jets (dashed curves).
The spread in values is due to retraining the models ten times.}
\label{fig:topdimsweep}
\end{figure}

Given the different topology typical of top jets compared to quark or gluon jets, we implement some additional preprocessing steps designed to help the model train more efficiently.
As with the quark and gluon jets of \Sec{sec:qg}, we center all of the jets in the rapidity-azimuth plane based on the four-momentum of the jet, and we normalize the transverse momenta of the particles to sum to one.
Models were trained with just this minimal preprocessing, as well as with additional rotation and reflection (r.r.) operations.
For the EFN-r.r.\ and PFN-r.r.\ models, rotations were performed to align the leading eigenvector of the two-dimensional moment of the radiation pattern along the vertical axis, and reflections were performed about the horizontal and vertical axes to place the highest-$p_T$ quadrant in a consistent quadrant.

\begin{figure}[t]
\centering
\subfloat[]{\includegraphics[width=0.5\columnwidth]{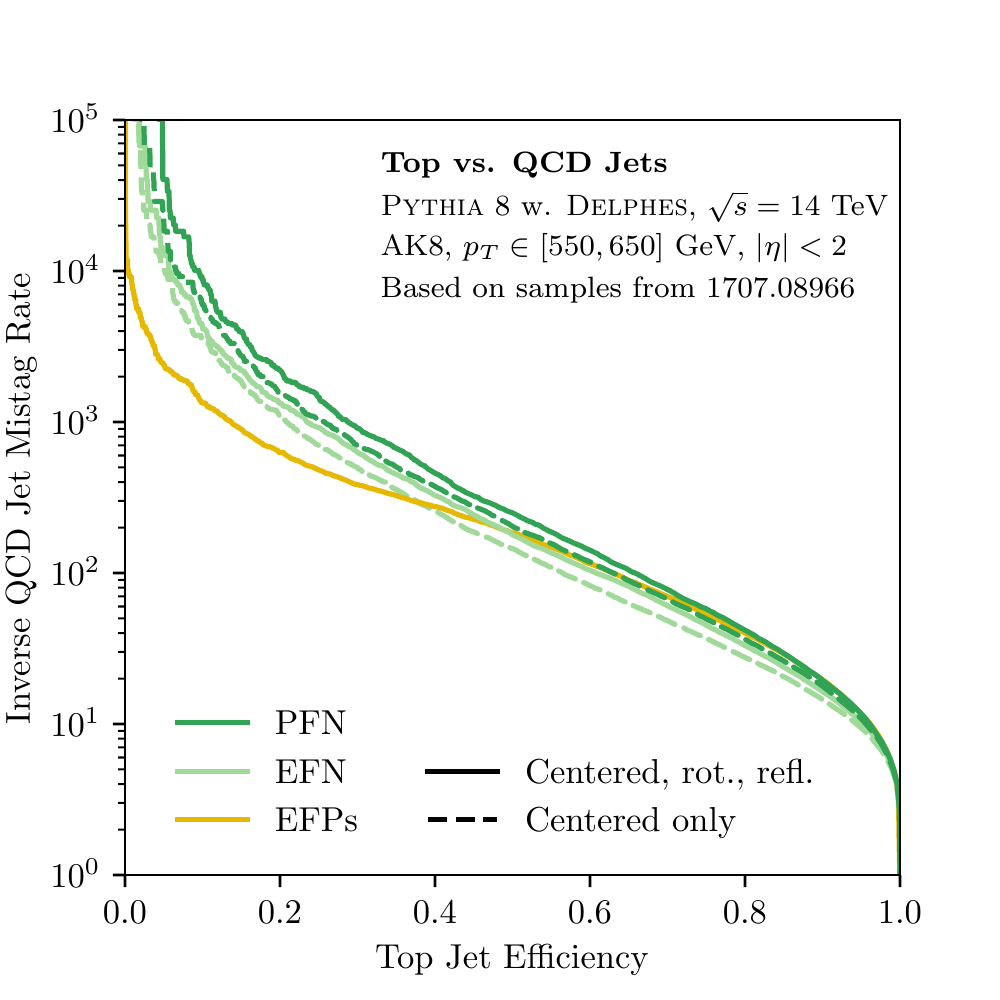}}
\subfloat[]{\includegraphics[width=0.5\columnwidth]{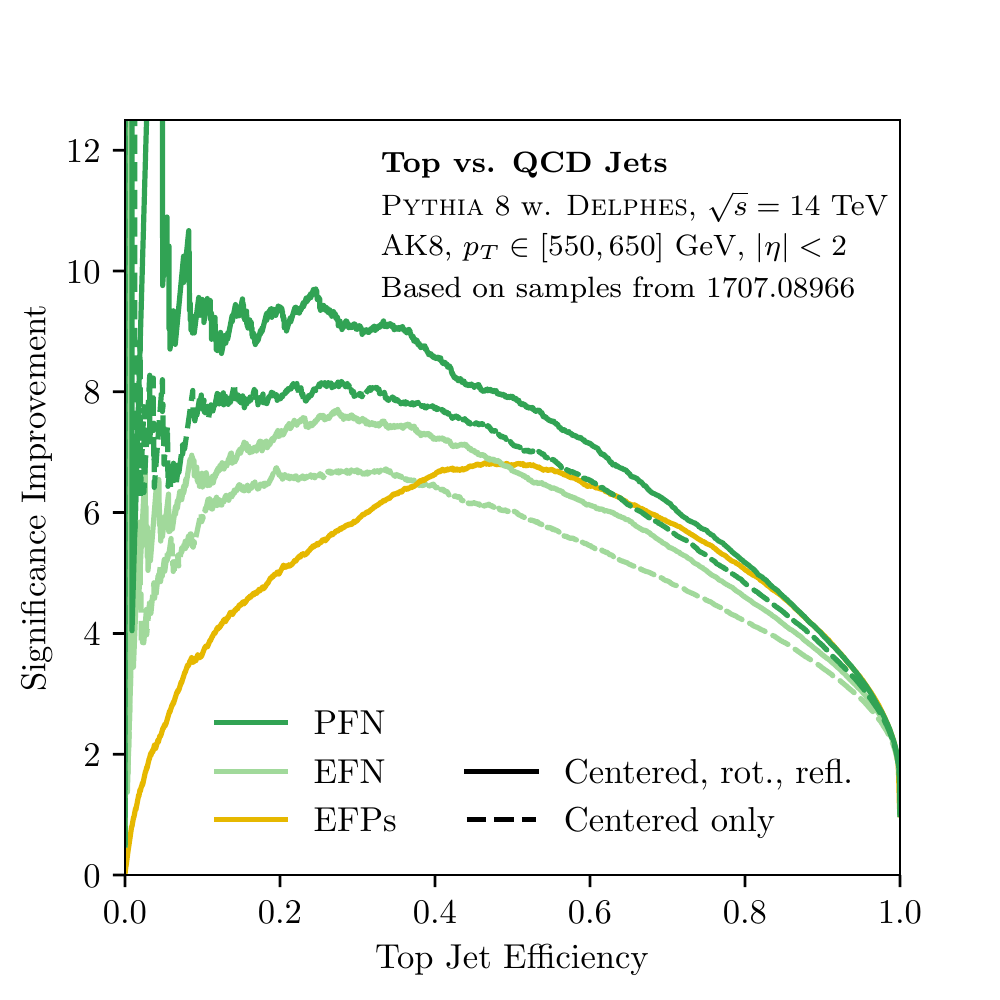}}
\caption{The (a) inverse ROC and (b) SI curves for each of the EFN/PFN top tagging models with latent dimension 256, compared to the EFP linear model.
The curve with the median AUC out of ten independent trainings is shown.
As expected, the PFN yields better performance than the EFN, with the additional rotation/reflection preprocessing steps providing a further boost in performance.}
\label{fig:toproccomp}
\end{figure}

The results of training EFN and PFN models for top tagging are shown in \Fig{fig:topdimsweep} with the latent dimension of the models varying from 8 to 256 in powers of 2.
ROC and SI curves of the trained models are shown in \Fig{fig:toproccomp} and also compared to the linear EFP model.
Performance competitive with the results in \Ref{Butter:2017cot} is achieved, particularly for the PFN models.
The preprocessing step of rotating and reflecting can be seen to notably improve both the EFN and PFN classification performance.
In order to facilitate future comparisons with other models trained on these samples, in \Tab{tab:topstats} we report the AUC and gluon background rejection factor at both 0.3 and 0.5 quark efficiency.

\begin{table}[t]
\centering
\begin{tabular}{|l|c|c|c|}
\hline
\multicolumn{1}{|c|}{\textbf{Model}}  & {\bf AUC} & {\bf $\boldsymbol{1/\varepsilon_b}$ at  $\boldsymbol{\varepsilon_s=50\%}$} & {\bf $\boldsymbol{1/\varepsilon_b}$ at  $\boldsymbol{\varepsilon_s=30\%}$} \\
\hline \hline
PFN-r.r. & ${\bf 0.9819}\pm0.0001$ & ${\bf 247}\pm3$ & ${\bf 888}\pm17$\\
PFN & $0.9801\pm0.0001$ & $203\pm1$ & $732\pm21$\\
\hline
EFN-r.r. & $0.9789\pm0.0001$ & $181\pm2$ & $619\pm23$\\ 
EFN & $0.9760\pm0.0001$ & $143\pm2$ & $481\pm12$\\ 
\hline
EFPs & 0.9803 & 184 & 384\\
\hline
\end{tabular}
\caption{Quantified classification performance (AUC, $1/\varepsilon_b$ at $\varepsilon_s=0.5$, $1/\varepsilon_b$ at $\varepsilon_s=0.3$) for each of the models trained in \Fig{fig:toproccomp}.
Reported uncertainties are half of the interquartile range over ten trainings.
The PFNs achieve the best performance, with improvements seen by including the rotation and reflection preprocessing.
The EFPs slightly outperform the EFNs in AUC and background rejection at 0.5 signal efficiency but perform more poorly in background rejection at 0.3 signal efficiency.}
\label{tab:topstats}
\end{table}

\begin{figure}[p]
\centering
\subfloat[]{\includegraphics[width=0.5\columnwidth]{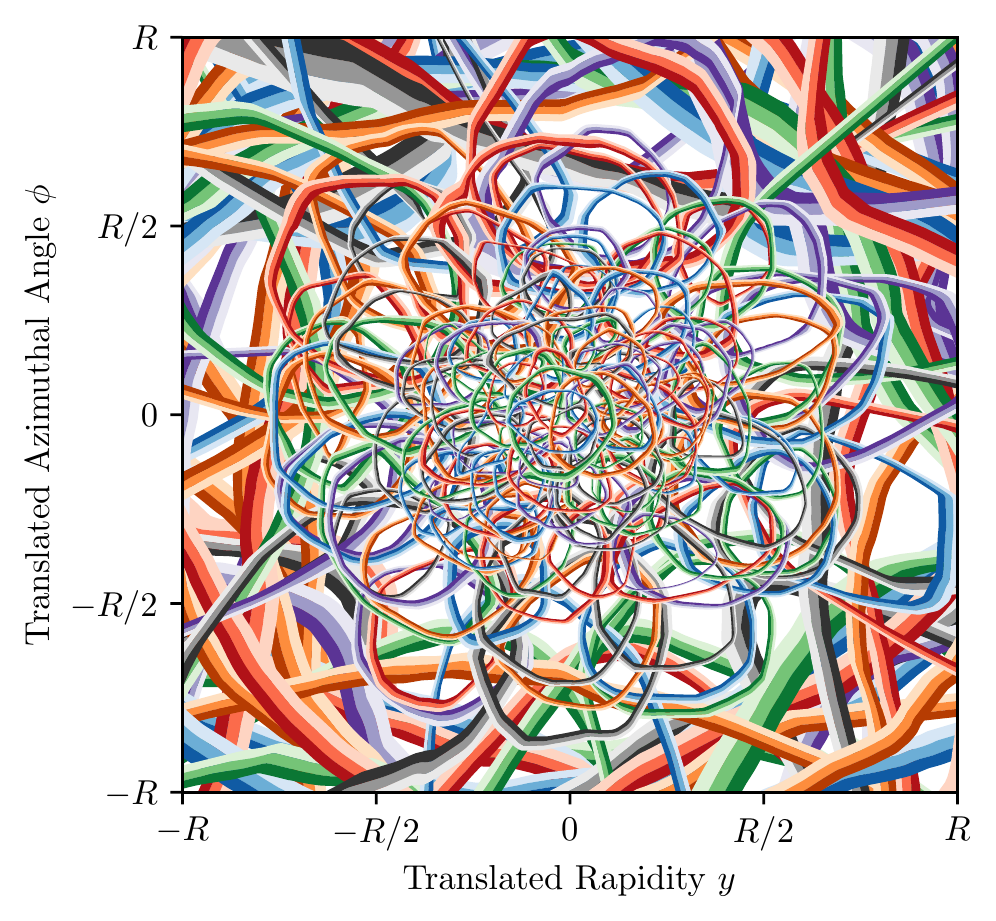}}
\subfloat[]{\includegraphics[width=0.5\columnwidth]{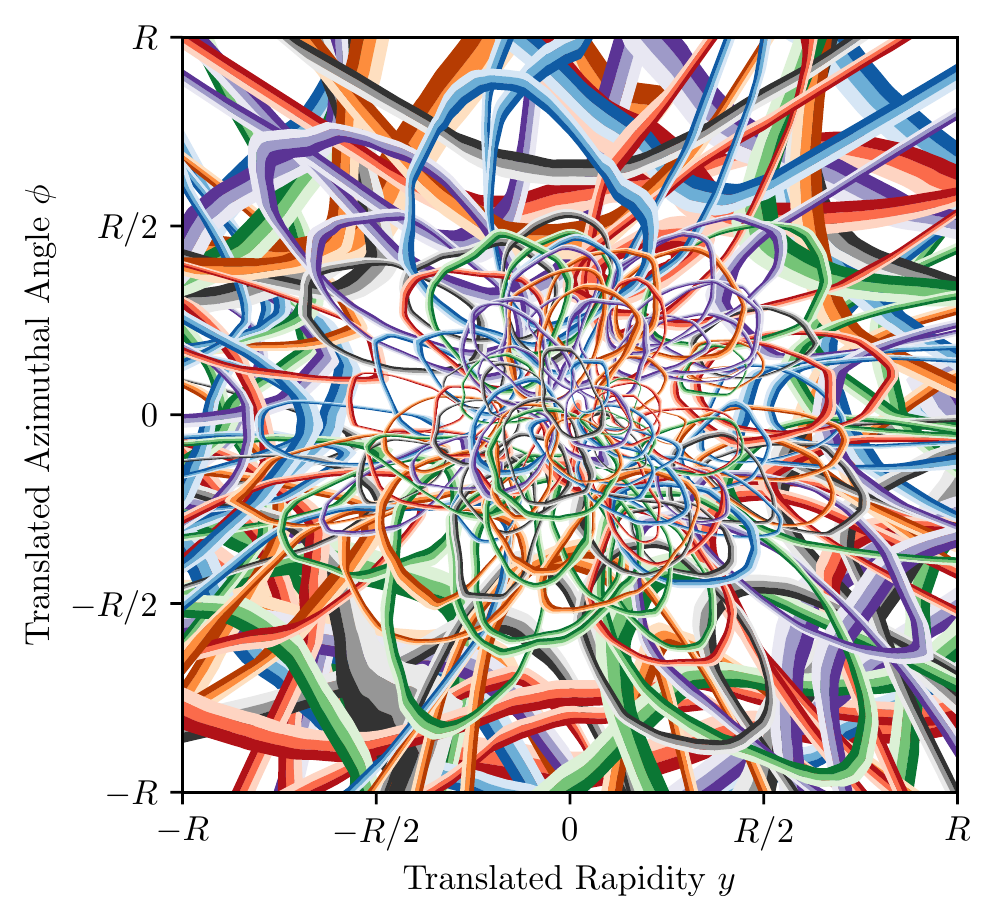}}
\caption{Visualization of the learned filters of a top tagging $\ell = 256$ EFN (a) without any additional preprocessing and (b) with additional rotation/reflection.
In (b), the rotational symmetry around the center is broken compared to (a).}
\label{fig:topvisual}
\end{figure}

\begin{figure}[p]
\centering
\subfloat[]{\includegraphics[width=0.5\columnwidth]{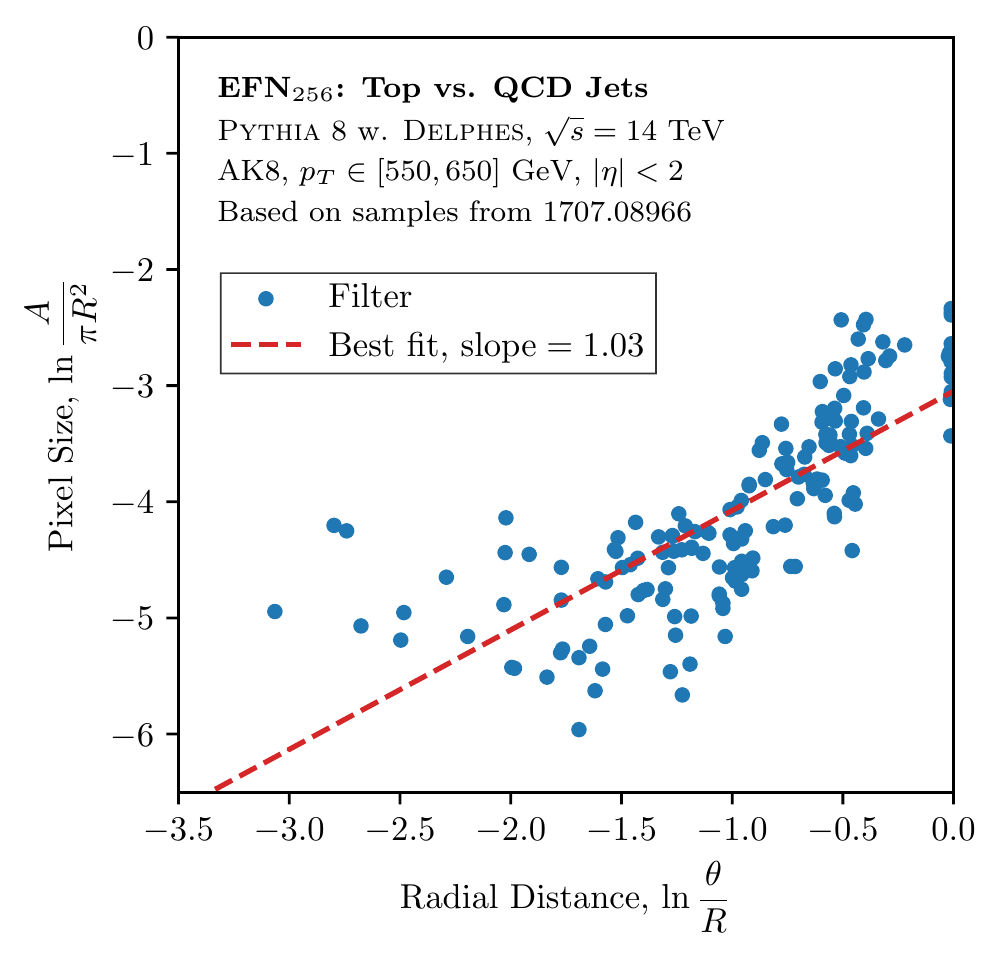}}
\subfloat[]{\includegraphics[width=0.5\columnwidth]{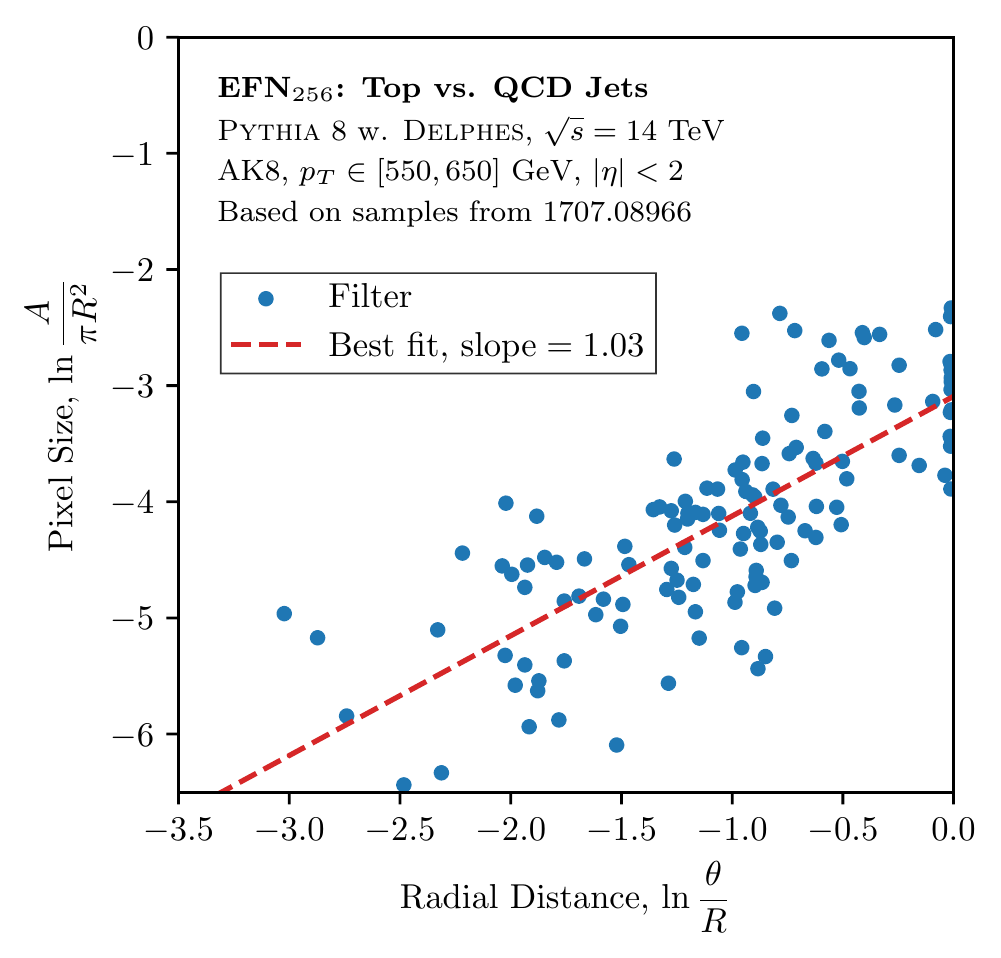}}
\caption{The filter sizes plotted as a function of their distance from the jet center for a top tagging $\ell = 256$ EFN (a) without any additional preprocessing and (b) with additional rotation/reflection.
Though a general trend is seen that pixels closer to the center are smaller, the points are generally not as well fit by a line as in the quark/gluon case in \Fig{fig:qgquant}.}
\label{fig:topquant}
\end{figure}

For the EFNs, we can also visualize the learned filters using the technique of \Sec{sec:visualizeqcd}.
The resulting filter visualizations for latent dimension 256 are shown in \Fig{fig:topvisual}.
The learned filters have some tendency to be smaller near the center and larger near the periphery, but not nearly as much as in the quark/gluon discrimination case in \Fig{fig:visualdemo}.
This is expected because given the typical three-prong topology of a boosted top jet, the jet axis does not have any distinguished radiation associated with it, unlike for a QCD jet where the jet axis tends to lie along a core of radiation.
To quantify this effect, we look at the size of the filters as a function of their distance from the origin, shown in \Fig{fig:topquant}.
The relationship in the top tagging case is much weaker than the linear relationship present in the quark/gluon discrimination study, with significantly worse linear fits.

\section{Additional visualizations}
\label{app:addvis}

\begin{figure}[p]
\centering
\includegraphics[width=0.7\columnwidth]{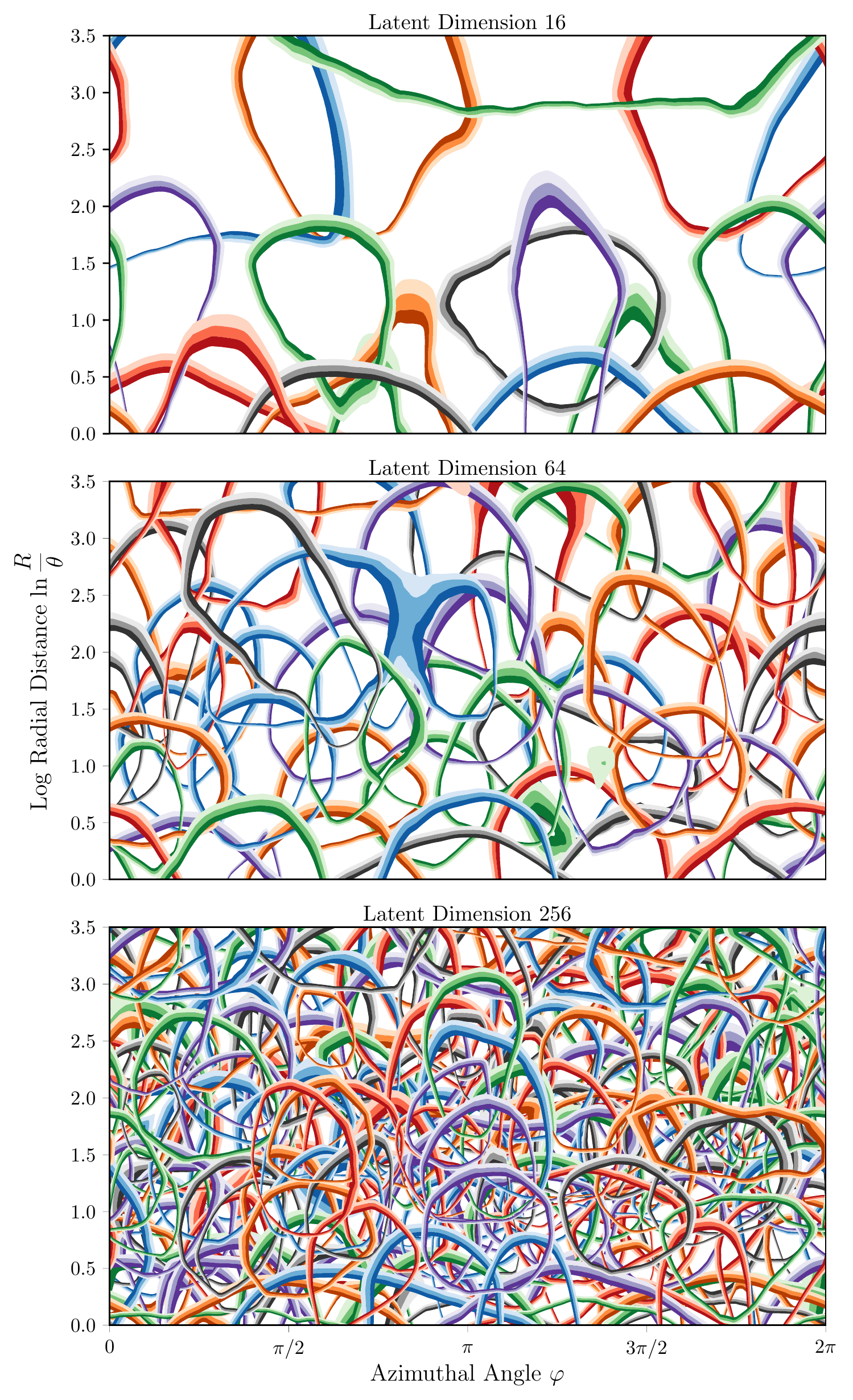}
\caption{Visualization of the learned filters of a quark/gluon EFN in the $(\ln\frac{R}{\theta},\varphi)$ plane with latent dimensions of (top) 16, (middle) 64, and (bottom) 256.
The pixelization learned by the network is much more regular and uniform in this space compared to the rapidity-azimuth plane in \Fig{fig:qgvisual}.
}
\label{fig:qglundvisual2}
\end{figure}

For the quark/gluon EFN model in \Sec{sec:qg}, the observed relationship between the size and location of the filters qualitatively (in \Fig{fig:qgvisual}) and quantitatively (in \Fig{fig:qgquant}) suggested that the model learned a uniform pixelization in the $\left(\ln\frac{R}{\theta},\varphi\right)$ emission plane.
(The top tagging EFN model in \App{app:toptag} did not exhibit as clear of a relationship in either \Fig{fig:topvisual} or \Fig{fig:topquant}.)
To directly visualize the EFN filters in the appropriate space, we implement a change of variables from Cartesian $(y,\,\phi)$ coordinates to polar $(\theta,\varphi)$ coordinates.
The visualizations in \Fig{fig:qglundvisual2} use the same contouring and overlaying technique of \Fig{fig:visualdemo} to demonstrate the roughly uniform pixelation in the emission plane for the quark/gluon EFN models with latent dimensions of 16, 64, and 256.
In this emission plane, the learned filters can be seen to be much more uniform in size and location compared to the rapidity-azimuth plane.
The uniformity as a function of $\varphi$ indicates that the set of filters approximately has rotational symmetry.
We also checked that the model output typically changed by less than 0.1 after applying a random rotation about the jet axis in the rapidity-azimuth plane.

\begin{figure}[t]
\centering
\includegraphics[width=\columnwidth]{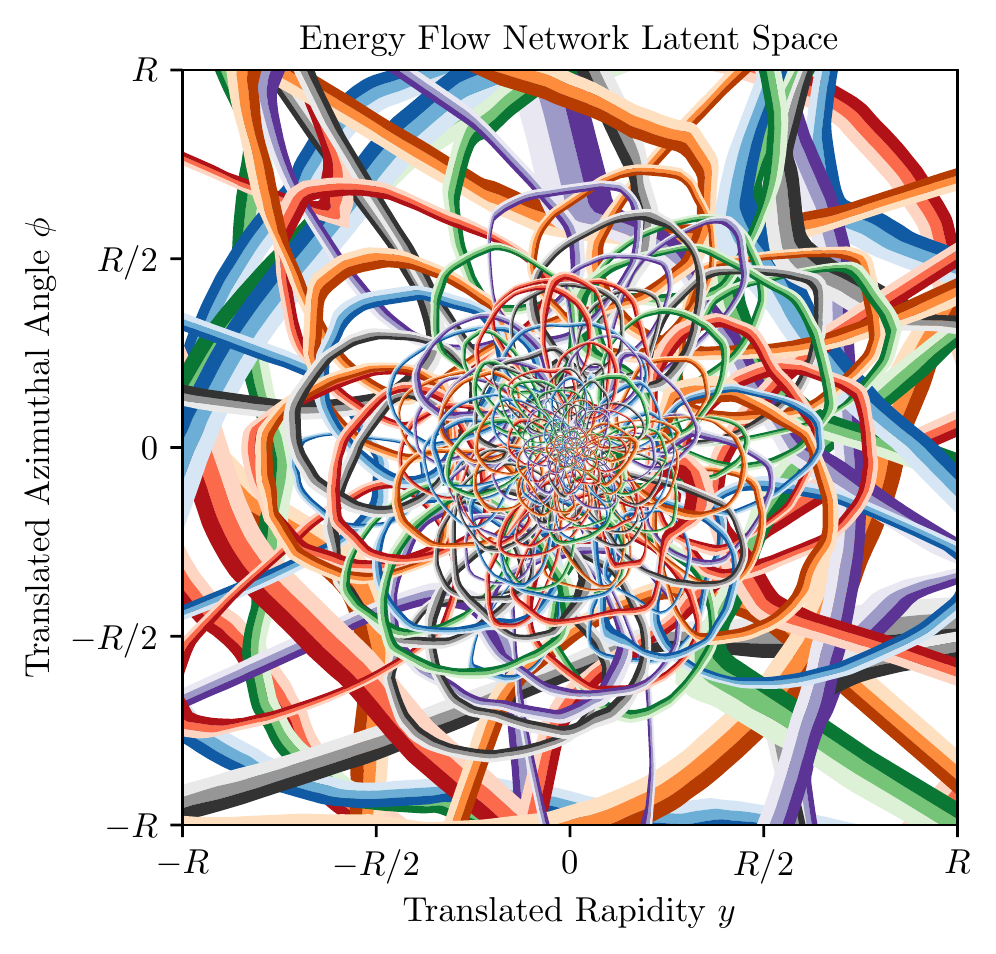}
\caption{A larger scale visualization of the filters for the quark/gluon $\ell=256$ EFN model, shown originally in \Fig{fig:qgvisual}.}
\label{fig:qg256visual}
\end{figure}

Finally, we show a larger version of the quark/gluon $\ell = 256$ EFN model from \Fig{fig:qgvisual} in \Fig{fig:qg256visual}, suitable for framing.

\afterpage{\clearpage}

\bibliography{efns}

\providecommand{\href}[2]{#2}\begingroup\raggedright\begin{thebibliography}{100}

\bibitem{Beaudette:2014cea}
{\scshape CMS} collaboration, F.~Beaudette, \emph{{The CMS Particle Flow
  Algorithm}},  in \emph{{Proceedings, International Conference on Calorimetry
  for the High Energy Frontier (CHEF 2013): Paris, France, April 22-25, 2013}},
  pp.~295--304, 2013, \href{https://arxiv.org/abs/1401.8155}{{\ttfamily
  1401.8155}}.

\bibitem{Sirunyan:2017ulk}
{\scshape CMS} collaboration, A.~M. Sirunyan et~al., \emph{{Particle-flow
  reconstruction and global event description with the CMS detector}},
  \href{https://doi.org/10.1088/1748-0221/12/10/P10003}{\emph{JINST} {\bfseries
  12} (2017) P10003} [\href{https://arxiv.org/abs/1706.04965}{{\ttfamily
  1706.04965}}].

\bibitem{Aaboud:2017aca}
{\scshape ATLAS} collaboration, M.~Aaboud et~al., \emph{{Jet reconstruction and
  performance using particle flow with the ATLAS Detector}},
  \href{https://doi.org/10.1140/epjc/s10052-017-5031-2}{\emph{Eur. Phys. J.}
  {\bfseries C77} (2017) 466}
  [\href{https://arxiv.org/abs/1703.10485}{{\ttfamily 1703.10485}}].

\bibitem{Cogan:2014oua}
J.~Cogan, M.~Kagan, E.~Strauss and A.~Schwarztman, \emph{{Jet-Images: Computer
  Vision Inspired Techniques for Jet Tagging}},
  \href{https://doi.org/10.1007/JHEP02(2015)118}{\emph{JHEP} {\bfseries 02}
  (2015) 118} [\href{https://arxiv.org/abs/1407.5675}{{\ttfamily 1407.5675}}].

\bibitem{deOliveira:2015xxd}
L.~de~Oliveira, M.~Kagan, L.~Mackey, B.~Nachman and A.~Schwartzman,
  \emph{{Jet-images --- deep learning edition}},
  \href{https://doi.org/10.1007/JHEP07(2016)069}{\emph{JHEP} {\bfseries 07}
  (2016) 069} [\href{https://arxiv.org/abs/1511.05190}{{\ttfamily
  1511.05190}}].

\bibitem{Baldi:2016fql}
P.~Baldi, K.~Bauer, C.~Eng, P.~Sadowski and D.~Whiteson, \emph{{Jet
  Substructure Classification in High-Energy Physics with Deep Neural
  Networks}}, \href{https://doi.org/10.1103/PhysRevD.93.094034}{\emph{Phys.
  Rev.} {\bfseries D93} (2016) 094034}
  [\href{https://arxiv.org/abs/1603.09349}{{\ttfamily 1603.09349}}].

\bibitem{Barnard:2016qma}
J.~Barnard, E.~N. Dawe, M.~J. Dolan and N.~Rajcic, \emph{{Parton Shower
  Uncertainties in Jet Substructure Analyses with Deep Neural Networks}},
  \href{https://doi.org/10.1103/PhysRevD.95.014018}{\emph{Phys. Rev.}
  {\bfseries D95} (2017) 014018}
  [\href{https://arxiv.org/abs/1609.00607}{{\ttfamily 1609.00607}}].

\bibitem{Komiske:2016rsd}
P.~T. Komiske, E.~M. Metodiev and M.~D. Schwartz, \emph{{Deep learning in
  color: towards automated quark/gluon jet discrimination}},
  \href{https://doi.org/10.1007/JHEP01(2017)110}{\emph{JHEP} {\bfseries 01}
  (2017) 110} [\href{https://arxiv.org/abs/1612.01551}{{\ttfamily
  1612.01551}}].

\bibitem{ATL-PHYS-PUB-2017-017}
{\scshape ATLAS} collaboration, \emph{{Quark versus Gluon Jet Tagging Using Jet
  Images with the ATLAS Detector}},  Tech. Rep. ATL-PHYS-PUB-2017-017, CERN,
  Geneva, Jul, 2017.

\bibitem{Kasieczka:2017nvn}
G.~Kasieczka, T.~Plehn, M.~Russell and T.~Schell, \emph{{Deep-learning Top
  Taggers or The End of QCD?}},
  \href{https://doi.org/10.1007/JHEP05(2017)006}{\emph{JHEP} {\bfseries 05}
  (2017) 006} [\href{https://arxiv.org/abs/1701.08784}{{\ttfamily
  1701.08784}}].

\bibitem{Bhimji:2017qvb}
W.~Bhimji, S.~A. Farrell, T.~Kurth, M.~Paganini, Prabhat and E.~Racah,
  \emph{{Deep Neural Networks for Physics Analysis on low-level whole-detector
  data at the LHC}},  in \emph{{18th International Workshop on Advanced
  Computing and Analysis Techniques in Physics Research (ACAT 2017) Seattle,
  WA, USA, August 21-25, 2017}}, 2017,
  \href{https://arxiv.org/abs/1711.03573}{{\ttfamily 1711.03573}}.

\bibitem{Macaluso:2018tck}
S.~Macaluso and D.~Shih, \emph{{Pulling Out All the Tops with Computer Vision
  and Deep Learning}},  \href{https://arxiv.org/abs/1803.00107}{{\ttfamily
  1803.00107}}.

\bibitem{Guo:2018hbv}
J.~Guo, J.~Li, T.~Li, F.~Xu and W.~Zhang, \emph{{Deep learning for the R-parity
  violating supersymmetry searches at the LHC}},
  \href{https://arxiv.org/abs/1805.10730}{{\ttfamily 1805.10730}}.

\bibitem{Dreyer:2018nbf}
F.~A. Dreyer, G.~P. Salam and G.~Soyez, \emph{{The Lund Jet Plane}},
  \href{https://arxiv.org/abs/1807.04758}{{\ttfamily 1807.04758}}.

\bibitem{Guest:2016iqz}
D.~Guest, J.~Collado, P.~Baldi, S.-C. Hsu, G.~Urban and D.~Whiteson, \emph{{Jet
  Flavor Classification in High-Energy Physics with Deep Neural Networks}},
  \href{https://doi.org/10.1103/PhysRevD.94.112002}{\emph{Phys. Rev.}
  {\bfseries D94} (2016) 112002}
  [\href{https://arxiv.org/abs/1607.08633}{{\ttfamily 1607.08633}}].

\bibitem{Louppe:2017ipp}
G.~Louppe, K.~Cho, C.~Becot and K.~Cranmer, \emph{{QCD-Aware Recursive Neural
  Networks for Jet Physics}},
  \href{https://arxiv.org/abs/1702.00748}{{\ttfamily 1702.00748}}.

\bibitem{Cheng:2017rdo}
T.~Cheng, \emph{{Recursive Neural Networks in Quark/Gluon Tagging}},
  \href{https://doi.org/10.1007/s41781-018-0007-y}{\emph{Comput. Softw. Big
  Sci.} {\bfseries 2} (2018) 3}
  [\href{https://arxiv.org/abs/1711.02633}{{\ttfamily 1711.02633}}].

\bibitem{Egan:2017ojy}
S.~Egan, W.~Fedorko, A.~Lister, J.~Pearkes and C.~Gay, \emph{{Long Short-Term
  Memory (LSTM) networks with jet constituents for boosted top tagging at the
  LHC}},  \href{https://arxiv.org/abs/1711.09059}{{\ttfamily 1711.09059}}.

\bibitem{Fraser:2018ieu}
K.~Fraser and M.~D. Schwartz, \emph{{Jet Charge and Machine Learning}},
  \href{https://arxiv.org/abs/1803.08066}{{\ttfamily 1803.08066}}.

\bibitem{Almeida:2015jua}
L.~G. Almeida, M.~Backović, M.~Cliche, S.~J. Lee and M.~Perelstein,
  \emph{{Playing Tag with ANN: Boosted Top Identification with Pattern
  Recognition}}, \href{https://doi.org/10.1007/JHEP07(2015)086}{\emph{JHEP}
  {\bfseries 07} (2015) 086}
  [\href{https://arxiv.org/abs/1501.05968}{{\ttfamily 1501.05968}}].

\bibitem{Pearkes:2017hku}
J.~Pearkes, W.~Fedorko, A.~Lister and C.~Gay, \emph{{Jet Constituents for Deep
  Neural Network Based Top Quark Tagging}},
  \href{https://arxiv.org/abs/1704.02124}{{\ttfamily 1704.02124}}.

\bibitem{Butter:2017cot}
A.~Butter, G.~Kasieczka, T.~Plehn and M.~Russell, \emph{{Deep-learned Top
  Tagging with a Lorentz Layer}},
  \href{https://doi.org/10.21468/SciPostPhys.5.3.028}{\emph{SciPost Phys.}
  {\bfseries 5} (2018) 028} [\href{https://arxiv.org/abs/1707.08966}{{\ttfamily
  1707.08966}}].

\bibitem{Roxlo:2018adx}
T.~Roxlo and M.~Reece, \emph{{Opening the black box of neural nets: case
  studies in stop/top discrimination}},
  \href{https://arxiv.org/abs/1804.09278}{{\ttfamily 1804.09278}}.

\bibitem{Datta:2017rhs}
K.~Datta and A.~Larkoski, \emph{{How Much Information is in a Jet?}},
  \href{https://doi.org/10.1007/JHEP06(2017)073}{\emph{JHEP} {\bfseries 06}
  (2017) 073} [\href{https://arxiv.org/abs/1704.08249}{{\ttfamily
  1704.08249}}].

\bibitem{Aguilar-Saavedra:2017rzt}
J.~A. Aguilar-Saavedra, J.~H. Collins and R.~K. Mishra, \emph{{A generic
  anti-QCD jet tagger}},
  \href{https://doi.org/10.1007/JHEP11(2017)163}{\emph{JHEP} {\bfseries 11}
  (2017) 163} [\href{https://arxiv.org/abs/1709.01087}{{\ttfamily
  1709.01087}}].

\bibitem{Luo:2017ncs}
H.~Luo, M.-x. Luo, K.~Wang, T.~Xu and G.~Zhu, \emph{{Quark jet versus gluon
  jet: deep neural networks with high-level features}},
  \href{https://arxiv.org/abs/1712.03634}{{\ttfamily 1712.03634}}.

\bibitem{Moore:2018lsr}
L.~Moore, K.~Nordstr{\"o}m, S.~Varma and M.~Fairbairn, \emph{{Reports of My
  Demise Are Greatly Exaggerated: $N$-subjettiness Taggers Take On Jet
  Images}},  \href{https://arxiv.org/abs/1807.04769}{{\ttfamily 1807.04769}}.

\bibitem{Datta:2017lxt}
K.~Datta and A.~J. Larkoski, \emph{{Novel Jet Observables from Machine
  Learning}}, \href{https://doi.org/10.1007/JHEP03(2018)086}{\emph{JHEP}
  {\bfseries 03} (2018) 086}
  [\href{https://arxiv.org/abs/1710.01305}{{\ttfamily 1710.01305}}].

\bibitem{Komiske:2017aww}
P.~T. Komiske, E.~M. Metodiev and J.~Thaler, \emph{{Energy flow polynomials: A
  complete linear basis for jet substructure}},
  \href{https://doi.org/10.1007/JHEP04(2018)013}{\emph{JHEP} {\bfseries 04}
  (2018) 013} [\href{https://arxiv.org/abs/1712.07124}{{\ttfamily
  1712.07124}}].

\bibitem{Komiske:2017ubm}
P.~T. Komiske, E.~M. Metodiev, B.~Nachman and M.~D. Schwartz, \emph{{Pileup
  Mitigation with Machine Learning (PUMML)}},
  \href{https://doi.org/10.1007/JHEP12(2017)051}{\emph{JHEP} {\bfseries 12}
  (2017) 051} [\href{https://arxiv.org/abs/1707.08600}{{\ttfamily
  1707.08600}}].

\bibitem{Collins:2018epr}
J.~H. Collins, K.~Howe and B.~Nachman, \emph{{CWoLa Hunting: Extending the Bump
  Hunt with Machine Learning}},
  \href{https://arxiv.org/abs/1805.02664}{{\ttfamily 1805.02664}}.

\bibitem{DAgnolo:2018cun}
R.~T. D'Agnolo and A.~Wulzer, \emph{{Learning New Physics from a Machine}},
  \href{https://arxiv.org/abs/1806.02350}{{\ttfamily 1806.02350}}.

\bibitem{DeSimone:2018efk}
A.~De~Simone and T.~Jacques, \emph{{Guiding New Physics Searches with
  Unsupervised Learning}},  \href{https://arxiv.org/abs/1807.06038}{{\ttfamily
  1807.06038}}.

\bibitem{Hajer:2018kqm}
J.~Hajer, Y.-Y. Li, T.~Liu and H.~Wang, \emph{{Novelty Detection Meets Collider
  Physics}},  \href{https://arxiv.org/abs/1807.10261}{{\ttfamily 1807.10261}}.

\bibitem{Farina:2018fyg}
M.~Farina, Y.~Nakai and D.~Shih, \emph{{Searching for New Physics with Deep
  Autoencoders}},  \href{https://arxiv.org/abs/1808.08992}{{\ttfamily
  1808.08992}}.

\bibitem{Heimel:2018mkt}
T.~Heimel, G.~Kasieczka, T.~Plehn and J.~M. Thompson, \emph{{QCD or What?}},
  \href{https://arxiv.org/abs/1808.08979}{{\ttfamily 1808.08979}}.

\bibitem{Brehmer:2018kdj}
J.~Brehmer, K.~Cranmer, G.~Louppe and J.~Pavez, \emph{{Constraining Effective
  Field Theories with Machine Learning}},
  \href{https://doi.org/10.1103/PhysRevLett.121.111801}{\emph{Phys. Rev. Lett.}
  {\bfseries 121} (2018) 111801}
  [\href{https://arxiv.org/abs/1805.00013}{{\ttfamily 1805.00013}}].

\bibitem{Brehmer:2018eca}
J.~Brehmer, K.~Cranmer, G.~Louppe and J.~Pavez, \emph{{A Guide to Constraining
  Effective Field Theories with Machine Learning}},
  \href{https://doi.org/10.1103/PhysRevD.98.052004}{\emph{Phys. Rev.}
  {\bfseries D98} (2018) 052004}
  [\href{https://arxiv.org/abs/1805.00020}{{\ttfamily 1805.00020}}].

\bibitem{DHondt:2018cww}
J.~D'Hondt, A.~Mariotti, K.~Mimasu, S.~Moortgat and C.~Zhang, \emph{{Learning
  to pinpoint effective operators at the LHC: a study of the $t\bar{t}b\bar{b}$
  signature}},  \href{https://arxiv.org/abs/1807.02130}{{\ttfamily
  1807.02130}}.

\bibitem{deOliveira:2017pjk}
L.~de~Oliveira, M.~Paganini and B.~Nachman, \emph{{Learning Particle Physics by
  Example: Location-Aware Generative Adversarial Networks for Physics
  Synthesis}}, \href{https://doi.org/10.1007/s41781-017-0004-6}{\emph{Comput.
  Softw. Big Sci.} {\bfseries 1} (2017) 4}
  [\href{https://arxiv.org/abs/1701.05927}{{\ttfamily 1701.05927}}].

\bibitem{Paganini:2017hrr}
M.~Paganini, L.~de~Oliveira and B.~Nachman, \emph{{Accelerating Science with
  Generative Adversarial Networks: An Application to 3D Particle Showers in
  Multilayer Calorimeters}},
  \href{https://doi.org/10.1103/PhysRevLett.120.042003}{\emph{Phys. Rev. Lett.}
  {\bfseries 120} (2018) 042003}
  [\href{https://arxiv.org/abs/1705.02355}{{\ttfamily 1705.02355}}].

\bibitem{deOliveira:2017rwa}
L.~de~Oliveira, M.~Paganini and B.~Nachman, \emph{{Controlling Physical
  Attributes in GAN-Accelerated Simulation of Electromagnetic Calorimeters}},
  in \emph{{18th International Workshop on Advanced Computing and Analysis
  Techniques in Physics Research (ACAT 2017) Seattle, WA, USA, August 21-25,
  2017}}, 2017, \href{https://arxiv.org/abs/1711.08813}{{\ttfamily
  1711.08813}}.

\bibitem{Paganini:2017dwg}
M.~Paganini, L.~de~Oliveira and B.~Nachman, \emph{{CaloGAN : Simulating 3D high
  energy particle showers in multilayer electromagnetic calorimeters with
  generative adversarial networks}},
  \href{https://doi.org/10.1103/PhysRevD.97.014021}{\emph{Phys. Rev.}
  {\bfseries D97} (2018) 014021}
  [\href{https://arxiv.org/abs/1712.10321}{{\ttfamily 1712.10321}}].

\bibitem{Andreassen:2018apy}
A.~Andreassen, I.~Feige, C.~Frye and M.~D. Schwartz, \emph{{JUNIPR: a Framework
  for Unsupervised Machine Learning in Particle Physics}},
  \href{https://arxiv.org/abs/1804.09720}{{\ttfamily 1804.09720}}.

\bibitem{Baldi:2014kfa}
P.~Baldi, P.~Sadowski and D.~Whiteson, \emph{{Searching for Exotic Particles in
  High-Energy Physics with Deep Learning}},
  \href{https://doi.org/10.1038/ncomms5308}{\emph{Nature Commun.} {\bfseries 5}
  (2014) 4308} [\href{https://arxiv.org/abs/1402.4735}{{\ttfamily 1402.4735}}].

\bibitem{Baldi:2014pta}
P.~Baldi, P.~Sadowski and D.~Whiteson, \emph{{Enhanced Higgs Boson to
  $\tau^+\tau^-$ Search with Deep Learning}},
  \href{https://doi.org/10.1103/PhysRevLett.114.111801}{\emph{Phys. Rev. Lett.}
  {\bfseries 114} (2015) 111801}
  [\href{https://arxiv.org/abs/1410.3469}{{\ttfamily 1410.3469}}].

\bibitem{Searcy:2015apa}
J.~Searcy, L.~Huang, M.-A. Pleier and J.~Zhu, \emph{{Determination of the $WW$
  polarization fractions in $pp \to W^\pm W^\pm jj$ using a deep machine
  learning technique}},
  \href{https://doi.org/10.1103/PhysRevD.93.094033}{\emph{Phys. Rev.}
  {\bfseries D93} (2016) 094033}
  [\href{https://arxiv.org/abs/1510.01691}{{\ttfamily 1510.01691}}].

\bibitem{Santos:2016kno}
R.~Santos, M.~Nguyen, J.~Webster, S.~Ryu, J.~Adelman, S.~Chekanov et~al.,
  \emph{{Machine learning techniques in searches for $t\bar{t}h$ in the $h
  \rightarrow b\bar{b}$ decay channel}},
  \href{https://doi.org/10.1088/1748-0221/12/04/P04014}{\emph{JINST} {\bfseries
  12} (2017) P04014} [\href{https://arxiv.org/abs/1610.03088}{{\ttfamily
  1610.03088}}].

\bibitem{Barberio:2017ngd}
E.~Barberio, B.~Le, E.~Richter-Was, Z.~Was, D.~Zanzi and J.~Zaremba,
  \emph{{Deep learning approach to the Higgs boson CP measurement in $H \to
  \tau \tau$ decay and associated systematics}},
  \href{https://doi.org/10.1103/PhysRevD.96.073002}{\emph{Phys. Rev.}
  {\bfseries D96} (2017) 073002}
  [\href{https://arxiv.org/abs/1706.07983}{{\ttfamily 1706.07983}}].

\bibitem{Duarte:2018ite}
J.~Duarte et~al., \emph{{Fast inference of deep neural networks in FPGAs for
  particle physics}},
  \href{https://doi.org/10.1088/1748-0221/13/07/P07027}{\emph{JINST} {\bfseries
  13} (2018) P07027} [\href{https://arxiv.org/abs/1804.06913}{{\ttfamily
  1804.06913}}].

\bibitem{Abdughani:2018wrw}
M.~Abdughani, J.~Ren, L.~Wu and J.~M. Yang, \emph{{Probing stop with graph
  neural network at the LHC}},
  \href{https://arxiv.org/abs/1807.09088}{{\ttfamily 1807.09088}}.

\bibitem{Lin:2018cin}
J.~Lin, M.~Freytsis, I.~Moult and B.~Nachman, \emph{{Boosting $H\to b\bar b$
  with Machine Learning}},  \href{https://arxiv.org/abs/1807.10768}{{\ttfamily
  1807.10768}}.

\bibitem{Lai:2018ixk}
Y.~S. Lai, \emph{{Automated Discovery of Jet Substructure Analyses}},
  \href{https://arxiv.org/abs/1810.00835}{{\ttfamily 1810.00835}}.

\bibitem{Larkoski:2017jix}
A.~J. Larkoski, I.~Moult and B.~Nachman, \emph{{Jet Substructure at the Large
  Hadron Collider: A Review of Recent Advances in Theory and Machine
  Learning}},  \href{https://arxiv.org/abs/1709.04464}{{\ttfamily 1709.04464}}.

\bibitem{Guest:2018yhq}
D.~Guest, K.~Cranmer and D.~Whiteson, \emph{{Deep Learning and its Application
  to LHC Physics}},  \href{https://arxiv.org/abs/1806.11484}{{\ttfamily
  1806.11484}}.

\bibitem{Albertsson:2018maf}
K.~Albertsson et~al., \emph{{Machine Learning in High Energy Physics Community
  White Paper}},  \href{https://arxiv.org/abs/1807.02876}{{\ttfamily
  1807.02876}}.

\bibitem{Radovic:2018dip}
A.~Radovic, M.~Williams, D.~Rousseau, M.~Kagan, D.~Bonacorsi, A.~Himmel et~al.,
  \emph{{Machine learning at the energy and intensity frontiers of particle
  physics}}, \href{https://doi.org/10.1038/s41586-018-0361-2}{\emph{Nature}
  {\bfseries 560} (2018) 41}.

\bibitem{Sadowski2018}
P.~Sadowski and P.~Baldi, \emph{Deep Learning in the Natural Sciences:
  Applications to Physics}, pp.~269--297.
\newblock Springer International Publishing, Cham, 2018.

\bibitem{DBLP:conf/acl/IyyerMBD15}
M.~Iyyer, V.~Manjunatha, J.~L. Boyd{-}Graber and H.~D. III, \emph{Deep
  unordered composition rivals syntactic methods for text classification},  in
  \emph{Proceedings of the 53rd Annual Meeting of the Association for
  Computational Linguistics and the 7th International Joint Conference on
  Natural Language Processing of the Asian Federation of Natural Language
  Processing, {ACL} 2015, July 26-31, 2015, Beijing, China, Volume 1: Long
  Papers}, pp.~1681--1691, 2015.

\bibitem{DBLP:conf/cvpr/QiSMG17}
C.~R. Qi, H.~Su, K.~Mo and L.~J. Guibas, \emph{Pointnet: Deep learning on point
  sets for 3d classification and segmentation},  in \emph{2017 {IEEE}
  Conference on Computer Vision and Pattern Recognition, {CVPR} 2017, Honolulu,
  HI, USA, July 21-26, 2017}, pp.~77--85, 2017,
  \href{https://doi.org/10.1109/CVPR.2017.16}{DOI}.

\bibitem{DBLP:conf/iccv/RezatofighiGMAD17}
S.~H. Rezatofighi, V.~K.~B. G, A.~Milan, E.~Abbasnejad, A.~R. Dick and I.~D.
  Reid, \emph{Deepsetnet: Predicting sets with deep neural networks},  in
  \emph{{IEEE} International Conference on Computer Vision, {ICCV} 2017,
  Venice, Italy, October 22-29, 2017}, pp.~5257--5266, 2017,
  \href{https://doi.org/10.1109/ICCV.2017.561}{DOI}.

\bibitem{DBLP:conf/nips/QiYSG17}
C.~R. Qi, L.~Yi, H.~Su and L.~J. Guibas, \emph{Pointnet++: Deep hierarchical
  feature learning on point sets in a metric space},  in \emph{Advances in
  Neural Information Processing Systems 30: Annual Conference on Neural
  Information Processing Systems 2017, 4-9 December 2017, Long Beach, CA,
  {USA}}, pp.~5105--5114, 2017,
  \href{https://arxiv.org/abs/1706.02413}{{\ttfamily 1706.02413}}.

\bibitem{DBLP:conf/nips/ZaheerKRPSS17}
M.~Zaheer, S.~Kottur, S.~Ravanbakhsh, B.~P{\'{o}}czos, R.~R. Salakhutdinov and
  A.~J. Smola, \emph{Deep sets},  in \emph{Advances in Neural Information
  Processing Systems 30: Annual Conference on Neural Information Processing
  Systems 2017, 4-9 December 2017, Long Beach, CA, {USA}}, pp.~3394--3404,
  2017, \href{https://arxiv.org/abs/1703.06114}{{\ttfamily 1703.06114}}.

\bibitem{DBLP:journals/corr/abs-1709-03019}
A.~Gardner, J.~Kanno, C.~A. Duncan and R.~R. Selmic, \emph{Classifying
  unordered feature sets with convolutional deep averaging networks},
  \href{https://arxiv.org/abs/1709.03019}{{\ttfamily 1709.03019}}.

\bibitem{DBLP:journals/corr/abs-1712-07262}
Y.~Yang, C.~Feng, Y.~Shen and D.~Tian, \emph{Foldingnet: Interpretable
  unsupervised learning on 3d point clouds},
  \href{https://arxiv.org/abs/1712.07262}{{\ttfamily 1712.07262}}.

\bibitem{DBLP:conf/aaai/RezatofighiMSD018}
S.~H. Rezatofighi, A.~Milan, Q.~Shi, A.~R. Dick and I.~D. Reid, \emph{Joint
  learning of set cardinality and state distribution},  in \emph{Proceedings of
  the Thirty-Second {AAAI} Conference on Artificial Intelligence, New Orleans,
  Louisiana, USA, February 2-7, 2018}, 2018,
  \href{https://arxiv.org/abs/1709.04093}{{\ttfamily 1709.04093}}.

\bibitem{DBLP:journals/corr/abs-1805-00613}
S.~H. Rezatofighi, R.~Kaskman, F.~T. Motlagh, Q.~Shi, D.~Cremers,
  L.~Leal{-}Taix{\'{e}} et~al., \emph{Deep perm-set net: Learn to predict sets
  with unknown permutation and cardinality using deep neural networks},
  \href{https://arxiv.org/abs/1805.00613}{{\ttfamily 1805.00613}}.

\bibitem{DBLP:journals/corr/abs-1806-00050}
A.~Cotter, M.~R. Gupta, H.~Jiang, J.~Muller, T.~Narayan, S.~Wang et~al.,
  \emph{Interpretable set functions},
  \href{https://arxiv.org/abs/1806.00050}{{\ttfamily 1806.00050}}.

\bibitem{Pumplin:1991kc}
J.~Pumplin, \emph{{How to tell quark jets from gluon jets}},
  \href{https://doi.org/10.1103/PhysRevD.44.2025}{\emph{Phys. Rev.} {\bfseries
  D44} (1991) 2025}.

\bibitem{Thaler:2010tr}
J.~Thaler and K.~Van~Tilburg, \emph{{Identifying Boosted Objects with
  N-subjettiness}}, \href{https://doi.org/10.1007/JHEP03(2011)015}{\emph{JHEP}
  {\bfseries 03} (2011) 015} [\href{https://arxiv.org/abs/1011.2268}{{\ttfamily
  1011.2268}}].

\bibitem{Thaler:2011gf}
J.~Thaler and K.~Van~Tilburg, \emph{{Maximizing Boosted Top Identification by
  Minimizing N-subjettiness}},
  \href{https://doi.org/10.1007/JHEP02(2012)093}{\emph{JHEP} {\bfseries 02}
  (2012) 093} [\href{https://arxiv.org/abs/1108.2701}{{\ttfamily 1108.2701}}].

\bibitem{Krohn:2012fg}
D.~Krohn, M.~D. Schwartz, T.~Lin and W.~J. Waalewijn, \emph{{Jet Charge at the
  LHC}}, \href{https://doi.org/10.1103/PhysRevLett.110.212001}{\emph{Phys. Rev.
  Lett.} {\bfseries 110} (2013) 212001}
  [\href{https://arxiv.org/abs/1209.2421}{{\ttfamily 1209.2421}}].

\bibitem{Chang:2013rca}
H.-M. Chang, M.~Procura, J.~Thaler and W.~J. Waalewijn, \emph{{Calculating
  Track-Based Observables for the LHC}},
  \href{https://doi.org/10.1103/PhysRevLett.111.102002}{\emph{Phys. Rev. Lett.}
  {\bfseries 111} (2013) 102002}
  [\href{https://arxiv.org/abs/1303.6637}{{\ttfamily 1303.6637}}].

\bibitem{Larkoski:2014pca}
A.~J. Larkoski, J.~Thaler and W.~J. Waalewijn, \emph{{Gaining (Mutual)
  Information about Quark/Gluon Discrimination}},
  \href{https://doi.org/10.1007/JHEP11(2014)129}{\emph{JHEP} {\bfseries 11}
  (2014) 129} [\href{https://arxiv.org/abs/1408.3122}{{\ttfamily 1408.3122}}].

\bibitem{Larkoski:2013eya}
A.~J. Larkoski, G.~P. Salam and J.~Thaler, \emph{{Energy Correlation Functions
  for Jet Substructure}},
  \href{https://doi.org/10.1007/JHEP06(2013)108}{\emph{JHEP} {\bfseries 06}
  (2013) 108} [\href{https://arxiv.org/abs/1305.0007}{{\ttfamily 1305.0007}}].

\bibitem{Moult:2016cvt}
I.~Moult, L.~Necib and J.~Thaler, \emph{{New Angles on Energy Correlation
  Functions}}, \href{https://doi.org/10.1007/JHEP12(2016)153}{\emph{JHEP}
  {\bfseries 12} (2016) 153}
  [\href{https://arxiv.org/abs/1609.07483}{{\ttfamily 1609.07483}}].

\bibitem{Larkoski:2014wba}
A.~J. Larkoski, S.~Marzani, G.~Soyez and J.~Thaler, \emph{{Soft Drop}},
  \href{https://doi.org/10.1007/JHEP05(2014)146}{\emph{JHEP} {\bfseries 05}
  (2014) 146} [\href{https://arxiv.org/abs/1402.2657}{{\ttfamily 1402.2657}}].

\bibitem{Frye:2017yrw}
C.~Frye, A.~J. Larkoski, J.~Thaler and K.~Zhou, \emph{{Casimir Meets Poisson:
  Improved Quark/Gluon Discrimination with Counting Observables}},
  \href{https://doi.org/10.1007/JHEP09(2017)083}{\emph{JHEP} {\bfseries 09}
  (2017) 083} [\href{https://arxiv.org/abs/1704.06266}{{\ttfamily
  1704.06266}}].

\bibitem{Abdesselam:2010pt}
A.~Abdesselam et~al., \emph{{Boosted objects: A Probe of beyond the Standard
  Model physics}},
  \href{https://doi.org/10.1140/epjc/s10052-011-1661-y}{\emph{Eur. Phys. J.}
  {\bfseries C71} (2011) 1661}
  [\href{https://arxiv.org/abs/1012.5412}{{\ttfamily 1012.5412}}].

\bibitem{Altheimer:2012mn}
A.~Altheimer et~al., \emph{{Jet Substructure at the Tevatron and LHC: New
  results, new tools, new benchmarks}},
  \href{https://doi.org/10.1088/0954-3899/39/6/063001}{\emph{J. Phys.}
  {\bfseries G39} (2012) 063001}
  [\href{https://arxiv.org/abs/1201.0008}{{\ttfamily 1201.0008}}].

\bibitem{Altheimer:2013yza}
A.~Altheimer et~al., \emph{{Boosted objects and jet substructure at the LHC.
  Report of BOOST2012, held at IFIC Valencia, 23rd-27th of July 2012}},
  \href{https://doi.org/10.1140/epjc/s10052-014-2792-8}{\emph{Eur. Phys. J.}
  {\bfseries C74} (2014) 2792}
  [\href{https://arxiv.org/abs/1311.2708}{{\ttfamily 1311.2708}}].

\bibitem{Adams:2015hiv}
D.~Adams et~al., \emph{{Towards an Understanding of the Correlations in Jet
  Substructure}},
  \href{https://doi.org/10.1140/epjc/s10052-015-3587-2}{\emph{Eur. Phys. J.}
  {\bfseries C75} (2015) 409}
  [\href{https://arxiv.org/abs/1504.00679}{{\ttfamily 1504.00679}}].

\bibitem{Asquith:2018igt}
L.~Asquith et~al., \emph{{Jet Substructure at the Large Hadron Collider :
  Experimental Review}},  \href{https://arxiv.org/abs/1803.06991}{{\ttfamily
  1803.06991}}.

\bibitem{Gallicchio:2011xq}
J.~Gallicchio and M.~D. Schwartz, \emph{{Quark and Gluon Tagging at the LHC}},
  \href{https://doi.org/10.1103/PhysRevLett.107.172001}{\emph{Phys. Rev. Lett.}
  {\bfseries 107} (2011) 172001}
  [\href{https://arxiv.org/abs/1106.3076}{{\ttfamily 1106.3076}}].

\bibitem{Gallicchio:2012ez}
J.~Gallicchio and M.~D. Schwartz, \emph{{Quark and Gluon Jet Substructure}},
  \href{https://doi.org/10.1007/JHEP04(2013)090}{\emph{JHEP} {\bfseries 04}
  (2013) 090} [\href{https://arxiv.org/abs/1211.7038}{{\ttfamily 1211.7038}}].

\bibitem{Aad:2014gea}
{\scshape ATLAS} collaboration, G.~Aad et~al., \emph{{Light-quark and gluon jet
  discrimination in $pp$ collisions at $\sqrt{s}=7\mathrm {\ TeV}$ with the
  ATLAS detector}},
  \href{https://doi.org/10.1140/epjc/s10052-014-3023-z}{\emph{Eur. Phys. J.}
  {\bfseries C74} (2014) 3023}
  [\href{https://arxiv.org/abs/1405.6583}{{\ttfamily 1405.6583}}].

\bibitem{Gras:2017jty}
P.~Gras, S.~H{\"o}che, D.~Kar, A.~Larkoski, L.~L{\"o}nnblad, S.~Pl{\"a}tzer
  et~al., \emph{{Systematics of quark/gluon tagging}},
  \href{https://doi.org/10.1007/JHEP07(2017)091}{\emph{JHEP} {\bfseries 07}
  (2017) 091} [\href{https://arxiv.org/abs/1704.03878}{{\ttfamily
  1704.03878}}].

\bibitem{Metodiev:2017vrx}
E.~M. Metodiev, B.~Nachman and J.~Thaler, \emph{{Classification without labels:
  Learning from mixed samples in high energy physics}},
  \href{https://doi.org/10.1007/JHEP10(2017)174}{\emph{JHEP} {\bfseries 10}
  (2017) 174} [\href{https://arxiv.org/abs/1708.02949}{{\ttfamily
  1708.02949}}].

\bibitem{Komiske:2018oaa}
P.~T. Komiske, E.~M. Metodiev, B.~Nachman and M.~D. Schwartz, \emph{{Learning
  to classify from impure samples with high-dimensional data}},
  \href{https://doi.org/10.1103/PhysRevD.98.011502}{\emph{Phys. Rev.}
  {\bfseries D98} (2018) 011502}
  [\href{https://arxiv.org/abs/1801.10158}{{\ttfamily 1801.10158}}].

\bibitem{Komiske:2018vkc}
P.~T. Komiske, E.~M. Metodiev and J.~Thaler, \emph{{An operational definition
  of quark and gluon jets}},
  \href{https://arxiv.org/abs/1809.01140}{{\ttfamily 1809.01140}}.

\bibitem{energyflow}
``{EnergyFlow}.'' \url{https://energyflow.network}.

\bibitem{Stone:1948gen}
M.~H. Stone, \emph{The generalized weierstrass approximation theorem},
  \href{https://doi.org/10.2307/3029750}{\emph{Mathematics Magazine} {\bfseries
  21} 237}.

\bibitem{CMS:2013kfa}
{\scshape CMS} collaboration, CMS, \emph{{Performance of quark/gluon
  discrimination in 8 TeV pp data}},  Tech. Rep. CMS-PAS-JME-13-002, CERN,
  2013.

\bibitem{Parisi:1978eg}
G.~Parisi, \emph{{Super Inclusive Cross-Sections}},
  \href{https://doi.org/10.1016/0370-2693(78)90061-8}{\emph{Phys. Lett.}
  {\bfseries 74B} (1978) 65}.

\bibitem{Kinoshita:1962ur}
T.~Kinoshita, \emph{{Mass singularities of Feynman amplitudes}},
  \href{https://doi.org/10.1063/1.1724268}{\emph{J. Math. Phys.} {\bfseries 3}
  (1962) 650}.

\bibitem{Lee:1964is}
T.~D. Lee and M.~Nauenberg, \emph{{Degenerate Systems and Mass Singularities}},
  \href{https://doi.org/10.1103/PhysRev.133.B1549}{\emph{Phys. Rev.} {\bfseries
  133} (1964) B1549}.

\bibitem{sterman1995handbook}
G.~Sterman, J.~Smith, J.~C. Collins, J.~Whitmore, R.~Brock, J.~Huston et~al.,
  \emph{Handbook of perturbative qcd}, {\emph{Reviews of Modern Physics}
  {\bfseries 67} (1995) 157}.

\bibitem{Weinberg:1995mt}
S.~Weinberg, \emph{{The Quantum theory of fields. Vol. 1: Foundations}}.
  Cambridge University Press, 2005.

\bibitem{Larkoski:2013paa}
A.~J. Larkoski and J.~Thaler, \emph{{Unsafe but Calculable: Ratios of
  Angularities in Perturbative QCD}},
  \href{https://doi.org/10.1007/JHEP09(2013)137}{\emph{JHEP} {\bfseries 09}
  (2013) 137} [\href{https://arxiv.org/abs/1307.1699}{{\ttfamily 1307.1699}}].

\bibitem{Larkoski:2015lea}
A.~J. Larkoski, S.~Marzani and J.~Thaler, \emph{{Sudakov Safety in Perturbative
  QCD}}, \href{https://doi.org/10.1103/PhysRevD.91.111501}{\emph{Phys. Rev.}
  {\bfseries D91} (2015) 111501}
  [\href{https://arxiv.org/abs/1502.01719}{{\ttfamily 1502.01719}}].

\bibitem{Tkachov:1995kk}
F.~V. Tkachov, \emph{{Measuring multi - jet structure of hadronic energy flow
  or What is a jet?}},
  \href{https://doi.org/10.1142/S0217751X97002899}{\emph{Int. J. Mod. Phys.}
  {\bfseries A12} (1997) 5411}
  [\href{https://arxiv.org/abs/hep-ph/9601308}{{\ttfamily hep-ph/9601308}}].

\bibitem{Sveshnikov:1995vi}
N.~A. Sveshnikov and F.~V. Tkachov, \emph{{Jets and quantum field theory}},
  \href{https://doi.org/10.1016/0370-2693(96)00558-8}{\emph{Phys. Lett.}
  {\bfseries B382} (1996) 403}
  [\href{https://arxiv.org/abs/hep-ph/9512370}{{\ttfamily hep-ph/9512370}}].

\bibitem{Cherzor:1997ak}
P.~S. Cherzor and N.~A. Sveshnikov, \emph{{Jet observables and energy momentum
  tensor}},  in \emph{{Quantum field theory and high-energy physics.
  Proceedings, Workshop, QFTHEP'97, Samara, Russia, September 4-10, 1997}},
  pp.~402--407, 1997, \href{https://arxiv.org/abs/hep-ph/9710349}{{\ttfamily
  hep-ph/9710349}}.

\bibitem{Fox:1978vu}
G.~C. Fox and S.~Wolfram, \emph{{Observables for the Analysis of Event Shapes
  in e+ e- Annihilation and Other Processes}},
  \href{https://doi.org/10.1103/PhysRevLett.41.1581}{\emph{Phys. Rev. Lett.}
  {\bfseries 41} (1978) 1581}.

\bibitem{Donoghue:1979vi}
J.~F. Donoghue, F.~E. Low and S.-Y. Pi, \emph{{Tensor Analysis of Hadronic Jets
  in Quantum Chromodynamics}},
  \href{https://doi.org/10.1103/PhysRevD.20.2759}{\emph{Phys. Rev.} {\bfseries
  D20} (1979) 2759}.

\bibitem{GurAri:2011vx}
G.~Gur-Ari, M.~Papucci and G.~Perez, \emph{{Classification of Energy Flow
  Observables in Narrow Jets}},
  \href{https://arxiv.org/abs/1101.2905}{{\ttfamily 1101.2905}}.

\bibitem{Tanabashi:2018oca}
{\scshape ParticleDataGroup} collaboration, M.~Tanabashi et~al., \emph{{Review
  of Particle Physics}},
  \href{https://doi.org/10.1103/PhysRevD.98.030001}{\emph{Phys. Rev.}
  {\bfseries D98} (2018) 030001}.

\bibitem{relu}
V.~Nair and G.~E. Hinton, \emph{Rectified linear units improve restricted
  boltzmann machines},  in \emph{Proceedings of the 27th international
  conference on machine learning (ICML-10)}, pp.~807--814, 2010.

\bibitem{heuniform}
K.~He, X.~Zhang, S.~Ren and J.~Sun, \emph{Delving deep into rectifiers:
  Surpassing human-level performance on imagenet classification},  in
  \emph{Proceedings of the IEEE international conference on computer vision},
  pp.~1026--1034, 2015.

\bibitem{Sjostrand:2006za}
T.~Sjostrand, S.~Mrenna and P.~Z. Skands, \emph{{PYTHIA 6.4 Physics and
  Manual}}, \href{https://doi.org/10.1088/1126-6708/2006/05/026}{\emph{JHEP}
  {\bfseries 05} (2006) 026}
  [\href{https://arxiv.org/abs/hep-ph/0603175}{{\ttfamily hep-ph/0603175}}].

\bibitem{Sjostrand:2014zea}
T.~Sj{\"o}strand, S.~Ask, J.~R. Christiansen, R.~Corke, N.~Desai, P.~Ilten
  et~al., \emph{{An Introduction to PYTHIA 8.2}},
  \href{https://doi.org/10.1016/j.cpc.2015.01.024}{\emph{Comput. Phys. Commun.}
  {\bfseries 191} (2015) 159}
  [\href{https://arxiv.org/abs/1410.3012}{{\ttfamily 1410.3012}}].

\bibitem{Cacciari:2008gp}
M.~Cacciari, G.~P. Salam and G.~Soyez, \emph{{The Anti-k(t) jet clustering
  algorithm}}, \href{https://doi.org/10.1088/1126-6708/2008/04/063}{\emph{JHEP}
  {\bfseries 04} (2008) 063} [\href{https://arxiv.org/abs/0802.1189}{{\ttfamily
  0802.1189}}].

\bibitem{Cacciari:2011ma}
M.~Cacciari, G.~P. Salam and G.~Soyez, \emph{{FastJet User Manual}},
  \href{https://doi.org/10.1140/epjc/s10052-012-1896-2}{\emph{Eur. Phys. J.}
  {\bfseries C72} (2012) 1896}
  [\href{https://arxiv.org/abs/1111.6097}{{\ttfamily 1111.6097}}].

\bibitem{Dery:2017fap}
L.~M. Dery, B.~Nachman, F.~Rubbo and A.~Schwartzman, \emph{{Weakly Supervised
  Classification in High Energy Physics}},
  \href{https://doi.org/10.1007/JHEP05(2017)145}{\emph{JHEP} {\bfseries 05}
  (2017) 145} [\href{https://arxiv.org/abs/1702.00414}{{\ttfamily
  1702.00414}}].

\bibitem{Cohen:2017exh}
T.~Cohen, M.~Freytsis and B.~Ostdiek, \emph{{(Machine) Learning to Do More with
  Less}}, \href{https://doi.org/10.1007/JHEP02(2018)034}{\emph{JHEP} {\bfseries
  02} (2018) 034} [\href{https://arxiv.org/abs/1706.09451}{{\ttfamily
  1706.09451}}].

\bibitem{blanchard2016classification}
G.~Blanchard, M.~Flaska, G.~Handy, S.~Pozzi and C.~Scott, \emph{Classification
  with asymmetric label noise: Consistency and maximal denoising},
  \href{https://doi.org/10.1214/16-EJS1193}{\emph{Electronic Journal of
  Statistics} {\bfseries 10} (2016) 2780}
  [\href{https://arxiv.org/abs/1303.1208}{{\ttfamily 1303.1208}}].

\bibitem{blanchard2018corrigendum}
G.~Blanchard and C.~Scott, \emph{Corrigendum to ``classification with
  asymmetric label noise: Consistency and maximal denoising''},
  \href{https://doi.org/10.1214/18-EJS1422}{\emph{Electronic Journal of
  Statistics} {\bfseries 12} (2018) 1779}.

\bibitem{Metodiev:2018ftz}
E.~M. Metodiev and J.~Thaler, \emph{{Jet Topics: Disentangling Quarks and
  Gluons at Colliders}},
  \href{https://doi.org/10.1103/PhysRevLett.120.241602}{\emph{Phys. Rev. Lett.}
  {\bfseries 120} (2018) 241602}
  [\href{https://arxiv.org/abs/1802.00008}{{\ttfamily 1802.00008}}].

\bibitem{Larkoski:2014uqa}
A.~J. Larkoski, D.~Neill and J.~Thaler, \emph{{Jet Shapes with the Broadening
  Axis}}, \href{https://doi.org/10.1007/JHEP04(2014)017}{\emph{JHEP} {\bfseries
  04} (2014) 017} [\href{https://arxiv.org/abs/1401.2158}{{\ttfamily
  1401.2158}}].

\bibitem{Likhomanenko:2015aba}
T.~Likhomanenko, P.~Ilten, E.~Khairullin, A.~Rogozhnikov, A.~Ustyuzhanin and
  M.~Williams, \emph{{LHCb Topological Trigger Reoptimization}},
  \href{https://doi.org/10.1088/1742-6596/664/8/082025}{\emph{J. Phys. Conf.
  Ser.} {\bfseries 664} (2015) 082025}
  [\href{https://arxiv.org/abs/1510.00572}{{\ttfamily 1510.00572}}].

\bibitem{scikit-learn}
F.~Pedregosa, G.~Varoquaux, A.~Gramfort, V.~Michel, B.~Thirion, O.~Grisel
  et~al., \emph{Scikit-learn: Machine learning in python}, {\emph{Journal of
  Machine Learning Research} {\bfseries 12} (2011) 2825}.

\bibitem{Larkoski:2014tva}
A.~J. Larkoski, I.~Moult and D.~Neill, \emph{{Toward Multi-Differential Cross
  Sections: Measuring Two Angularities on a Single Jet}},
  \href{https://doi.org/10.1007/JHEP09(2014)046}{\emph{JHEP} {\bfseries 09}
  (2014) 046} [\href{https://arxiv.org/abs/1401.4458}{{\ttfamily 1401.4458}}].

\bibitem{Procura:2018zpn}
M.~Procura, W.~J. Waalewijn and L.~Zeune, \emph{{Joint resummation of two
  angularities at next-to-next-to-leading logarithmic order}},
  \href{https://arxiv.org/abs/1806.10622}{{\ttfamily 1806.10622}}.

\bibitem{Bertolini:2014bba}
D.~Bertolini, P.~Harris, M.~Low and N.~Tran, \emph{{Pileup Per Particle
  Identification}}, \href{https://doi.org/10.1007/JHEP10(2014)059}{\emph{JHEP}
  {\bfseries 10} (2014) 059} [\href{https://arxiv.org/abs/1407.6013}{{\ttfamily
  1407.6013}}].

\bibitem{fjcontrib}
``Fastjet contrib.'' \url{https://fastjet.hepforge.org/contrib/}.

\bibitem{keras}
F.~Chollet, ``Keras.'' \url{https://github.com/fchollet/keras}, 2015.

\bibitem{tensorflow}
M.~Abadi, P.~Barham, J.~Chen, Z.~Chen, A.~Davis, J.~Dean et~al.,
  \emph{Tensorflow: A system for large-scale machine learning.},  in
  \emph{OSDI}, vol.~16, pp.~265--283, 2016.

\bibitem{numpy}
T.~Oliphant, \emph{Guide to {NumPy}}. Trelgol Publishing, 2006.

\bibitem{adam}
D.~Kingma and J.~Ba, \emph{Adam: A method for stochastic optimization},
  \href{https://arxiv.org/abs/1412.6980}{{\ttfamily 1412.6980}}.

\bibitem{deFavereau:2013fsa}
{\scshape DELPHES 3} collaboration, J.~de~Favereau, C.~Delaere, P.~Demin,
  A.~Giammanco, V.~Lema{\^\i}tre, A.~Mertens et~al., \emph{{DELPHES 3, A
  modular framework for fast simulation of a generic collider experiment}},
  \href{https://doi.org/10.1007/JHEP02(2014)057}{\emph{JHEP} {\bfseries 02}
  (2014) 057} [\href{https://arxiv.org/abs/1307.6346}{{\ttfamily 1307.6346}}].

\end{thebibliography}\endgroup

\end{document}